\documentclass[paperpaper,superscriptaddress,english,floatfix, showpacs, amsfonts, amssymb]{revtex4}
\usepackage{epsfig,graphicx,psfrag,amsmath,amssymb,float}
\usepackage[T1]{fontenc}
\usepackage[latin9]{inputenc}
\usepackage{amsmath}
\usepackage{amssymb}
\usepackage{bbold}
\usepackage{amscd}
\usepackage{bm}
\usepackage{psfrag}
\usepackage{epsfig}
\usepackage{graphicx}
\usepackage{dcolumn}
\usepackage{bm}
\usepackage{subfigure}

\usepackage{babel}

\begin{document}
\title{One-particle spectral function singularities in a one-dimensional gas of spin-$1/2$ fermions with repulsive delta-function interaction}
\author{T. \v{C}ade\v{z}}
\affiliation{Beijing Computational Science Research Center, Beijing 100193, China}
\affiliation{Center of Physics of University of Minho and University of Porto, P-4169-007 Oporto, Portugal}
\author{S. Nemati}
\affiliation{Beijing Computational Science Research Center, Beijing 100193, China}
\affiliation{Center of Physics of University of Minho and University of Porto, P-4169-007 Oporto, Portugal}
\affiliation{University of Aveiro, Department of Physics, P-3810193 Aveiro, Portugal}
\author{J. M. P. Carmelo}
\affiliation{Boston University, Department of Physics, 590 Commonwealth Ave, Boston, MA 02215, USA}
\affiliation{Massachusetts Institute of Technology, Department of Physics, Cambridge, MA 02139, USA}
\affiliation{Center of Physics of University of Minho and University of Porto, P-4169-007 Oporto, Portugal}
\affiliation{Department of Physics, University of Minho, Campus Gualtar, P-4710-057 Braga, Portugal}

\date{25 March 2019}

\begin{abstract}
The momentum, fermionic density, spin density, and interaction dependencies of the exponents
that control the $(k,\omega)$-plane singular features of the one-fermion spectral functions
of a one-dimensional gas of spin-$1/2$ fermions with repulsive delta-function interaction 
both at zero and finite magnetic field are studied in detail. 
Our results refer to energy scales beyond the reach of the low-energy Tomonaga-Luttinger liquid and
rely on the pseudofermion dynamical theory for integrable models. The one-fermion spectral weight 
distributions associated with the spectral functions studied in this paper may be observed in systems 
of spin-$1/2$ ultra-cold fermionic atoms in optical lattices.
\end{abstract}

\maketitle

\section{Introduction}
\label{Introduction}

The one-dimensional (1D) continuous fermionic gas with repulsive delta-function interaction,
which in this paper we call 1D repulsive fermion model, was 
one of the first quantum problems solved by the Bethe ansatz (BA) \cite{Bethe_31}.
This was achieved by Yang \cite{Yang_67} and by Gaudin \cite{Gaudin_67}. Yang's solution of the 1D repulsive fermion model
was actually the precursor of the BA solution of the lattice 1D Hubbard model 
by Lieb and Wu \cite{Lieb,Lieb-03,Takahashi,Martins}. 
That the latter is the simplest condensed-matter toy model for the description of the role of correlations 
in the exotic properties of 1D and quasi-1D lattice condensed matter systems 
\cite{spectral0,Dionys-87} justifies why for several decades 
it had attracted more attention than its continuous cousin, the fermionic gas with repulsive delta-function interaction.
This refers both to its metallic and Mott-Hubbard insulator phases \cite{DSF-n1}, the latter not existing
in the case of the continuous 1D repulsive fermion model.

However, in the last years the interest in that Yang-Baxter integrable model has been renewed by
its new found impact on experiments in both condensed matter physics 
and ultra-cold atomic gases \cite{Batchelor_16}. The latter have provided new opportunities
for studying 1D systems of spin-$1/2$ fermions with repulsive interaction \cite{Guan_13,Zinner_16}.
The present model can indeed be implemented with ultra-cold atoms 
\cite{Batchelor_16,Guan_13,Zinner_16,Dao_2007,Stewart_08,Clement_09,Febbri_12}. 
Ultra-cold Fermi gases trapped inside a tight atomic waveguide offer for instance the opportunity
to measure the spin-drag relaxation rate that controls the broadening of a spin packet.
It has been found that while the propagation of long-wavelength charge excitations is essentially ballistic, 
spin propagation is intrinsically damped and diffusive \cite{Polini_07,Rainis_08}.
A related interesting problem is the force applied to a spin-flipped fermion in a gas, which may lead to Bloch oscillations
of the fermion's position and velocity. The existence of such oscillations has been found crucially 
to depend on the viscous friction force exerted by the rest of the gas on the spin excitation \cite{Gangardt_09}.

The ground-state energy of the relative motion of a system of two fermions with spin up and spin down 
interacting via a delta-function potential in a 1D harmonic trap has been calculated by
combining the BA with the variational principle \cite{Rubeni_12}. Recently,
related ground-state properties of a 1D repulsive Fermi gas subjected to a commensurate periodic optical lattice
of arbitrary intensity have been investigated by the use of continuous-space quantum Monte-Carlo 
simulations \cite{Pilati_17}. The thermodynamic properties of the model have also been recently studied using a specific lattice
embedding and the quantum transfer matrix. That allowed the derivation of an exact system 
of only two nonlinear integral equations for the thermodynamics of the homogeneous model,
which is valid for all temperatures and values of the chemical potential, magnetic field, and 
repulsive interaction \cite{Patu_16}.

Another issue that has contributed to the renewed interest in the 1D repulsive fermion model
is the relation of integrable Yang-Baxter equation fermionic models to topology and
quantum computing \cite{Kauffman_18}. That equation can act as a parametric 
two-body quantum gate \cite{Zhang_12}. An experimental realization
of the Yang-Baxter equation through a Nuclear Magnetic Resonance interferometric setup
has actually verified its validity \cite{Vind_16}. 

The model dynamical properties is another problem of scientific interest.
The behavior of dynamic structure factors of fermionic models
differs dramatically for integrable and non-integrable models \cite{Tatche_11}.
The mobile quantum impurity model (MQIM) has been used to derive
dynamic response functions of interacting one-dimensional spin-1/2 fermions
\cite{Schmidt_10,Imambekov_12}. An approximation relying on the bosonization technique 
and diagonalizing the model to two Tomonaga-Luttinger liquid (TLL) Hamiltonians, was used in Ref. \cite{Orignac_11} to obtain 
some general expressions for the spectral function at zero spin density, 
expressed in terms of the Gauss hypergeometric function. The up-spin and down-spin one-fermion spectral functions of 
the present model in zero magnetic field and in a finite field have not been detailed studied though. 

In this paper a systematic and detailed study of the momentum dependent exponents and
energy spectra that control the line shape near the high-energy singularities of {\it both} (i) the one-fermion removal 
and addition spectral functions at zero magnetic field and (ii) the up-spin and down-spin 
one-fermion removal and addition spectral functions at finite magnetic field is conducted.
(Our designation {\it high energy} refers to energy scales beyond the
reach of the low-energy TLL \cite{Tomonaga-50,Luttinger-63,Solyom-79,Voit,Sutherland-04}.)

The 1D repulsive fermion model describes $N=N_{\uparrow}+N_{\downarrow}$ spin-$1/2$ fermions, $N_{\uparrow}$ 
with up-spin projection and $N_{\downarrow}$ with down-spin projection, which in real space have a repulsive delta-function interaction.
The model Hamiltonian in a chemical potential $\mu$ and magnetic field $h$ is in units of $\hbar =1$ and bare mass $m=1/2$ given by,
\begin{equation}
\hat{H} = - \sum_{j=1}^N{\partial^2\over \partial x_j^2} + 2\,{\cal{C}}\sum_{j'>j}\delta (x_j - x_{j'}) - \mu\,N 
+ 2\mu_B h\,{\hat{S}}^{z}\hspace{0.20cm}{\rm where}\hspace{0.20cm}
{\hat{S}}^{z} = -{1\over 2}({\hat{N}}_{\uparrow}-{\hat{N}}_{\downarrow}) \, .
\label{H}
\end{equation}
Here $\delta (x)$ denotes the Dirac delta-function distribution,
$x_j$ is the position of the $j$-$th$ fermion, ${\cal{C}}>0$ gives the strength of the repulsive interaction,
$\mu_B$ is the Bohr magneton, and the fermion number operator reads 
${\hat{N}}=\sum_{\sigma=\uparrow ,\downarrow }\,\hat{N}_{\sigma}$. Moreover, ${\hat{S}}^{z}$ is the 
diagonal generator of the Hamiltonian ${\hat{H}}$ global spin $SU(2)$ symmetry algebra.
The lowest-weight states (LWSs) and highest-weight states (HWSs) of that
$SU(2)$ symmetry algebra have numbers $S = - S^{z}$ and $S = S^{z}$, respectively, where
$S$ is the states spin and $S^{z}$ is the corresponding projection. The latter is an 
eigenvalue of the spin operator given in Eq. (\ref{H}). 

On the one hand, at zero magnetic field, $h=0$, and thus zero spin density, $m=0$, our study focuses on the one-fermion 
spectral function,
\begin{equation}
B_{\gamma} (k,\omega) = \sum_{\sigma=\uparrow,\downarrow} B_{\sigma,\gamma} (k,\omega) 
\hspace{0.20cm}{\rm for}\hspace{0.20cm} \gamma\,\omega \geq 0 \, ,
\label{Bkomega-m0}
\end{equation}
where,
\begin{equation}
\gamma = -1 \hspace{0.2cm}{\rm for}\hspace{0.2cm}{\rm fermion}\hspace{0.2cm}{\rm removal}
\hspace{0.20cm}{\rm and}\hspace{0.20cm}\gamma = + 1\hspace{0.20cm}{\rm for}\hspace{0.2cm}
{\rm fermion}\hspace{0.2cm}{\rm addition} \, .
\label{c0RA}
\end{equation}

On the other hand, for $h\neq 0$ and $m> 0$ it addresses the up-spin and down-spin one-fermion removal
and addition spectral functions 
$B_{\sigma,\gamma} (k,\,\omega)$ on the right-hand side of Eq. (\ref{Bkomega-m0}), which read,
\begin{eqnarray}
B_{\sigma,-1} (k,\,\omega) & = & \sum_{\nu^-}
\vert\langle\nu^-\vert\, c_{k,\sigma} \vert \,GS\rangle\vert^2 \,\delta (\omega
+ (E_{\nu^-}^{N_{\sigma}-1}-E_{GS}^{N_{\sigma}})) \hspace{0.20cm}{\rm for}\hspace{0.20cm} \omega \leq 0
\nonumber \\
B_{\sigma,+1} (k,\,\omega) & = & \sum_{\nu^+}
\vert\langle\nu^+\vert\, c^{\dagger}_{k,\sigma} \vert
\,GS\rangle\vert^2 \,\delta (\omega - (E_{\nu^+}^{N_{\sigma}+1}-E_{GS}^{N_{\sigma}}))  
\hspace{0.20cm}{\rm for}\hspace{0.20cm} \omega \geq 0 \, ,
\label{Bkomega}
\end{eqnarray}
Here $c_{k,\sigma}$ and $c^{\dagger}_{k,\sigma}$ are up-spin and down-spin fermion
annihilation and creation operators, respectively, of momentum $k$ and $\vert GS\rangle$ denotes the
initial $N_{\sigma}$-fermion ground state of energy $E_{GS}^{N_{\sigma}}$. The $\nu^-$ and $\nu^+$
summations run over the $N_{\sigma}-1$ and $N_{\sigma}+1$-fermion excited 
energy eigenstates, respectively, and $E_{\nu^-}^{N_{\sigma}-1}$ and 
$E_{\nu^+}^{N_{\sigma}+1}$ are the corresponding energies.

Our main goal is deriving the $(k,\omega)$-plane line shape near the singularities of the spectral functions
in Eq. (\ref{Bkomega-m0}) at zero spin density, $m=0$, and in Eq. (\ref{Bkomega}) for $m> 0$. 
This includes the detailed study of the dependence of the exponents that control that line shape on the 
excitation momentum, repulsive interaction ${\cal{C}}$, fermionic density $n\in [0,\infty[$, and spin-density $m \in [0,n]$. 
For such spin densities, the model ground states 
are LWSs of the spin $SU(2)$ symmetry algebra. Hence we use the LWS 
formulation of the model BA solution. 

The high-energy dynamical correlation functions of some integrable models 
\cite{Altshuler,Konik,Fuksa_17,Pakuliaka_15} can be studied by the form-factor approach.
Form factors of the 1D repulsive fermion model up-spin and down-spin fermion creation and 
annihilation operators involved in the spectral functions studied here remains though 
an unsolved problem. 

The present study of the momentum dependent exponents that control the line shape near the singularities
of the one-fermion spectral functions, Eqs. (\ref{Bkomega-m0}) and (\ref{Bkomega}), relies on the 
pseudofermion dynamical theory (PDT) introduced in Ref. \cite{Carmelo_05} for the related lattice 
1D Hubbard model, which applies to other integrable systems as well
\cite{Carmelo_18,Carmelo_16,Carmelo_15}, including the present 1D repulsive fermion model. 
For the latter we use in our study an exact representation suitable to the PDT in terms of {\it pseudofermions} of that
model BA solution in the subspace spanned by the ground state and one-fermion excited energy eigenstates. 
The pseudofermions are generated by a unitary transformation from corresponding pseudoparticles \cite{Carmelo_18,Carmelo_17}. 
For simplicity, in this article the pseudofermions are
called charge or spin particles, depending on the BA branch they refer to. 

The MQIM \cite{Imambekov_12} applies both to integrable and non-integrable models.
The previously introduced PDT  \cite{Carmelo_05} applies only to integrable models. In the case of the
latter models, the PDT and MQIM lead to exactly the same momentum dependent exponents in
the power-law expressions of the spectral functions near their edges of support.
Indeed, for integrable models the two methods have been shown to describe exactly the same fractionalized particles 
microscopic mechanisms \cite{Carmelo_18,Carmelo_16}.

The remainder of the paper is organized as follows. The related $c$ and $s$ pseudoparticle and $c$ and $s$ particle representations,
respectively, and corresponding BA and PDT basic quantities needed for the study of the up-spin and down-spin 
one-fermion spectral weights is the topic addressed in Section \ref{BAPDT}. 
In Section \ref{SFEX} the general types of one-fermion spectral singularities studied in this paper are reported.
The line shape near specific $(k,\omega)$-plane one-fermion removal and addition branch and boundary lines
singularities of the spectral functions, Eqs. (\ref{Bkomega-m0}) and (\ref{Bkomega}), is then studied in Section \ref{Specific}.
The low-energy TLL limit of the PDT one-fermion spectral function expressions near their singularities is the issue 
addressed in Section \ref{LESF}. Finally, the discussion of the relevance and consequences of the results
and the concluding remarks are presented in Section \ref{Disconclu}.

\section{The $c$ and $s$ pseudoparticle and $c$ and $s$ particle representations}
\label{BAPDT} 

\subsection{The BA equations and quantum numbers}
\label{BAeqQN} 

Let $\{\vert l_{\rm r},l_{s},{\cal{C}}\rangle\}$ be the complete set of energy eigenstates of the
Hamiltonian $\hat{H}$, Eq. (\ref{H}), associated with the BA solution for ${\cal{C}}>0$. 
We call a {\it Bethe state} an energy eigenstate that is a LWS of the spin $SU(2)$ symmetry algebra,
which is here denoted by $\vert l_{\rm r},l_{s}^0,{\cal{C}}\rangle$. The ${\cal{C}}$-independent label $l_{s}$ in 
general energy eigenstates $\{\vert l_{\rm r},l_{s},{\cal{C}}\rangle\}$ is a short notation for the set of quantum numbers, 
\begin{equation}
l_{s} = S,n_{s}\hspace{0.20cm}{\rm where}\hspace{0.20cm}n_{s} = 
S+S^{z} = 0,1,..., 2S \, .
\label{etas-states-ll}
\end{equation}
For a Bethe state one then has that $n_s=0$, so that $l_{s}^0$ stands for $S,0$. 
Furthermore, the label $l_{\rm r}$ refers to the set of all remaining ${\cal{C}}$-independent quantum numbers needed
to uniquely specify an energy eigenstate $\vert l_{\rm r},l_{s},{\cal{C}}\rangle$. This refers to occupancy
configurations of BA momentum quantum numbers $q_j = {2\pi\over L}\,I^{\beta}_j$. Here $I^{\beta}_j$ are 
successive integers, $I^{\beta}_j=0,\pm 1,\pm 2,...$, or half-odd integers, $I^{\beta}_j=\pm 1/2,\pm 3/2,\pm 5/2,...$, 
according to well-defined boundary conditions. Their allowed occupancies are zero and one. The index
$\beta$ denotes several BA branches of quantum numbers. 

In the case of the up-spin and down-spin
one-fermion removal and addition spectral functions, the line-shape near their singularities does not involve at finite
magnetic field excited states described by spin complex BA rapidities. Fortunately, the finite-field quantities
lead to correct zero-spin density results in the limit of zero spin density. Hence for our study only the charge 
$\beta =c$ band and spin $\beta =s$ band momentum $q_j = {2\pi\over L}\,I^{\beta}_j$ 
branches described by real BA rapidities are needed. 

The general non-LWSs $\vert l_{\rm r},l_{s},{\cal{C}}\rangle$ for which $n_s>0$ can be generated from the corresponding 
Bethe states $\vert l_{\rm r},l_{s}^0,{\cal{C}}\rangle$ as,
\begin{equation}
\vert l_{\rm r},l_{s},{\cal{C}}\rangle = \left(\frac{1}{
\sqrt{{\cal{C}}_{s}}}({\hat{S}}^{+})^{n_s}\right)\vert l_{\rm r},l_{s}^0,{\cal{C}}\rangle
\hspace{0.20cm}{\rm where}\hspace{0.20cm}  
{\cal{C}}_{s} = (n_{s}!)\prod_{j=1}^{n_{s}}(\,2S+1-j\,) 
\hspace{0.20cm}{\rm and}\hspace{0.20cm}  n_{s}=1,...,2S \, .
\nonumber
\end{equation}
Here ${\hat{S}}^{+} = {\hat{S}}^{x}+i{\hat{S}}^{y}$, ${\hat{S}}^{x}$ and ${\hat{S}}^{y}$ are usual spin component
operators, and ${\cal{C}}_{s}$ is a  normalization constant. 

The BA equations of the 1D repulsive fermion model, Eq. (\ref{H}), in the subspace spanned by the
ground state and the one-fermion excited energy eigenstates that contribute to the spectral
weight in the vicinity of the spectral functions singularities are of the form \cite{Yang_67},
\begin{equation}
k_j  = q_j - {2\over L}\sum_{j' =1}^{N_{\downarrow}} \arctan\left({2k_j - 2\Lambda_{j'}\over {\cal{C}}}\right) 
\hspace{0.20cm}{\rm where}\hspace{0.20cm}q_j = {2\pi\over L}\,I_j^c 
\hspace{0.20cm}{\rm and}\hspace{0.20cm} j = 1,...,\infty \, ,
\label{kj}
\end{equation}
and
\begin{equation}
{2\over L}\sum_{j' =1}^{N} \arctan \left({2\Lambda_j - 2k_{j'}\over {\cal{C}}}\right) 
= q_j - {2\over L}\sum_{j' =1}^{N_{\downarrow}} \arctan \left({\Lambda_j  - \Lambda_{j'}\over {\cal{C}}}\right) 
\hspace{0.20cm}{\rm where}\hspace{0.20cm}q_j = {2\pi\over L}\,I_j^s 
\hspace{0.20cm}{\rm and}\hspace{0.20cm} j = 1,...,N_{\uparrow} \, .
\label{Lambdaj}
\end{equation}
The $c$ and $s$ band discrete momentum values in those equations,
\begin{equation}
q_j = {2\pi\over L}\,I_j^c\hspace{0.20cm}{\rm where}\hspace{0.20cm}\hspace{0.20cm} j=1,...,\infty 
\hspace{0.20cm}{\rm and}\hspace{0.20cm}  
q_j = {2\pi\over L}\,I_j^{s}\hspace{0.20cm}{\rm where}\hspace{0.20cm}j=1,...,N_{\uparrow} \, ,
\label{q-j}
\end{equation} 
respectively, are directly related to the BA solution quantum numbers $I_j^c$ and $I_j^{s}$, respectively. 
Those are such discrete momentum values in units of $2\pi/L$. They are
integers or half-odd integers according to the following boundary conditions,
\begin{eqnarray}
I_j^c & = & 0, \pm 1, \pm 2, ... \hspace{0.20cm}{\rm for}\hspace{0.20cm} N_{s} =N_{\downarrow} \hspace{0.2cm}{\rm even} 
\nonumber \\
& = & \pm 1/2, \pm 3/2, 5/2, ...  \hspace{0.20cm}{\rm for}\hspace{0.20cm}  N_{s} = N_{\downarrow} \hspace{0.2cm}{\rm odd}  \, ,
\label{I-j-c}
\end{eqnarray} 
and
\begin{eqnarray}
I_j^{s} & = & 0, \pm 1, \pm 2, ... \hspace{0.20cm}{\rm for}\hspace{0.20cm} N_c - N_{s} = N_{\uparrow}\hspace{0.2cm}{\rm odd} 
\nonumber \\
& = & \pm 1/2, \pm 3/2, 5/2, ... \hspace{0.20cm}{\rm for}\hspace{0.20cm}  N_c  - N_{s} = N_{\uparrow}\hspace{0.2cm}{\rm even}  \, ,
\label{I-j-s}
\end{eqnarray} 
respectively. Hence under transitions from the ground state to one-fermion removal or addition excited energy eigenstates  
there may occur shakeup effects involving overall $\beta =c,s$ band discrete momentum shifts, 
$q_j\rightarrow q_j + 2\pi\Phi_{\beta}^0/L$. It follows directly from the boundary conditions, Eqs. (\ref{I-j-c}) and  (\ref{I-j-s}),
that the non-scattering phase shift $2\pi\Phi_{\beta}^0$ is given by,
\begin{eqnarray}
2\pi\Phi_{c}^0 & = & 0\hspace{0.20cm}{\rm for}\hspace{0.20cm}\delta N_s = \delta N_{\downarrow}\hspace{0.20cm} {\rm even} 
\hspace{0.20cm}{\rm and}\hspace{0.20cm}2\pi\Phi_{c}^0=\pm\pi\hspace{0.20cm}{\rm for}\hspace{0.20cm}
\delta N_s = \delta N_{\downarrow}\hspace{0.20cm} {\rm odd}
\nonumber \\
2\pi\Phi_{s}^0 & = & 0\hspace{0.20cm}{\rm for}\hspace{0.20cm}\delta N_{c}-\delta N_s = \delta N_{\uparrow} 
\hspace{0.20cm}{\rm even}\hspace{0.20cm}{\rm and}\hspace{0.20cm}2\pi\Phi_{s}^0=\pm\pi\hspace{0.20cm}{\rm for}\hspace{0.20cm}
\delta N_{c}-\delta N_s = \delta N_{\uparrow}\hspace{0.20cm}{\rm odd} \, .
\label{pican}
\end{eqnarray}

The complete set of the model energy eigenstates involves those whose spin rapidities $\Lambda_j$
in Eqs. (\ref{kj}) and (\ref{Lambdaj}) are complex numbers. Fortunately, as mentioned in Sec. \ref{Introduction},
such states do not contribute to the expressions of the up-spin and down-spin one-fermion spectral functions 
in the vicinity of the singularities studied in this paper.

\subsection{The $c$ and $s$ pseudoparticle representation}
\label{PR} 

Within the pseudoparticle representation \cite{Carmelo_18}, 
the energy eigenstates are generated by exclusion-principle occupancy configurations
of $N_c=N$ charge $c$ pseudoparticles over $j=1,...,\infty$ discrete $c$ band momentum values $q_j$
and $N_s=N_{\downarrow}$ spin $s$ pseudoparticles over $j=1,...,N_{\uparrow}$ discrete $s$ band momentum values $q_j$
in Eq. (\ref{q-j}). Hence within that representation each occupied $\beta =c,s$ band discrete momentum $q_j$ corresponds to
one $\beta =c,s$ pseudoparticle. 

The $\beta =c,s$ band momentum distribution function $N_{\beta} (q_j)$ reads $N_{\beta} (q_j)=1$ and 
$N_{\beta} (q_j)=0$ for occupied and unoccupied discrete momentum values $q_j$, respectively. The BA 
equations, Eqs. (\ref{kj}) and (\ref{Lambdaj}), can then be written in a corresponding functional form as,
\begin{equation}
k (q_j)  = q_j - {2\over L}\sum_{j' =1}^{N_{\uparrow}}N_{s} (q_{j'})\arctan\left({2k (q_j) - 2\Lambda (q_{j'})\over {\cal{C}}}\right) 
\hspace{0.20cm}{\rm where}\hspace{0.20cm}q_j = {2\pi\over L}\,I_j^c 
\hspace{0.20cm}{\rm and}\hspace{0.20cm} j = 1,...,\infty \, .
\label{FUkj}
\end{equation}
and
\begin{eqnarray}
{2\over L}\sum_{j' =1}^{\infty}N_c (q_{j'}) \arctan \left({2\Lambda (q_j) - 2k (q_{j'})\over {\cal{C}}}\right) 
& = & q_j - {2\over L}\sum_{j' =1}^{N_{\uparrow}}N_{s} (q_{j'}) \arctan \left({\Lambda (q_j)  - \Lambda (q_{j'})\over {\cal{C}}}\right) 
\nonumber \\
{\rm where} & & q_j = {2\pi\over L}\,I_j^s \hspace{0.20cm}{\rm and}\hspace{0.20cm} j = 1,...,N_{\uparrow} \, .
\label{FULambdaj}
\end{eqnarray}
respectively.

For the excited energy eigenstates that contribute to the one-fermion spectral weights distributions
near singularities studied below in Section \ref{SFEX}
the numbers $N_c$ and $N_{s}$ of $c$ and $s$ pseudoparticles, respectively, and the number $N_{s}^h$
of $s$ band holes are related to those of the 
spin-$1/2$ fermions $N=N_{\uparrow}+N_{\downarrow}$, $N_{\uparrow}$, and $N_{\downarrow}$ as follows,
\begin{equation}
N_c = N\, ,\hspace{0.20cm}N_s = N_{\downarrow}\hspace{0.20cm}{\rm and}\hspace{0.20cm}
N_s^h = N_c - 2N_s = N_{\uparrow} - N_{\downarrow} \, ,
\nonumber
\end{equation}
so that,
\begin{equation}
N = N_c\, ,\hspace{0.20cm}N_{\uparrow} - N_{\downarrow} = N_s^h\, ,\hspace{0.20cm}
N_{\uparrow} = N_s + N_{s}^h\hspace{0.20cm}{\rm and}\hspace{0.20cm}
N_{\downarrow} = N_s \, .
\nonumber
\end{equation}

The general energy spectrum of the one-fermion excited energy eigenstates generated by $c$ and $s$ particle 
occupancy configurations reads,
\begin{equation}
\delta E = \sum_{\beta=c,s}\sum_{j=1}^{L_{\beta}}\varepsilon_{\beta} (q_j)\delta N_{\beta} (q_j) 
+ {1\over L}\sum_{\beta =c,s}\sum_{\beta'=c,s}\sum_{j=1}^{L_{\beta}}\sum_{j'=1}^{L_{\beta'}}
{1\over 2}\,f_{\beta\,\beta'} (q_j,q_{j'})\,\delta N_{\beta} (q_j)\delta N_{\beta'} (q_{j'})  \, .
\label{DE-fermions0}
\end{equation}
Here $L_c = L$ is the system length that is given by $L\rightarrow\infty$ in the thermodynamic limit,
$L_s = N_{\uparrow}$, and the $\beta =c,s$ momentum band distribution function deviations read,
\begin{equation}
\delta N_{\beta} (q_j)  = N_{\beta} (q_j) - N^0_{\beta} (q_j) \hspace{0.20cm}{\rm where}\hspace{0.20cm}
j = 1,...,L_{\beta} \hspace{0.20cm}{\rm and}\hspace{0.20cm} \beta = c,s \, .
\label{DNq}
\end{equation}
$N_{\beta} (q_j)$ is in this equation the $\beta =c,s$ band pseudoparticle momentum distribution function
for excited states for which the deviation $\delta N_{\beta} (q_j)$ is small and thus
in the thermodynamic limit involve a vanishing density of $\beta$ pseudoparticles.
The ground-state $\beta = c,s$ band pseudoparticle momentum distribution functions $N^0_{\beta} (q_j)$
also appearing in Eq. (\ref{DNq}) are given by,
\begin{equation}
N_c^0 (q_j) = \theta (q_j - q_{Fc}^{-})\,\theta (q_{Fc}^{+} - q_j)  
\hspace{0.20cm}{\rm and}\hspace{0.20cm}
N_{s}^0 (q_j) = \theta (q_j - q_{Fs}^{-})\,\theta (q_{Fs}^{+} - q_j)  \, ,
\label{N0q1DHm}
\end{equation}
where the distribution $\theta (x)$ reads $\theta (x)=1$ for $x> 0$ and 
$\theta (x)=0$ for $x\leq 0$. The $\beta =c,s$ Fermi points of the compact and symmetrical occupancy configurations,
Eq. (\ref{N0q1DHm}), are associated with the Fermi momentum values $q_{F\beta}^{\pm}$.
If within the thermodynamic limit we ignore unimportant $1/L$ corrections, one may consider that
$q_{F\beta}^{\pm}=\pm q_{F\beta}$ and thus that $N_{\beta}^0 (q_j) = \theta (q_{F\beta} - \vert q_j\vert)$ 
for $\beta =c,s$. For densities $0<n<\infty$ and $0<m<n$ the $\beta =c,s$ Fermi momenta
$q_{F\beta}$ are given by,
\begin{equation}
q_{Fc} = 2k_F = {\pi\over L}(N-1) \approx \pi\,n\hspace{0.20cm}{\rm and}\hspace{0.20cm}
q_{Fs} =k_{F\downarrow} = {\pi\over L}(N_{\downarrow} -1) \approx
\pi\,n_{\downarrow} = {\pi\over 2}(n-m) \, .
\label{q0Fcs}
\end{equation}
Within the thermodynamic limit, the $c$ and $s$ band discrete momentum values, Eq. (\ref{q-j}),
such that $q_{j+1}-q_j=2\pi/L$, may be replaced by $c$ and $s$ band continuum momentum
variables $q$ and $q'$, respectively. (In some cases the $s$ band momentum $q'$ may be denoted by $q$ yet in general is called $q'$.)
The ground-state rapidity functions $k_0 (q) \in [-\infty,\infty]$ and $\Lambda_0 (q') \in [-\infty,\infty]$ whose domains are 
$q \in [-\infty,\infty]$ and $q' \in [-k_{F\uparrow},k_{F\uparrow}]$, respectively,
are defined in Eqs. (\ref{eq-cont1})-(\ref{Gexp}) of Appendix \ref{UBAQ}.
In the case of the ground state, the BA equations, Eqs. (\ref{FUkj}) and (\ref{FULambdaj}), and
corresponding BA distributions are given in Eqs. (\ref{eq-cont1})-(\ref{eq-cont-q}) of that Appendix.

Moreover, the $c$ and $s$ pseudoparticle energy dispersions in Eq. (\ref{DE-fermions0}) are defined as follows,
\begin{eqnarray}
\varepsilon_c (q) & = & {\bar{\varepsilon}}_c (k_0 (q)) 
\hspace{0.20cm}{\rm and}\hspace{0.20cm} \varepsilon_{s} (q') = {\bar{\varepsilon}}_{s} (\Lambda_0 (q'))
\hspace{0.20cm}{\rm where}
\nonumber \\
{\bar{\varepsilon}}_c (k) & = & \int_Q^{k}dk^{\prime}\,\eta_c (k^{\prime}) 
\hspace{0.20cm}{\rm and}\hspace{0.20cm}
{\bar{\varepsilon}}_{s} (\Lambda) = \int_B^{\Lambda}d\Lambda^{\prime}\,\eta_{s} (\Lambda^{\prime}) \, .
\label{varepsilon-c-s}
\end{eqnarray}
The distributions $\eta_c (\Lambda)$ and $\eta_{s} (\Lambda)$ appearing here are solutions of the integral equations
given in Eqs. (\ref{ceta})-(\ref{eta-epsolin}) of Appendix \ref{UBAQ}.
In the ${\cal{C}}\rightarrow 0$ limit the $c$ and $s$ pseudoparticle energy dispersions read,
\begin{eqnarray}
\varepsilon_c (q) & = & {q^2\over 2} - k_{F\uparrow}^2 - k_{F\downarrow}^2 
\hspace{0.20cm} {\rm for}\hspace{0.20cm} \vert q\vert \leq 2k_{F\downarrow}
\nonumber \\
& = & (\vert q\vert - k_{F\downarrow})^2 - k_{F\uparrow}^2 
\hspace{0.20cm} {\rm for}\hspace{0.20cm} \vert q\vert \geq 2k_{F\downarrow} 
\nonumber \\
\varepsilon_{s} (q') & = & (q')^2 - k_{F\downarrow}^2 \, , 
\label{varepsilon-c-s-C0}
\end{eqnarray}
whereas in the ${\cal{C}}\rightarrow\infty$ limit they are given by,
\begin{eqnarray}
\varepsilon_c (q) & = & q^2 - (2k_F)^2
\nonumber \\
\varepsilon_{s} (q') & = & 0 \, .
\label{varepsilon-c-s-inf}
\end{eqnarray}

The $c$ and $s$ pseudoparticles have energy residual interactions associated with the $f$ functions 
in the second-order terms of the energy functional, Eq. (\ref{DE-fermions0}). Such $f$ functions expression
given below involves the $c$ and $s$ bands group velocities,
\begin{equation}
v_c (q) = {\partial\varepsilon_c (q)\over \partial q} 
\hspace{0.20cm}{\rm and}\hspace{0.20cm}
v_{s} (q') = {\partial\varepsilon_{s} (q')\over \partial q'} \, ,
\label{vcvs}
\end{equation}
associated with the energy dispersions, Eq. (\ref{varepsilon-c-s}), respectively. 
They can be expressed in terms of BA distributions, as given
in Eq. (\ref{vcsBAR}) of Appendix \ref{UBAQ}.

The $c$ and $s$ band group velocities at the corresponding Fermi points,
\begin{equation}
\pm v_c \equiv v (\pm q_{Fc}) = v (\pm 2k_F) 
\hspace{0.20cm}{\rm and}\hspace{0.20cm}
\pm v_{s} \equiv v (\pm q_{Fs}) =v_{s} (\pm k_{F\downarrow}) \, ,
\label{vFcvFs}
\end{equation}
play an important role, as they are the velocities of the low-energy particle-hole processes
near the $c$ and $s$ bands Fermi points. Their expression in terms of BA distributions
is provided in Eq. (\ref{vcsBARF}) of Appendix \ref{UBAQ}.

Moreover, the $f$ functions expression involves the functions 
$2\pi\Phi_{\beta,\beta'} (q_j,q_{j'})$ defined below in Section \ref{PRPS}. The latter are related
to the residual interactions of the $\beta$ pseudoparticle or pseudohole of momentum $q_j$ with
a $\beta'$ pseudoparticle or pseudohole created at momentum $q_{j'}$ under a transition from
the ground state to an excited energy eigenstate. Those processes are behind the 
momentum function deviations, Eq. (\ref{DNq}), in the energy functional, Eq. (\ref{DE-fermions0}). 
Specifically, the $f$ functions expression reads,
\begin{eqnarray}
f_{\beta\,\beta'}(q_j,q_{j'}) & = & v_{\beta}(q_{j})\,2\pi \,\Phi_{\beta,\beta'}(q_{j},q_{j'})+
v_{\beta'}(q_{j'})\,2\pi \,\Phi_{\beta',\beta}(q_{j'},q_{j}) 
\nonumber \\
& + & {1\over 2\pi}\sum_{\beta''=c,s} \sum_{\iota =\pm 1} v_{\beta''}\,
2\pi\Phi_{\beta'',\beta}(\iota q_{F\beta''},q_{j})\,2\pi\Phi_{\beta'',\beta'} (\iota q_{F\beta''},q_{j'}) \, .
\nonumber
\end{eqnarray}

The $c$ and $s$ pseudoparticle energy dispersions, Eq. (\ref{varepsilon-c-s}), can be written as,
\begin{eqnarray}
\varepsilon_c (q) & = & \varepsilon_c^0 (q) - \varepsilon_c^0 (2k_F) \hspace{0.20cm}{\rm for}\hspace{0.20cm}
q \in ]-\infty,\infty[\hspace{0.20cm}{\rm and}
\nonumber \\
\varepsilon_{s} (q') & = & \varepsilon_{s}^0 (q') - \varepsilon_{s}^0 (k_{F\downarrow}) \hspace{0.20cm}{\rm for}\hspace{0.20cm}
q \in [-k_{F\uparrow},k_{F\uparrow}] \, ,
\label{varepsilon}
\end{eqnarray}
respectively. The energy dispersions $\varepsilon_c^0 (q)$ and $\varepsilon_{s}^0 (q)$ in this equation fully control
the dependence of the magnetic field $h$ on the spin density $m$ and chemical
potential $\mu$ on the fermionic density $n$. Such dependencies are contained in the
following expressions of the energy scales $2\mu$ and $2\mu_B\,h$,
\begin{equation}
2\mu = 2\varepsilon_c^0 (2k_F) - \varepsilon_{s}^0 (k_{F\downarrow}) 
\hspace{0.20cm}{\rm and}\hspace{0.20cm}
2\mu_B\,h = - \varepsilon_{s}^0 (k_{F\downarrow}) = \varepsilon_{s} (k_{F\uparrow}) \, ,
\label{mu-muBH0}
\end{equation}
where $\mu_B$ is the Bohr magneton. (See also Eqs. (\ref{mu-muBH}) Appendix \ref{UBAQ}.)

For the present spin density interval, $m\in [0,n]$, the magnetic field varies in the 
domain $h\in [0,h_c]$ where $h_c$ is the critical field for fully polarized ferromagnetism 
achieved in the $m\rightarrow n$ and $k_{F\downarrow}\rightarrow 0$ limits. 
An analytical expression for the energy scale $2\mu_B\,h_c$ associated with the critical field $h_c$
can be derived from the use of the expressions of ${\bar{\varepsilon}}_{s}^{\,0} (\Lambda)$ and $\varepsilon_{s}^{\,0} (q)$
in the $m\rightarrow n$ limit given in Eqs. (\ref{varepsilonsB0}) and (\ref{varepsilonsqB0}) of Appendix \ref{EDPS}, respectively. It reads,
\begin{equation}
2\mu_B h_c = - \varepsilon_{s}^0 (0)\vert_{k_{F\downarrow}=0} = {1\over 2\pi}\left({\cal{C}}^2 + (2\pi n)^2\right)\arctan\left({2\pi n\over {\cal{C}}}\right)
-  {\cal{C}}\,n \, .
\label{hc}
\end{equation}
Its limiting behaviors are given in Eqs. (\ref{hclimitsn}) and (\ref{hclimitsC}) of Appendix \ref{EDPS}.
(In that Appendix simplified expressions in the fully polarized ferromagnetism
limit of several physical quantities are provided.)

\subsection{The related $c$ and $s$ particle representation and corresponding phase shifts}
\label{PRPS} 

For the 1D repulsive fermion model in the subspace populated only by $c$ and $s$
pseudoparticles considered here, the BA $c$ and $s$ rapidity functions $k (q_j)$ and $\Lambda (q_j)$
of the excited energy eigenstates, which are solutions of the BA equations, Eqs. (\ref{FUkj}) and (\ref{FULambdaj}),
can be expressed in terms of those of the corresponding initial ground state, $k_0 (q)$ 
and $\Lambda_0 (q)$, respectively, defined in Eqs. (\ref{eq-cont1})-(\ref{Gexp}) of Appendix \ref{UBAQ}.
Specifically, $k (q_j) = k_0 ({\bar{q}}_j)$ and $\Lambda (q_j) = \Lambda_0 ({\bar{q}}_j)$.

The set of $j=1,...,L_{\beta}$ values ${\bar{q}}_j = {\bar{q}} (q_j)$ in such excited energy eigenstates rapidity 
expressions $k (q_j) = k_0 ({\bar{q}} (q_j))$ and $\Lambda (q_j) = \Lambda_0 ({\bar{q}} (q_j))$
are the $\beta = c,s$ band discrete {\it canonical momentum} values. They read,
\begin{equation}
{\bar{q}}_j = {\bar{q}} (q_j) = q_j + {2\pi\Phi_{\beta} (q_j)\over L} = {2\pi\over L}\left(I^{\beta}_j + \Phi_{\beta} (q_j)\right) 
\hspace{0.20cm}{\rm where}\hspace{0.20cm} j=1,...,L_{\beta}\hspace{0.20cm}{\rm and}\hspace{0.20cm}\beta = c,s \, .
\label{barqan}
\end{equation}
Here ${\bar{q}}_{j+1}-{\bar{q}}_{j}= 2\pi/L + {\rm h.o.}$ where h.o. stands for contributions of second order in $1/L$.
The function $2\pi\Phi_{\beta} (q_j)$ in Eq. (\ref{barqan}) is defined below. 

We call a {\it $\beta =c,s$ particle} each of the $N_{\beta}$ occupied $\beta$-band discrete canonical momentum values ${\bar{q}}_j$
\cite{Carmelo_05,LE}. We call a {\it $\beta$ hole} the remaining $N_{\beta}^h$ unoccupied 
$\beta$-band discrete canonical momentum values ${\bar{q}}_j$ of an excited energy eigenstate. (In the case
of the related 1D Hubbard model PDT, such particles were rather called pseudofermions \cite{Carmelo_05,Carmelo_18,Carmelo_17}.) 
There is a $c$ and $s$ particle representation 
for each initial ground state and its excited states. This holds for all fermionic and spin densities. 

The set of $L_{\beta}$ discrete $\beta =c,s$ bare momentum values $\{q_j\}$, Eq. (\ref{q-j}), and the corresponding set of
$L_{\beta}$ discrete $\beta =c,s$ canonical momentum values $\{{\bar{q}}_j\}$, Eq. (\ref{barqan}), are equally ordered.
This is because $I^{\beta}_{j+1}-I^{\beta}_j = 1$ and $\Phi_{\beta} (q_{j+1})-\Phi_{\beta} (q_j)={\cal{O}} (1/L)$
in ${\bar{q}}_j = {2\pi\over L}(I^{\beta}_j + \Phi_{\beta} (q_j))$. For simplicity, in the case of some $q_j$ dependent physical
quantities one then often associates in the thermodynamic limit the bare momentum $q_j$ to the $\beta =c,s$ particle 
of canonical momentum ${\bar{q}}_j = {2\pi\over L}(I^{\beta}_j + \Phi_{\beta} (q_j))$. (This is in spite of $q_j$ being the
discrete momentum value of the $\beta =c,s$ pseudoparticle which is transformed into the $\beta =c,s$ particle
under the $\beta$ pseudoparticle - $\beta$ particle unitary transformation.) Moreover, if in the
thermodynamic limit one replaces the sets of discrete $c$ and $s$ bare momentum values $\{q_j\}$ and $\{q_j'\}$ by
continuous momentum variables $q\in [-\infty,\infty]$ and $q'\in [-k_{F\uparrow},k_{F\uparrow}]$, respectively, the corresponding
sets of discrete $c$ and $s$ canonical momentum values $\{{\bar{q}}_j\}$ and $\{{\bar{q}}_j'\}$ are replaced by 
undistinguishable continuous momentum variables. For the excited states considered here, the exception is at the $c$ and $s$ 
bands Fermi points, which are slightly shifted under the $q_j\rightarrow {\bar{q}}_{j}$ and $q_{j}'\rightarrow{\bar{q}}_{j}'$ 
unitary transformation, respectively. (See Eq. (\ref{DeviqFb}) below.)

An example of such $q_j$ dependent physical quantities is $2\pi\Phi_{\beta} (q_j)$ in Eq. (\ref{barqan}). 
It is a functional that involves the deviations $\delta N_{\beta'}(q_{j'})$ defined in Eq. (\ref{DNq}) and reads,
\begin{equation}
2\pi\Phi_{\beta} (q_j) = \sum_{\beta'=c,s}\,\sum_{j'=1}^{L_{\beta'}}\,2\pi\Phi_{\beta,\beta'}(q_j,q_{j'})\, \delta N_{\beta'}(q_{j'}) 
\hspace{0.20cm}{\rm where}\hspace{0.20cm} j=1,...,L_{\beta}\hspace{0.20cm}{\rm and}\hspace{0.20cm}\beta = c,s \, ,
\label{Phibetaq}
\end{equation}
and $2\pi\Phi_{\beta,\beta'}(q_j,q_{j'})$ is for $\beta =c,s$ and $\beta' =c,s$ given by,
\begin{eqnarray}
2\pi\Phi_{s,s}\left(q,q'\right) & = & 2\pi\bar{\Phi }_{s,s} \left({2\Lambda (q)\over {\cal{C}}},{2\Lambda (q')\over {\cal{C}}}\right) \, ;
\hspace{0.40cm}
2\pi\Phi_{s,c}\left(q,q'\right) = 2\pi\bar{\Phi }_{s,c} \left({2\Lambda (q)\over {\cal{C}}},{2k (q')\over {\cal{C}}}\right) 
\nonumber \\
2\pi\Phi_{c,c}\left(q,q'\right) & = & 2\pi\bar{\Phi }_{c,c} \left({2k (q)\over {\cal{C}}},{2k (q')\over {\cal{C}}}\right) \, ;
\hspace{0.40cm}
2\pi\Phi_{c,s}\left(q,q'\right) = 2\pi\bar{\Phi }_{c,s} \left({2k (q)\over {\cal{C}}},{2\Lambda (q')\over {\cal{C}}}\right) \, . 
\label{Phis-all-qq}
\end{eqnarray}
The quantities on the right-hand side of these equations are functions of the rapidity-related variables
$r=2k/{\cal{C}}$ for the $c$ band and $r=2\Lambda/{\cal{C}}$ for the $s$ band. They are 
uniquely defined by the integral equations given in Eqs. (\ref{Phissn-m})-(\ref{Phisc-m2}) of Appendix \ref{UBAQ}.
(In such equations they appear in units of $2\pi$.)

In the $c$ and $s$ particle representation, $2\pi\Phi_{\beta,\beta'}(q_j,q_{j'})$, Eq. (\ref{Phis-all-qq}), 
has a precise physical meaning: $2\pi\Phi_{\beta,\beta'}(q_j,q_{j'})$ (and $-2\pi\Phi_{\beta,\beta'}(q_j,q_{j'})$) 
is the phase shift acquired by a $\beta=c,s$ particle or
hole of canonical momentum ${\bar{q}}_j={\bar{q}} (q_j)$ 
upon scattering off a $\beta'=c,s$ particle (and $\beta'=c,s$ hole) 
of canonical momentum value ${\bar{q}}_{j'}={\bar{q}} (q_{j'})$
created under a transition from the ground state to an excited energy eigenstate. 
For simplicity, in the thermodynamic limit one often says it to be the phase shift acquired by a $\beta=c,s$ particle or
hole of momentum $q_j$ upon scattering off a $\beta'=c,s$ particle (and $\beta'=c,s$ hole) 
of momentum $q_{j'}$. Indeed, $2\pi\Phi_{\beta,\beta'}(q_j,q_{j'})$ is expressed in terms of those
bare momentum values.

Such a phase shift is thus imposed to the $\beta=c,s$ particle or hole scatterer 
by the $\beta'=c,s$ particle or hole created under such a transition, which plays
the role of mobile scattering center. Within the MQIM the latter is called a mobile quantum
impurity.

It then follows that the functional $2\pi\Phi_{\beta} (q_j)$, Eq. (\ref{Phibetaq}), in the $\beta =c,s$ canonical momentum
expression ${\bar{q}}_j = q_j + {2\pi\over L}\Phi_{\beta} (q_j)$, Eq. (\ref{barqan}), is the phase
shift acquired by a $\beta$ particle or hole of canonical momentum value ${\bar{q}}_j={\bar{q}} (q_j)$ 
(or momentum value $q_j$) upon scattering off the set of $\beta'$ particles and $\beta'$ holes created under such a transition. 
Hence the $\beta$ particle phase shift $2\pi\Phi_{\beta} (q_j)$ has a specific value for each 
ground-state - excited-state transition. 

The overall phase shift,
\begin{equation}
2\pi\Phi_{\beta}^T (q_j) = 2\pi\Phi_{\beta}^0 + 2\pi\Phi_{\beta} (q_j) 
\hspace{0.20cm}{\rm where}\hspace{0.20cm}j=1,...,L_{\beta} \hspace{0.20cm}{\rm and}\hspace{0.20cm}\beta = c,s \, ,
\nonumber
\end{equation}
involves both a non-scattering term $2\pi\Phi_{\beta}^0$, Eq. (\ref{pican}), and 
the scattering term $2\pi\Phi_{\beta} (q_j)$, Eq. (\ref{Phibetaq}).

The scattering functional $2\pi\Phi_{\beta} (q_j)$, Eq. (\ref{Phibetaq}), and the overall functional $2\pi\Phi_{\beta}^T (q_j)$
fully determine the deviations of the $\beta =c,s$ Fermi canonical momentum values under transitions from the
ground state to one-fermion excited states as follows,
\begin{equation}
\delta {\bar{q}}_{F\beta}^{\iota} =  \left(\iota\,\delta N^{0,F}_{\beta,\iota}+\Phi_{\beta}^T (\iota q_{F\beta})\right){2\pi\over L}
= \left(\iota\,\delta N^F_{\beta,\iota}+\Phi_{\beta} (\iota q_{F\beta})\right){2\pi\over L}
\hspace{0.20cm}{\rm where}\hspace{0.20cm}\beta = c, s\hspace{0.20cm}{\rm and}\hspace{0.20cm}\iota =\pm 1 \, . 
\label{DeviqFb}
\end{equation}
Here such functionals appear in units of $2\pi$, $q_{Fc}=2k_F$ and $q_{Fs}=k_{F\downarrow}$ as given in Eq. (\ref{q0Fcs}),
$\delta N^F_{\beta,\iota}=\delta N^{0,F}_{\beta,\iota}+\iota\,\Phi_{\beta}^0$ is the deviation in the
number of right $(\iota =1)$ and left $(\iota =-1)$ $\beta$ particles at the corresponding
$\beta,\iota$ Fermi point, and $\delta N^{0,F}_{\beta,\iota}$ is such a deviation without accounting for
the effects of the non-scattering phase shift $2\pi\Phi_{\beta}^0$, Eq. (\ref{pican}). Hence 
while $\delta N^{0,F}_{\beta,\iota}$ either vanishes or is a positive or negative integer number, 
the deviation $\delta N^F_{\beta,\iota}$ may be a positive or negative half-odd integer number.
Indeed, $2\pi\Phi_{\beta}^0$ has in units of $2\pi$ the values $\Phi_{\beta}^0=0,\pm 1/2$.

The exponents in the one-fermion spectral functions power-law expressions given below in 
Sections \ref{SFEX} and \ref{LESF} have different expressions in the high-energy regime and
in the low-energy TLL regime, respectively. On the one hand, the TLL 
and the crossover to TLL regimes involve processes in the $c$ and $s$ bands whose continuum momentum absolute values
are in the intervals $\vert q\vert\in [2k_F- k_{Fc}^0,2k_F+k_{Fc}^0]$ and 
$\vert q'\vert\in [k_{F\downarrow}-k_{Fs}^0,k_{F\downarrow}+k_{Fs}^0]$, respectively. 
Here $k_{Fc}^0/2k_F\ll 1$ and $k_{Fs}^0/k_{F\downarrow}\ll 1$. On the other hand,
the high-energy regime involves processes in the complementary $c$ band momentum
intervals $q\in [-2k_F+k_{Fc}^0,2k_F-k_{Fc}^0]$, $q\in ]-\infty,-2k_F-k_{Fc}^0]$, 
and $q\in [2k_F+k_{Fc}^0,\infty[$ and $s$ band momentum intervals $q'\in [-k_{F\downarrow}+k_{Fs}^0,k_{F\downarrow}-k_{Fs}^0]$, 
$q'\in [-k_{F\uparrow},-k_{F\downarrow}-k_{Fs}^0]$, and $q\in [k_{F\downarrow}+k_{Fs}^0,k_{F\uparrow}]$.

The one-fermion spectral functions exponents expressions studied below in Section \ref{SFEX} involve
the following general functionals, which are merely the square of the Fermi canonical momentum value
deviations, Eq. (\ref{DeviqFb}), in units of $2\pi/L$,
\begin{equation}
2\Delta_{\beta}^{\iota} = \left({\delta {\bar{q}}_{F\beta}^{\iota}\over (2\pi/L)}\right)^2 
= \left(\iota\delta N^{F}_{\beta,\iota} + \Phi_{\beta}(\iota q_{F\beta})\right)^2 
\hspace{0.20cm}{\rm where}\hspace{0.20cm} \beta = c, s\hspace{0.20cm}{\rm and}\hspace{0.20cm}\iota =\pm 1 \, . 
\label{a10DP-iota}
\end{equation}

Finally, expression of the energy functional, Eq. (\ref{DE-fermions0}), in the $c$ and $s$ particle representation involves
the $\beta =c,s$ bands discrete canonical momentum values 
${\bar{q}}_j = {\bar{q}} (q_j)$, Eq. (\ref{barqan}). One finds after some algebra
that in such a representation it reads up to ${\cal{O}}(1/L)$ order,
\begin{equation}
\delta E = \sum_{\beta=c,s}\sum_{j=1}^{L_{\beta}}\varepsilon_{\beta} ({\bar{q}}_j)\,\delta {\cal{N}}_{\beta}({\bar{q}}_j) \, . 
\label{DE}
\end{equation}
Here $\delta {\cal{N}}_{\beta}({\bar{q}}_j) = N_{\beta} (q_j)$ and
the $\beta =c,s$ particle energy dispersions $\varepsilon_{\beta} ({\bar{q}}_j)$
have exactly the same form as those given in Eq. (\ref{varepsilon-c-s})
with the bare momentum, $q_j$, replaced by the corresponding canonical momentum, ${\bar{q}}_j= {\bar{q}} (q_j)$.

In contrast to the equivalent pseudoparticle energy functional, Eq. (\ref{DE-fermions0}), that in Eq. (\ref{DE}) has no energy 
interaction terms of second-order in the deviations $\delta {\cal{N}}_{\beta}({\bar{q}}_j)$. This has a deep physical meaning:
The $\beta =c,s$ particles generated from corresponding $\beta =c,s$ pseudoparticles by a $q_j\rightarrow {\bar{q}}_j$
uniquely defined unitary transformation have no such interactions up to ${\cal{O}}(1/L)$ order.

Within the present thermodynamic limit, only finite-size corrections up to that order are relevant for the spectral functions expressions. 
The property that the excitation energy spectrum, Eq. (\ref{DE}), 
has no $c$ and $s$ particle energy interactions plays a key role in the derivation by the PDT of the general one-fermion
spectral functions used below in our studies of Section \ref{SFEX}.
Indeed it allows them to be expressed in terms of a sum of convolutions of $c$ and $s$ particle spectral functions.  
Moreover, the spectral weights of the latter spectral functions 
can be expressed as Slater determinants of $c$ and $s$ particles operators. 

Such spectral weights involve the functionals, Eq. (\ref{a10DP-iota}), determined by the
Fermi canonical momentum value deviations, Eq. (\ref{DeviqFb}). Since the derivation within the PDT of the general 
one-fermion spectral functions used below in Section \ref{SFEX} is similar to that of other integrable models
\cite{Carmelo_18,Carmelo_16,Carmelo_15}, it is not reported in this paper.

\section{General types of one-fermion spectral singularities}
\label{SFEX}

\subsection{The two-dimensional $(k,\omega)$-plane spectra where the one-fermion
spectral singularities are contained}
\label{OES2D}

The two-parametric excitation processes that are behind $(k,\omega)$-plane one-fermion spectral weight 
distribution near the singularity {\it branch lines} and {\it boundary lines} defined below involve both creation of one charge 
or hole particle and creation of one spin or hole particle.

In contrast to the related lattice 1D Hubbard model \cite{Carmelo_18}, all charge excitations of the present
continuous model only involve real BA rapidities. At finite magnetic field the same applies to spin part
of the one-fermion excitations studied in this paper
whereas at zero magnetic field they involve as well complex BA spin rapidities.
Fortunately, due to both the spin $SU(2)$ symmetry and the lack of a spin energy gap at zero spin density, 
the same zero-spin-density spectral-function expressions 
in the vicinity of the singularities are reached by taking the limit of zero  magnetic field in the corresponding 
suitable finite-field expressions or by their direct derivation at zero magnetic field. The former method used
in this paper has the advantage of only involving real BA spin rapidities.

At zero spin density, $m=0$, the transitions from the ground state under one-fermion removal lead to excited energy eigenstates associated 
with the two-parametric processes whose number deviations relative to those of the initial ground state read,
\begin{equation}
\delta N_c = -1 \, ; \hspace{0.20cm} \delta J_c^F = \pm 1/2 \, ; \hspace{0.20cm} \delta N_s = -1 \, ; \hspace{0.20cm} \delta J_s^F = 0 \, .
\label{dN2P-DR}
\end{equation}

Both the $\beta =c,s$ Fermi points current number deviations $\delta J^F_{\beta}$ in this equation and
the $\beta =c,s$ Fermi points number deviations $\delta N^F_{\beta}$ are defined in terms of
$\beta =c,s$ left ($\iota =-1$) and right ($\iota =+1$) Fermi points number deviations $\delta N^{F}_{\beta,\iota}$
as follows,
\begin{equation}
\delta N^F_{\beta} = \sum_{\iota=\pm 1}\delta N^{F}_{\beta,\iota}\hspace{0.20cm}{\rm and}\hspace{0.20cm}
\delta J^F_{\beta} = {1\over 2}\sum_{\iota=\pm 1}(\iota)\delta N^{F}_{\beta,\iota}
\hspace{0.20cm}{\rm for}\hspace{0.20cm}\beta = c,s
\label{NFJF}
\end{equation}

The number deviations $\delta N^{F}_{c}$ and $\delta N^{F}_{s}$ are particular cases of those given in Eq. (\ref{dN2P-DR}).
The latter refer to $c$ and $s$ band momentum values, respectively, at and away of the Fermi points.
While the $\beta =c,s$ Fermi points number deviations and current number deviations $\delta N^F_{\beta}$ and
$\delta J^F_{\beta}$, respectively, defined in Eq. (\ref{NFJF}) correspond to number fluctuations 
at the $\beta =c,s$ Fermi points, one denotes by  $\delta N_{\beta }^{NF}$ 
the $\beta =c,s$ the number deviations that refer to creation of $\beta =c,s$ particles or holes away from 
those Fermi points.

The $\omega <0$ energy spectrum of such excitations
is of the form $-\omega = \omega_{R} (k) = -\varepsilon_c (q) -\varepsilon_s (q')$.
It has the following two branches, 
\begin{eqnarray}
\omega_{R} (k) & = & -\varepsilon_c (q) - \varepsilon_s (q')\hspace{0.20cm}{\rm where}\hspace{0.20cm}k = \pm 2k_F - q - q' 
\hspace{0.20cm}{\rm for}
\nonumber \\
q & \in & [-2k_F,2k_F]\hspace{0.20cm}{\rm and}\hspace{0.20cm}q' \in [-k_F,k_F]\hspace{0.20cm}{\rm with}\hspace{0.20cm}{\rm intervals}
\nonumber \\
k & = & 2k_F - q - q' \in [-k_F, 5k_F] \hspace{0.20cm}{\rm for}\hspace{0.20cm}{\rm branch}\hspace{0.20cm}A \, ,
\nonumber \\
k & = & - 2k_F - q - q' \in [-5k_F, k_F] \hspace{0.20cm}{\rm for}\hspace{0.20cm}{\rm branch}\hspace{0.20cm}B  \, .
\label{SpElremo-m0}
\end{eqnarray}
Here the two alternative contributions $\pm 2k_F$ to the excitation momentum $k = \pm 2k_F - q - q'$ result
from $c$ band momentum shifts $2\pi\Phi_{c}^0=\pm\pi$ in Eq. (\ref{pican}), such that
${2\pi\over L}\Phi_{c}^0\times N_c=\pm\pi n =\pm 2k_F$ where $N_c$ is the number of $c$ band
Fermi sea occupied discrete momentum values $q_j$, Eq. (\ref{q-j}). In contrast, $2\pi\Phi_{s}^0=0$ for the present 
one-fermion removal excitations.

At zero spin density, $m=0$, the transitions under one-fermion addition lead to excited energy eigenstates associated 
with the two-parametric processes whose number deviations relative to those of the initial ground state are given by,
\begin{equation}
\delta N_c = 1 \, ; \hspace{0.20cm} \delta J_c^F = 0 \, ; \hspace{0.20cm} \delta N_s = 0 \, ; \hspace{0.20cm} \delta J_s^F = \pm 1/2 \, .
\label{dN2P-UA}
\end{equation}

The $\omega >0$ energy spectrum of such excitations is given by
$\omega = \omega_{A} (k) = \varepsilon_c (q) -\varepsilon_s (q')$. It has again two branches, 
\begin{eqnarray}
\omega_{A} (k) & = & \varepsilon_c (q) - \varepsilon_s (q') \hspace{0.20cm}{\rm where}\hspace{0.20cm}k = q - q'
\hspace{0.20cm}{\rm and}
\nonumber \\
k & \in & [k_F,\infty]  \hspace{0.20cm}{\rm for}\hspace{0.20cm}q \in [2k_F,\infty]
\hspace{0.20cm}{\rm and}\hspace{0.20cm}q' \in [-k_F,k_F]  \hspace{0.20cm}{\rm for}
\hspace{0.20cm}{\rm branch}\hspace{0.20cm}A 
\nonumber \\
k & \in & [-\infty,-k_F]  \hspace{0.20cm}{\rm for}\hspace{0.20cm}q \in [-\infty,-2k_F]
\hspace{0.20cm}{\rm and}\hspace{0.20cm}q' \in [-k_F,k_F] \hspace{0.20cm}{\rm for}
\hspace{0.20cm}{\rm branch}\hspace{0.20cm}B \, .
\label{SpEladd-m0}
\end{eqnarray}

The transitions under up-spin one-fermion removal at spin density $m> 0$ 
lead to excited energy eigenstates associated with the two-parametric processes 
whose number deviations relative to those of the initial ground state read,
\begin{equation}
\delta N_c = -1 \, ; \hspace{0.20cm} \delta J_c^F = 0 \, ; \hspace{0.20cm} \delta N_s = 0 \, ; \hspace{0.20cm} \delta J_s^F = \pm 1/2 \, .
\label{dN2P-UR}
\end{equation}
Moreover, in general $\delta N_s^F = -1$ and $\delta N_s^{NF} = 1$ with a limiting case being $\delta N_s^F = \delta N_s^{NF} = 0$.

The corresponding $\omega <0$ energy spectrum of such excitations reads 
$-\omega = \omega_{R}^{\uparrow} (k) = -\varepsilon_c (q) + \varepsilon_s (q')$. It has the following two branches, 
\begin{eqnarray}
\omega_{R}^{\uparrow} (k) & = & -\varepsilon_c (q) +\varepsilon_s (q') \hspace{0.20cm}{\rm where}\hspace{0.20cm} k = - q + q' 
\hspace{0.20cm}{\rm and}
\nonumber \\
k & \in & [-k_{F\uparrow}, (2k_F + k_{F\uparrow})]  \hspace{0.20cm}{\rm for}\hspace{0.20cm}q \in [-2k_F,2k_F]
\hspace{0.20cm}{\rm and}\hspace{0.20cm}q' \in [k_{F\downarrow},k_{F\uparrow}]
\hspace{0.20cm}{\rm for}\hspace{0.20cm}{\rm branch}\hspace{0.20cm}A 
\nonumber \\
k & \in & [-(2k_F + k_{F\uparrow}),k_{F\uparrow}]  \hspace{0.20cm}{\rm for}\hspace{0.20cm}q \in [-2k_F,2k_F]
\hspace{0.20cm}{\rm and}\hspace{0.20cm}q' \in [-k_{F\uparrow},-k_{F\downarrow}] 
\hspace{0.20cm}{\rm for}\hspace{0.20cm}{\rm branch}\hspace{0.20cm}B \, .
\label{SpupElremo}
\end{eqnarray}

The transitions under up-spin one-fermion addition at spin density $m> 0$ 
give rise to excited energy eigenstates associated with the two-parametric processes 
whose number deviations relative to those of the initial ground state are given in Eq. (\ref{dN2P-UA}).

The $\omega >0$ energy spectrum of such excitations spectrum is given by
$\omega = \omega_{A}^{\uparrow} (k) = \varepsilon_c (q) -\varepsilon_s (q')$. It has again two branches, 
\begin{eqnarray}
\omega_{A}^{\uparrow} (k) & = & \varepsilon_c (q) - \varepsilon_s (q') 
\hspace{0.20cm}{\rm for}\hspace{0.20cm}k = q - q' 
\hspace{0.20cm}{\rm and}
\nonumber \\
k & \in & [k_{F\uparrow},\infty]\hspace{0.20cm}{\rm for}\hspace{0.20cm}q \in [2k_F,\infty]
\hspace{0.20cm}{\rm and}\hspace{0.20cm}q' \in [-k_{F\downarrow},k_{F\downarrow}]  
\hspace{0.20cm}{\rm for}\hspace{0.20cm}{\rm branch}\hspace{0.20cm}A 
\nonumber \\
k & \in & [-\infty,-k_{F\uparrow}]\hspace{0.20cm}{\rm for}\hspace{0.20cm}q \in [-\infty,-2k_F]
\hspace{0.20cm}{\rm and}\hspace{0.20cm}q' \in [-k_{F\downarrow},k_{F\downarrow}] 
\hspace{0.20cm}{\rm for}\hspace{0.20cm}{\rm branch}\hspace{0.20cm}B \, .
\label{SpupEladd}
\end{eqnarray}

The transitions under down-spin one-fermion removal lead at spin density $m> 0$ 
to excited energy eigenstates associated with the two-parametric processes 
whose number deviations relative to those of the initial ground state are given in Eq. (\ref{dN2P-DR}).

The $\omega <0$ energy spectrum of such excitations
is of the form $-\omega = \omega_{R}^{\downarrow} (k) = -\varepsilon_c (q) -\varepsilon_s (q')$.
It has two branches corresponding to $\iota = \pm 1$, 
\begin{eqnarray}
\omega_{R}^{\downarrow} (k) & = & -\varepsilon_c (q) - \varepsilon_s (q')
\hspace{0.20cm}{\rm for}\hspace{0.20cm}k = \pm 2k_F - q - q' 
\hspace{0.20cm}{\rm where}\hspace{0.20cm}q \in [-2k_F,2k_F]
\hspace{0.20cm}{\rm and}\hspace{0.20cm}q' \in [-k_{F\downarrow},k_{F\downarrow}]
\hspace{0.20cm}{\rm with}\hspace{0.20cm}{\rm intervals}
\nonumber \\
k & = & 2k_F - q - q' \in [-k_{F\downarrow}, (4k_F + k_{F\uparrow})] 
\hspace{0.20cm}{\rm for}\hspace{0.20cm}{\rm branch}\hspace{0.20cm}A 
\nonumber \\
k & = & - 2k_F - q - q' \in [-(4k_F + k_{F\uparrow}), k_{F\downarrow}] 
\hspace{0.20cm}{\rm for}\hspace{0.20cm}{\rm branch}\hspace{0.20cm}B  \, .
\label{SpdownElremo}
\end{eqnarray}

Finally, the transitions under down-spin one-fermion addition give rise at spin density $m> 0$ 
to excited energy eigenstates associated with the two-parametric processes 
whose number deviations relative to those of the initial ground state are given by,
\begin{equation}
\delta N_c = 1 \, ; \hspace{0.20cm} \delta J_c^F = \pm 1/2 \, ; \hspace{0.20cm} \delta N_s = 1 \, ; \hspace{0.20cm} \delta J_s^F = 0 \, .
\nonumber
\end{equation}

The $\omega >0$ energy spectrum of such excitations
reads $\omega = \omega_{A}^{\downarrow} (k) = \varepsilon_c (q) + \varepsilon_{s} (q')$. It has four branches, 
\begin{eqnarray}
\omega_{A}^{\downarrow} (k) & = & \varepsilon_c (q) + \varepsilon_{s} (q') \hspace{0.20cm}{\rm for}\hspace{0.20cm}k = \pm 2k_F + q + q' 
\hspace{0.20cm}{\rm and}\hspace{0.20cm}{\rm sgn}\{q'\} = \pm\hspace{0.20cm}{\rm for}\hspace{0.20cm}q'\neq 0
\hspace{0.20cm}{\rm with}\hspace{0.20cm}{\rm intervals}
\nonumber \\
k & = & 2k_F + q + q' \in [(4k_F +k_{F\uparrow}),\infty]\hspace{0.20cm}{\rm for}\hspace{0.20cm}{\rm branch}\hspace{0.20cm}A 
\nonumber \\
& & \hspace{0.20cm}{\rm where}\hspace{0.20cm}q \in [2k_F,\infty] \hspace{0.20cm}{\rm and}\hspace{0.20cm}q' \in [k_{F\downarrow},k_{F\uparrow}] 
\nonumber \\
k & = & 2k_F + q + q' \in [ -\infty,k_{F\uparrow}] \hspace{0.20cm}{\rm for}\hspace{0.20cm}{\rm branch}\hspace{0.20cm}B 
\nonumber \\
& & \hspace{0.20cm}{\rm where}\hspace{0.20cm}q \in [-\infty,-2k_F]\hspace{0.20cm}{\rm and}\hspace{0.20cm}q' \in [k_{F\downarrow},k_{F\uparrow}] 
\nonumber \\ 
k & = & - 2k_F + q + q' \in [-\infty, -(4k_F +k_{F\uparrow})] \hspace{0.20cm}{\rm for}\hspace{0.20cm}{\rm branch}\hspace{0.20cm}A' 
\nonumber \\
& & \hspace{0.20cm}{\rm where}\hspace{0.20cm}q \in [-\infty,-2k_F]\hspace{0.20cm}{\rm and}\hspace{0.20cm}
q' \in [-k_{F\uparrow},-k_{F\downarrow}] 
\nonumber \\
k & = & - 2k_F + q + q' \in [-k_{F\uparrow},\infty] \hspace{0.20cm}{\rm for}\hspace{0.20cm}{\rm branch}\hspace{0.20cm}B' 
\nonumber \\
& & \hspace{0.20cm}{\rm where}\hspace{0.20cm}q \in [2k_F,\infty]\hspace{0.20cm}{\rm and}\hspace{0.20cm}
q' \in [-k_{F\uparrow},-k_{F\downarrow}] \, .
\label{SpdownEladd}
\end{eqnarray}

In the present case of the one-fermion spectral functions, Eq. (\ref{Bkomega-m0}), and of the
up-spin and down-spin one-fermion spectral functions, Eq. (\ref{Bkomega}), the 
one-parametric branch lines that for some momentum subdomains correspond to singularities 
are contained in the two-parametric spectra, Eqs. (\ref{dN2P-DR})-(\ref{SpEladd-m0}), and,
Eqs. (\ref{dN2P-UR})-(\ref{SpdownEladd}), respectively. Spectra that do not contain singularities 
generated by higher-order $c$ and $s$ particle processes and/or complex spin rapidities are not 
considered in our present study.

\subsection{The $(k,\omega)$-plane one-fermion spectral singularities on the $c$ and $s$ branch lines}
\label{OESblq}

A branch line results from transitions to a well-defined subclass of the excited energy eigenstates
associated with such spectra. At zero spin density, $m=0$, the one-parametric $(k,\omega)$-plane $\beta =c,s$ 
branch line spectrum has the general form,
\begin{equation}
\omega_{\beta} (k) = \pm\varepsilon_{\beta} (q)\geq 0\hspace{0.20cm}{\rm where}\hspace{0.20cm}
k = k_0 \pm q\hspace{0.20cm}{\rm for}\hspace{0.20cm}\delta N_{\beta} (q)=\pm 1\hspace{0.20cm}{\rm and}\hspace{0.20cm}
\beta = c,s \, . 
\label{dE-dP-bl-m0}
\end{equation}
Here $\varepsilon_{\beta} (q)$ is the $\beta =c,s$ band energy dispersion, Eq. (\ref{varepsilon-c-s}), 
the momentum distribution function deviation, Eq. (\ref{DNq}), reads $\delta N_{\beta} (q) = +1$ and 
$\delta N_{\beta} (q) = -1$ for a particle and hole $\beta$ branch 
line, respectively. The momentum $k_0$ in Eq. (\ref{dE-dP-bl-m0}) is given by,
\begin{equation}
k_0 = 4k_F\,\delta J_{c}^F + 2k_F\,\delta J_s^F \, .
\label{1el-omega0-m0}
\end{equation}

At zero spin density the branch-line singularities are found below to occur at some $k$ intervals of two $c$ branch 
lines called $c^{+}$ and $c^{-}$ branch lines, respectively, which refer to different subdomains of $q$ in
$k = k_{0} \pm q$, and at some $k$ intervals of one $s$ branch line. 

The one-parametric $(k,\omega)$-plane $\beta =c,s$ branch line spectrum has for spin densities $m> 0$
a similar general form,
\begin{equation}
\omega_{\beta}^{\sigma} (k) = \pm\varepsilon_{\beta} (q)\geq 0\hspace{0.20cm}{\rm where}\hspace{0.20cm}
k = k_0 \pm q\hspace{0.20cm}{\rm for}\hspace{0.20cm}\delta N_{\beta} (q)=\pm 1\hspace{0.20cm}{\rm and}\hspace{0.20cm}
\beta = c,s \, ,
\label{dE-dP-bl}
\end{equation}
where now $\sigma =\uparrow$ and $\sigma =\downarrow$ refers to the up-spin and down-spin one-fermion spectral function, respectively,
and the momentum $k_0$, Eq. (\ref{1el-omega0-m0}), more generally reads,
\begin{equation}
k_0 = 4k_F\,\delta J_{c}^F + 2k_{F\downarrow}\,\delta J_s^F \, .
\label{1el-omega0}
\end{equation}

At finite spin density the branch-line singularities of the up-spin and down-spin one-fermion spectral functions
are found to occur at some $k$ intervals of two $c$ branch lines called again $c^+$ and $c^-$ branch lines, respectively, 
which correspond to different subdomains of the $c$ band momentum $q$ in 
the excitation momentum expression $k = k_0 \pm q$, and at some $k$ intervals of one $s$ branch line. 

For one-fermion excitations at zero spin density, the use of the PDT leads in the case of the present model 
to the following general high-energy behavior in the vicinity of a $\beta = c,s$ branch line,
\begin{eqnarray}
B_{\gamma} (k,\omega) & = & C_{\gamma,\beta}\,
\Bigl(\gamma\,\omega - \omega_{\beta} (k)\Bigr)^{\xi_{\beta} (k)}\hspace{0.20cm}{\rm for}\hspace{0.20cm} 
(\gamma\,\omega - \omega_{\beta} (k)) \geq 0\hspace{0.20cm}{\rm and}\hspace{0.20cm}\gamma = \pm 1
\hspace{0.20cm}{\rm where}
\nonumber \\
\xi_{\beta} (k) & = & -1 + 2\Delta_{c}^{+1} + 2\Delta_{c}^{-1} + 2\Delta_{s}^{+1} + 2\Delta_{s}^{-1}
= -1+\sum_{\iota =\pm 1}(2\Delta_{c}^{\iota} (q)\vert_{q=\pm (k-k_0)} + 2\Delta_{s}^{\iota}) \, .
\label{branch-l-m0}
\end{eqnarray}
Here the momentum $k_0$ is provided in Eq. (\ref{1el-omega0-m0}) and $\gamma = -1$ for
fermion removal and $\gamma = +1$ for fermion addition, as given in Eq. (\ref{c0RA}).
The simplified expressions of the functionals $2\Delta_{c}^{\iota} (q)$ and $2\Delta_{s}^{\iota}$
appearing in Eq. (\ref{branch-l-m0}) that are specific to a $\beta = c,s$ branch line involve a 
summation $\sum_{j'=1}$ in the general phase-shift functional expression, 
Eq. (\ref{Phibetaq}), that refers to creation of a single $\beta'=c,s$ particle or hole. Moreover, at zero spin 
density, $m=0$, such functionals read,
\begin{eqnarray}
2\Delta^{\iota}_{c} (q) & = & \left({\iota\over\xi_0}{\delta N^F_c\over 2} 
+ \xi_0\left[\delta J^F_c + {\delta J^F_s\over 2}\right] + c_{\beta''}\,\Phi_{c,\beta''}(\iota 2k_F,q)\right)^2  \hspace{0.20cm}{\rm and}
\nonumber \\
2\Delta^{\iota}_{s} & = & {1\over 2}\left({\iota\over 2}\left[2\delta N^F_s - \delta N^F_c
+ c_{\beta''} (-1)^{\delta_{\beta'',c}}\right] + \delta J^F_s\right)^2 \, .
\label{OESFfunctional-m0}
\end{eqnarray}
In this equation $\xi_0$ is the parameter defined in Eq. (\ref{xi-0}) of Appendix \ref{UBAQ} whose
limiting values are $\xi_0=\sqrt{2}$ for ${\cal{C}}\rightarrow 0$ and
$\xi_0=1$ for ${\cal{C}}\rightarrow\infty$. 

On the one hand, the two $\iota = \pm 1$ functionals $2\Delta^{\iota}_{c} (q)$ depend on the excitation
momentum $k = k_0 \pm q$, Eq. (\ref{dE-dP-bl-m0}), through the $c$ band momentum $q$
where the momentum $k_0$ is given in Eq. (\ref{1el-omega0-m0}). On the other hand,
that the two $\iota = \pm 1$ functionals $2\Delta^{\iota}_{s}$
in Eq. (\ref{OESFfunctional-m0}) do not depend on the $s$ band momentum of the $\beta''$ particle
or hole created under the one-fermion excitation follows from the spin SU(2) symmetry. Indeed, that
symmetry is behind the very simple behavior of the $s$ particle phase shifts in units of $2\pi$ given
in Eq. (\ref{PhiscssFq}) of Appendix \ref{UBAQ} whose use in the general expression
of the $\iota = \pm 1$ functionals $2\Delta^{\iota}_{s}$, Eq. (\ref{a10DP-iota}), leads to the present simple expressions.

For up-spin and down-spin one-fermion excitations at spin densities $m> 0$, the use of the PDT 
leads to the following general high-energy behavior near a $\beta = c,s$ branch line,
\begin{eqnarray}
B_{\sigma,\gamma} (k,\omega) & = & C_{\sigma,\gamma,\beta}\,
\Bigl(\gamma\,\omega - \omega_{\beta}^{\sigma} (k)\Bigr)^{\xi_{\beta}^{\sigma} (k)}\hspace{0.20cm}{\rm for}\hspace{0.20cm} 
(\gamma\,\omega - \omega_{\beta}^{\sigma} (k)) \geq 0\hspace{0.20cm}{\rm and}\hspace{0.20cm}\gamma = \pm 1
\hspace{0.20cm}{\rm where}
\nonumber \\
\xi_{\beta}^{\sigma} (k) & = & -1 + 2\Delta_{c}^{+1} + 2\Delta_{c}^{-1} + 2\Delta_{s}^{+1} + 2\Delta_{s}^{-1}
= -1+\sum_{\beta' = c,s}\sum_{\iota =\pm 1}2\Delta_{\beta'}^{\iota} (q)\vert_{q=\pm (k-k_0)}  \, .
\label{branch-l}
\end{eqnarray}
At $m>0$ all four $\beta =c,s$ and $\iota=\pm 1$ functionals $2\Delta_{\beta}^{\iota} (q)$ in the exponent
expression depend on the excitation
momentum $k = k_0 \pm q$, Eq. (\ref{dE-dP-bl}), through the $\beta = c,s$ band momentum $q$
where $k_0$ is given in Eq. (\ref{1el-omega0}).

The four $\beta' =c,s$ and $\iota=\pm 1$ functionals $2\Delta_{\beta'}^{\iota} (q)$
in the exponent $\xi_{\beta}^{\sigma} (k)$ expression, Eq. (\ref{branch-l}), specific
to a $\beta = c,s$ branch line are again such that the summation $\sum_{j'=1}$ in Eq. (\ref{Phibetaq})
corresponds to creation of a single $\beta'=c,s$ particle or hole. Hence from the use of their general expression in
Eq. (\ref{a10DP-iota}) one finds,
\begin{eqnarray}
2\Delta^{\iota}_{c} (q) & = & \left(\sum_{\beta'=c,s}\left(\iota\, \xi^0_{c\,\beta'}\,{\delta N^F_{\beta'}\over 2} 
+ \xi^1_{c\,\beta'}\,\delta J^F_{\beta'}\right) + c_{\beta''}\,\Phi_{c,\beta''}(\iota 2k_F,q)\right)^2  \hspace{0.20cm}{\rm and}
\nonumber \\
2\Delta^{\iota}_{s} (q') & = & \left(\sum_{\beta'=c,s}\left(\iota\, \xi^0_{s\,\beta'}\,{\delta N^F_{\beta'}\over 2} 
+ \xi^1_{s\,\beta'}\,\delta J^F_{\beta'}\right) 
+ c_{\beta''}\,\Phi_{s,\beta''}(\iota k_{F\downarrow},q')\right)^2 \, .
\label{OESFfunctional}
\end{eqnarray}
Here $c_{\beta''}=1$ and $c_{\beta''}=-1$ for creation of one $\beta'' $ particle and of one $\beta'' $ hole, respectively.
The band momenta $q$ and $q'$ belong to the intervals $q\in [-2k_F+k_{Fc}^0,2k_F-k_{Fc}^0]$ for creation of one $\beta''=c$
hole, $q\in ]-\infty,-2k_F-k_{Fc}^0]$ and $q\in [2k_F+k_{Fc}^0,\infty[$ for creation of one $\beta''=c$ 
particle, $q'\in [-k_{F\downarrow}+k_{Fs}^0,k_{F\downarrow}-k_{Fs}^0]$ for creation of one $\beta''=s$
hole and $q'\in [-k_{F\uparrow},-k_{F\downarrow}-k_{Fs}^0]$ and $q\in [k_{F\downarrow}+k_{Fs}^0,k_{F\uparrow}]$
for creation of one $\beta''=s$ particle.

The definition of the phase shifts in Eqs. (\ref{OESFfunctional-m0}) and (\ref{OESFfunctional}) in units of $2\pi$ 
involves their expression provided in Eq. (\ref{PhiqFq}) of Appendix \ref{UBAQ} in terms of the corresponding phase-shift 
functions of rapidity variables. The latter are uniquely defined by solution of the coupled integral equations, 
Eqs. (\ref{Phissn-m})-(\ref{Phisc-m2}) of that Appendix.

Furthermore, the parameters $\xi^{j}_{\beta\,\beta'}$ appearing in Eq. (\ref{OESFfunctional})
are the following phase-shift superpositions,
\begin{equation}
\xi^{j}_{\beta\,\beta'} = \delta_{\beta,\beta'} 
+ \sum_{\iota=\pm 1} (\iota)^j\,\Phi_{\beta,\beta'}\left(q_{F\beta},\iota q_{F\beta'}\right)
\hspace{0.20cm}{\rm where}\hspace{0.20cm}
\beta = c, s \, , \hspace{0.20cm} \beta' = c, s \, ,\hspace{0.20cm}{\rm and}\hspace{0.20cm} j = 0, 1 \, .
\label{x-aa}
\end{equation}
(When $\beta=\beta'$ and $\iota =1$ the second momentum $q_{F\beta'}$ in
$\Phi_{\beta,\beta'}\left(q_{F\beta},\iota q_{F\beta'}\right)$ reads $q_{F\beta}-2\pi/L$.)
The behaviors of these parameters and their limiting values are given in Eqs. (\ref{ZZ-gen})-(\ref{ZZ-gen-m0u0}) of Appendix \ref{UBAQ}.

Importantly, the one-electron spectral functions expressions in the vicinity of a 
$\beta =c,s$ branch line, Eqs. (\ref{branch-l-m0}) and (\ref{branch-l}), are valid 
provided that $\xi_{\beta} (k)>-1$ and $\xi_{\beta}^{\sigma} (k)>-1$, respectively.
That for a given $\beta =c,s$ branch line $k$ range that exponents read $\xi_{\beta} (k)=-1$
and $\xi_{\beta}^{\sigma} (k)=-1$, respectively,
means that the exact expression of the spectral function is not that given in those
equations. For these $k$ ranges 
the four functionals $2\Delta^{\iota}_{\beta}$ in Eqs. (\ref{OESFfunctional-m0}) and (\ref{OESFfunctional})
vanish. In this case the PDT also provides the corresponding behavior of the
one-electron spectral functions in Eqs. (\ref{Bkomega-m0}) and (\ref{Bkomega}), which is $\delta$-function-like and given by,
\begin{equation}
B_{\gamma} (k,\omega) = \delta \Bigl(\gamma\,\omega - \omega_{\beta} (k)\Bigr)
\hspace{0.20cm}{\rm and}\hspace{0.20cm}
B_{\sigma,\gamma} (k,\omega) = \delta \Bigl(\gamma\,\omega - \omega_{\beta}^{\sigma} (k)\Bigr) \, ,
\label{branch-lexp-1}
\end{equation}
respectively.

On the one hand, the $s$ branch line studied below corresponds to edges of support of the one-fermion spectral functions,
{\it i.e} it separates two regions with finite and vanishing spectral weight, respectively.
The underlying physical mechanism behind the line shape near it
follows from the requirement of energy and momentum conservation. The excitation leads to 
creation of the $s$ hole or $s$ particle on its energy dispersion and thus mass shell, which carries almost the entire energy. 
The remaining momentum is absorbed by a dressing of low-energy particle-hole processes near 
the Fermi points. The expression of the PDT $s$ branch line momentum dependent exponent is exact.

On the other hand, the $c^{\pm}$ branch lines also studied in the following run within the spectral-weight distribution continuum.
In non-integrable models their power-law singularities are broadened or even progressively washed 
by relaxation processes of the $c$ band particle or  $c$ band hole created away from its band Fermi points \cite{Imambekov_12}.
However, in the present solvable model, its integrability is associated with the occurrence of an infinite
number of conservation laws. They ensure that the $c$ band multi-particle scattering 
factorizes into two-particle scattering processes. This prevents relaxation processes, so that 
the line shape near the $c^{\pm}$ branch lines remains power-law like.
\begin{figure}
\includegraphics[scale=0.50]{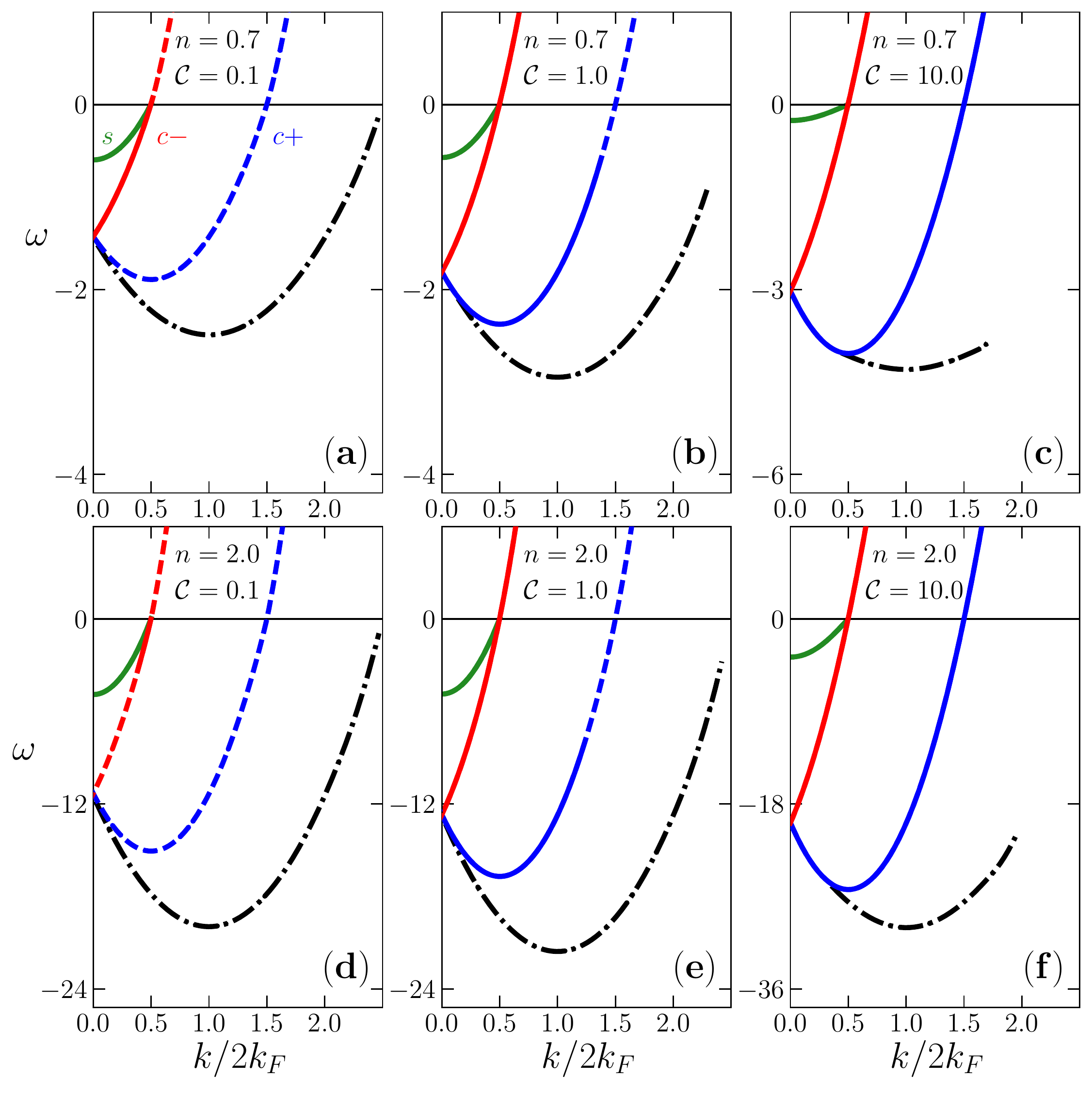}
\caption{The $\beta =c^+,c^-,s$ singular branch lines $k$ intervals (solid lines) and other branch lines
$k$ intervals (dashed lines) in units of $2k_F = \pi\,n$ for which the corresponding 
exponent $\xi_{\beta} (k)$, Eq. (\ref{branch-l-m0}), is negative and positive,
respectively, and the singular boundary lines (dashed-dotted lines) of the weight distribution associated with the
one-fermion spectral function are plotted in the $(k,\omega)$ plane. The curves refer to fermionic density $n=0.7$ for
interaction (a) ${\cal{C}}=0.1$, (b) ${\cal{C}}=1.0$, and (c) ${\cal{C}}=10.0$ and 
fermionic density $n=2.0$ for interaction (d) ${\cal{C}}=0.1$, (e) ${\cal{C}}=1.0$, and (f) ${\cal{C}}=10.0$.
The branch line spectra plotted here are defined in Section \ref{SFEX}. 
(Online, the $c^+$, $c^-$, and $s$ branch lines appear blue, red, and green, respectively.)}
\label{figure1} 
\end{figure}

Furthermore, in the present model there is very little continuum spectral weight 
just above for fermion removal and just below for fermion addition
the $c^{\pm}$ branch lines. For finite repulsive interaction ${\cal{C}}$ the $c^{\pm}$ branch lines exponents 
in Eqs. (\ref{branch-l-m0}) and (\ref{branch-l}) are the leading zero-order term of an expansion 
whose very small parameter is the coupling to that small spectral weight. Their expressions
are exact for large values of ${\cal{C}}$ and their use leads to exact
spectral-function expressions in the ${\cal{C}}\rightarrow 0$ limit. At intermediate
${\cal{C}}$ values the higher-order terms are extremely small and vanish
when the exponents vanish. This means that the exact exponents are negative and positive when their
leading terms are negative and positive, respectively. Otherwise, the $c^{\pm}$ branch lines exponents 
in Eqs. (\ref{branch-l-m0}) and (\ref{branch-l}) are a very good approximation.

\subsection{The $(k,\omega)$-plane one-fermion removal spectral singularities on boundary lines}
\label{OESRBL}

There is a second type of high-energy $(k,\omega)$-plane feature in the vicinity of which the PDT provides an analytical expression of 
the one-fermion spectral functions. It is called a boundary line. In the case of the present repulsive fermion model, such boundary lines exist in the 
one-fermion removal spectral function at zero magnetic field and in the
up-spin and down-spin one-fermion removal spectral functions at finite field. It is generated by a subclass of the
general processes behind the branch $A$ or B of the $m=0$ one-fermion removal two-parametric spectrum, Eq. (\ref{SpElremo-m0}),
and $m> 0$ up-spin and down-spin one-fermion removal two-parametric spectra, Eqs. (\ref{SpupElremo}) 
and (\ref{SpdownElremo}), respectively. 

Specifically, the up-spin one-fermion removal boundary line is generated by processes where one 
$c$ band hole is created at a momentum value $q$ and one $s$ band particle is created at a momentum value $q'$ 
such that their group velocities, Eq. (\ref{vcvs}), obey the equality $v_{c}(q) = v_s(q')$.
Similarly, both the $m=0$ one-fermion removal boundary line and the $m> 0$ down-spin one-fermion removal 
boundary line are generated by processes where one 
$c$ band hole is created at a momentum value $q$ and one $s$ band hole is created at a momentum value $q'$ 
such that again $v_{c}(q) = v_s(q')$.
\begin{figure}
\includegraphics[scale=0.50]{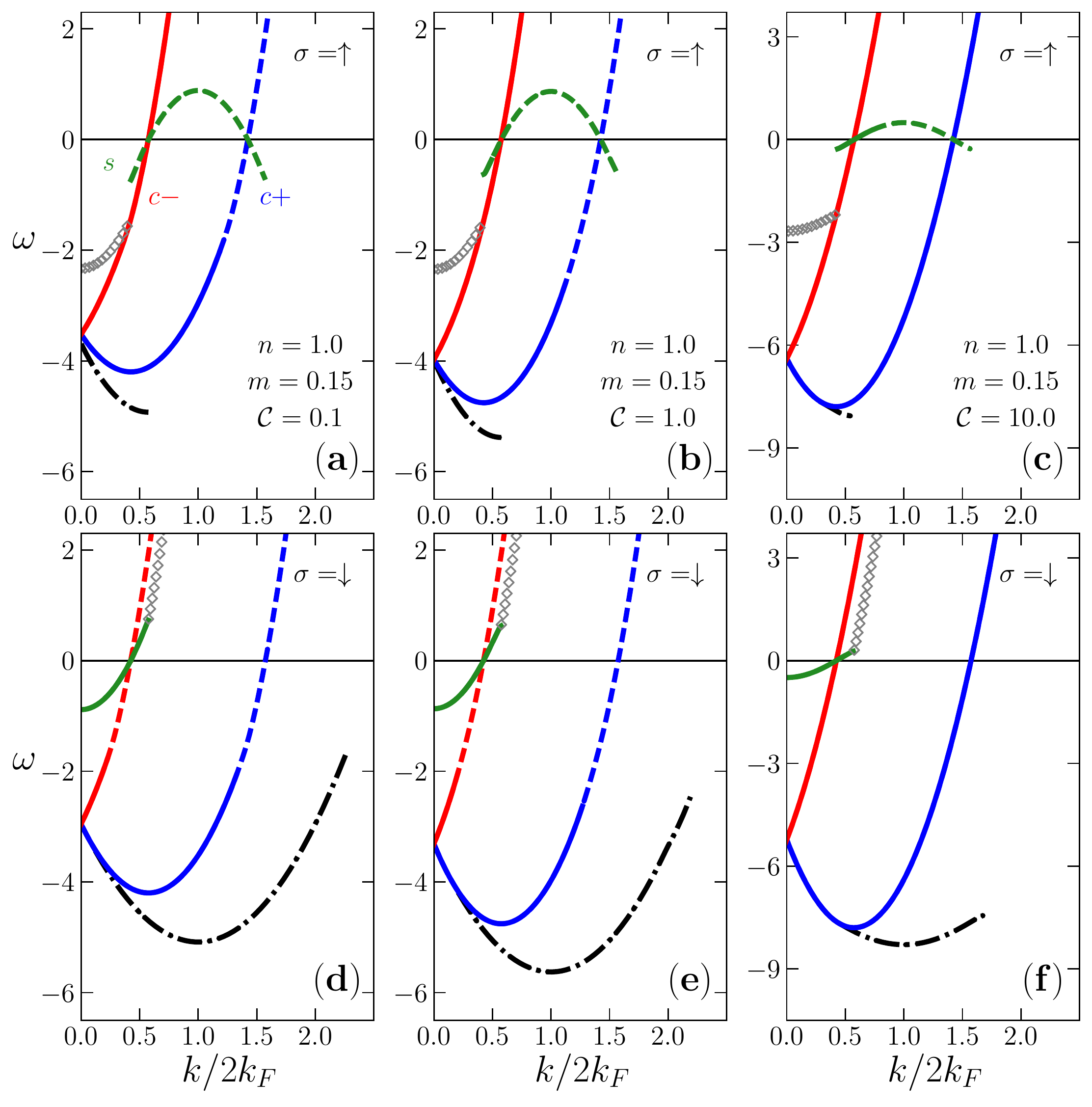}
\caption{The singular branch lines $k$ ranges (solid lines) and other branch lines
$k$ ranges (dashed lines) in units of $2k_F = \pi\,n$ for which the corresponding 
exponent $\xi_{\beta}^{\sigma} (k)$, Eq. (\ref{branch-l}), is negative and positive,
respectively, and the singular boundary lines (dashed-dotted lines) of the weight distribution associated with the up-spin and 
down-spin one-fermion spectral function are plotted in the $(k,\omega)$ plane. 
The curves refer to fermionic density $n=1.0$ and spin density $m=0.15$ for up spin and interaction 
(a) ${\cal{C}}=0.1$, (b) ${\cal{C}}=1.0$, and (c) ${\cal{C}}=10.0$ and for down spin and interaction
(d) ${\cal{C}}=0.1$, (e) ${\cal{C}}=1.0$, and (f) ${\cal{C}}=10.0$.
The branch line spectra plotted here are defined in Section \ref{SFEX}. 
(Online, the $c^+$, $c^-$, and $s$ branch lines appear blue, red, and green, respectively.)
The lines represented by sets of diamond symbols contribute to the ${\cal{C}}\rightarrow 0$ one-fermion spectrum yet are
not branch lines.}
\label{figure2} 
\end{figure} 

That such a feature is a $(k,\omega)$-plane line results from the $c$ band and $s$ band momentum values 
$q$ and $q'$, respectively, not being independent of each other because of the boundary-line velocity equality 
constraint, $v_{c}(q) = v_s(q')$. For the momentum $k$ domains for which those one-fermion removal boundary 
lines exist, they are {\it part} of the limiting line of the corresponding two-dimensional $(k,\omega)$ plane branches.
However and as further discussed below in Sections \ref{OneFBLs-m0} and \ref{UpDownOneFBLs},
most of such a limiting line {\it is not} a boundary line and thus does not correspond to a singularity. 

At zero spin density, $m=0$, the removal one-fermion boundary line $(k,\omega)$-plane 
spectrum has the following general form,
\begin{equation}
\omega_{BL} (k) = \left(-\varepsilon_{c}(q) - \varepsilon_s(q')\right)\,\delta_{v_{c}(q),v_s(q')}
\hspace{0.20cm}{\rm where}\hspace{0.20cm}
k = \pm 2k_F - q  - q' \, .
\label{dE-dP-c-s-m0}
\end{equation}
Near such a line at small energy deviation $(\omega +\omega_{BL} (k))$ values, 
the one-fermion removal spectral function has the following behavior,
\begin{equation}
B_{\gamma} (k,\omega) \propto \Bigl(\omega +\omega_{BL} (k)\Bigr)^{-1/2} \, .
\label{B-bol-m0}
\end{equation}
This expression is determined by the density of the two-parametric states generated upon varying $q$ and $q'$
within the corresponding $c$ and $s$ band values, respectively. 

The up-spin $(\sigma =\uparrow)$ and down-spin $(\sigma =\downarrow)$ one-fermion removal
boundary line $(k,\omega)$-plane spectrum has at spin density $m> 0$ the following general form,
\begin{equation}
\omega_{BL}^{\sigma} (k) = \left(-\varepsilon_{c}(q) + \gamma_{\sigma}\,\varepsilon_s(q')\right)\,\delta_{v_{c}(q),v_s(q')}
\hspace{0.20cm}{\rm where}\hspace{0.20cm}
k = \pm[1-\gamma_{\sigma}]\,k_F - q + \gamma_{\sigma}\,q' \, ,
\label{dE-dP-c-s}
\end{equation}
where,
\begin{equation}
\gamma_{\uparrow} = + 1 \hspace{0.20cm}{\rm and}\hspace{0.20cm} 
\gamma_{\downarrow} = - 1 \, .
\label{c0RAUD}
\end{equation}

In the vicinity of such lines at small energy deviation $(\omega +\omega_{BL}^{\sigma} (k))$ values
the up-spin and down-spin one-fermion removal spectral functions have the following behavior,
\begin{equation}
B_{\sigma,\gamma} (k,\omega) \propto \Bigl(\omega +\omega_{BL}^{\sigma} (k)\Bigr)^{-1/2} \, .
\label{B-bol}
\end{equation}
Again, this expression is determined by the density of the two-parametric states generated upon varying $q$ and $q'$
within the corresponding $c$ and $s$ band values, respectively. 
\begin{figure}
\includegraphics[scale=0.50]{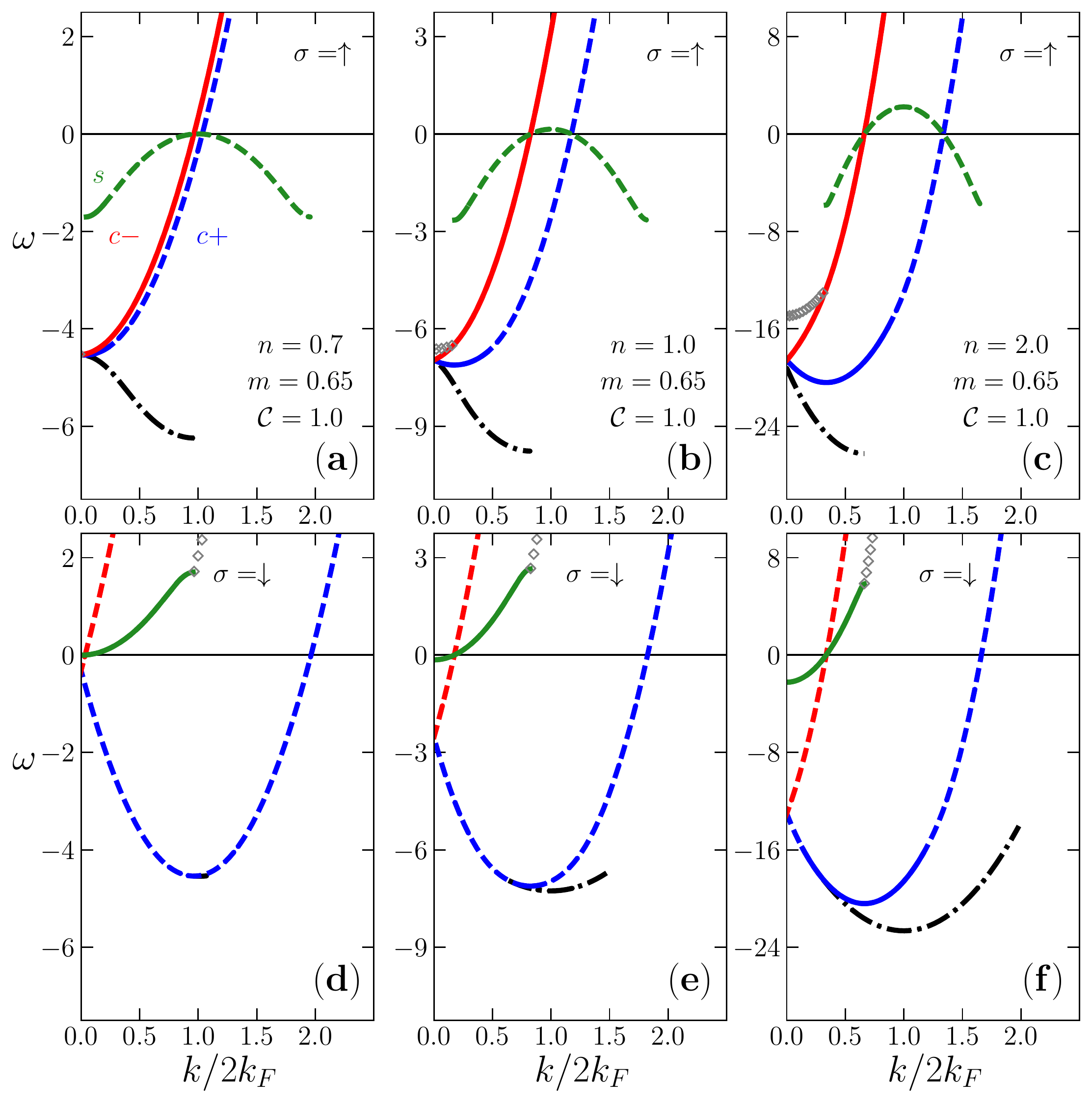}
\caption{The same as Fig. \ref{figure2} for spin density $m=0.65$ and interaction
${\cal{C}}=1.0$ for up spin and fermionic densities
(a) $n=0.7$, (b) $n=1.0$, and (c) $n=2.0$ and for down spin and fermionic densities
(d) $n=0.7$, (e) $n=1.0$, and (f) $n=2.0$.}
\label{figure3} 
\end{figure}

\section{Specific one-fermion removal and addition spectral singularities on branch and boundary lines}
\label{Specific}

In the following, the line shape behavior of the one-fermion spectral function, Eq. (\ref{Bkomega-m0}),
near the branch lines and boundary lines at zero spin density and line shape behavior
of the up-spin and down-spin one-fermion spectral functions, Eq. (\ref{Bkomega}), in the vicinity of the branch lines and boundary lines 
is studied. Such lines are plotted in the $(k,\omega)$-plane in Figs. \ref{figure1}-\ref{figure3}.
The curves refer to repulsive interactions ${\cal{C}}=0.1$, ${\cal{C}}=1.0$, ${\cal{C}}=10.0$, fermionic densities 
$n=0.7$, $n=1.0$, $n=2.0$, and corresponding spin densities $m=0$, $m=0.15$, $m=0.65$ such that $m<n$.
The $c^+,c^-,s$ branch lines are the only branch lines whose exponent is negative for at least some $k$ 
interval and ${\cal{C}}$, $n$, and $m$ ranges. At those $k$ intervals there are singularity cusps in the corresponding 
one-fermion spectral functions. Those branch lines are in Figs. \ref{figure1}-\ref{figure3}
represented by solid lines and dashed lines for the $k$ ranges for which the 
corresponding momentum dependent exponent is negative and positive,
respectively. The one-fermion removal boundary lines also refer to singularity cusps 
and are represented by dashed-dotted lines. 

At zero spin density, $m=0$, {\it all} ${\cal{C}}=0$ non-interacting $\delta$-function like one-fermion spectrum $k$ ranges 
are recovered from specific branch lines in the ${\cal{C}}\rightarrow 0$ limit. For $m>0$ this applies
to most of the ${\cal{C}}=0$ non-interacting $\delta$-function like up-spin and down-spin one-fermion spectrum $k$ ranges. 
The exceptions refer to the ${\cal{C}}=0$ non-interacting up-spin one-fermion removal 
spectrum for the momentum interval $k \in [-k_{F\downarrow},k_{F\downarrow}]$
and to the ${\cal{C}}=0$ non-interacting down-spin one-fermion addition spectrum 
for the momentum intervals $k\in [-\infty,-k_{F\uparrow}]$ and
$k\in [k_{F\uparrow},\infty]$. The corresponding ${\cal{C}}=0$ non-interacting $\delta$-function like one-fermion spectra are in these
$k$ intervals recovered in the ${\cal{C}}\rightarrow 0$ limit from well-defined ${\cal{C}}>0$ spectral features that are 
here called {\it non-branch lines}. Those are represented for $m>0$ in Figs. \ref{figure2} and \ref{figure3} by sets 
of diamond symbols. 

\subsection{The one-fermion removal and addition $c^{\pm}$ branch lines at zero magnetic field}
\label{upRccm0}

At zero magnetic field and thus zero spin density, $m=0$, the
one-fermion removal and addition $c^{\pm}$ branch lines
are generated by one-parametric processes that correspond to particular cases of the
two-parametric processes that generate the spectra
in Eqs. (\ref{SpElremo-m0}) and (\ref{SpEladd-m0}), respectively.
These lines one-parametric spectra are plotted in Fig. \ref{figure1}
where they are contained within such two-parametric spectra.
(Online, the $c^+$ and $c^-$ branch lines are blue and red, respectively, in these figures.)

The one-parametric spectra $\omega_{c^{\pm}} (k)$ and the corresponding exponents 
$\xi_{c^{\pm}} (k)$ associated with these branch lines are related by the following symmetry,
\begin{equation}
\omega_{c^{+}} (k) = \omega_{c^{-}} (-k)\hspace{0.20cm}{\rm and}\hspace{0.20cm}
\xi_{c^{+}} (k) = \xi_{c^{-}} (-k) \, .
\label{c+-rela-m0}
\end{equation}
Considering both the $c^{+}$ and $c^{-}$ branch lines for
$k \in [0,\infty]$ or only the $c^{+}$ branch line for $k \in [-\infty,\infty]$ contains exactly the same information.
Here we chose the latter option. 

The one-fermion removal and addition $c^{+}$ branch line refers to excited energy 
eigenstates with the following number deviations relative to those of the initial ground state,
\begin{equation}
\delta N_c^F = 0 \, ; \hspace{0.20cm} \delta J_c^F = \delta_{\gamma,-1}/2 \, ; \hspace{0.20cm} 
\delta N_c^{NF} = \gamma \, ; \hspace{0.20cm} \delta N_s^F = - \delta_{\gamma,-1}\, ; \hspace{0.20cm} 
\delta J_s^F = \gamma/2 \, .
\nonumber
\end{equation}
\begin{figure}
\includegraphics[scale=0.30]{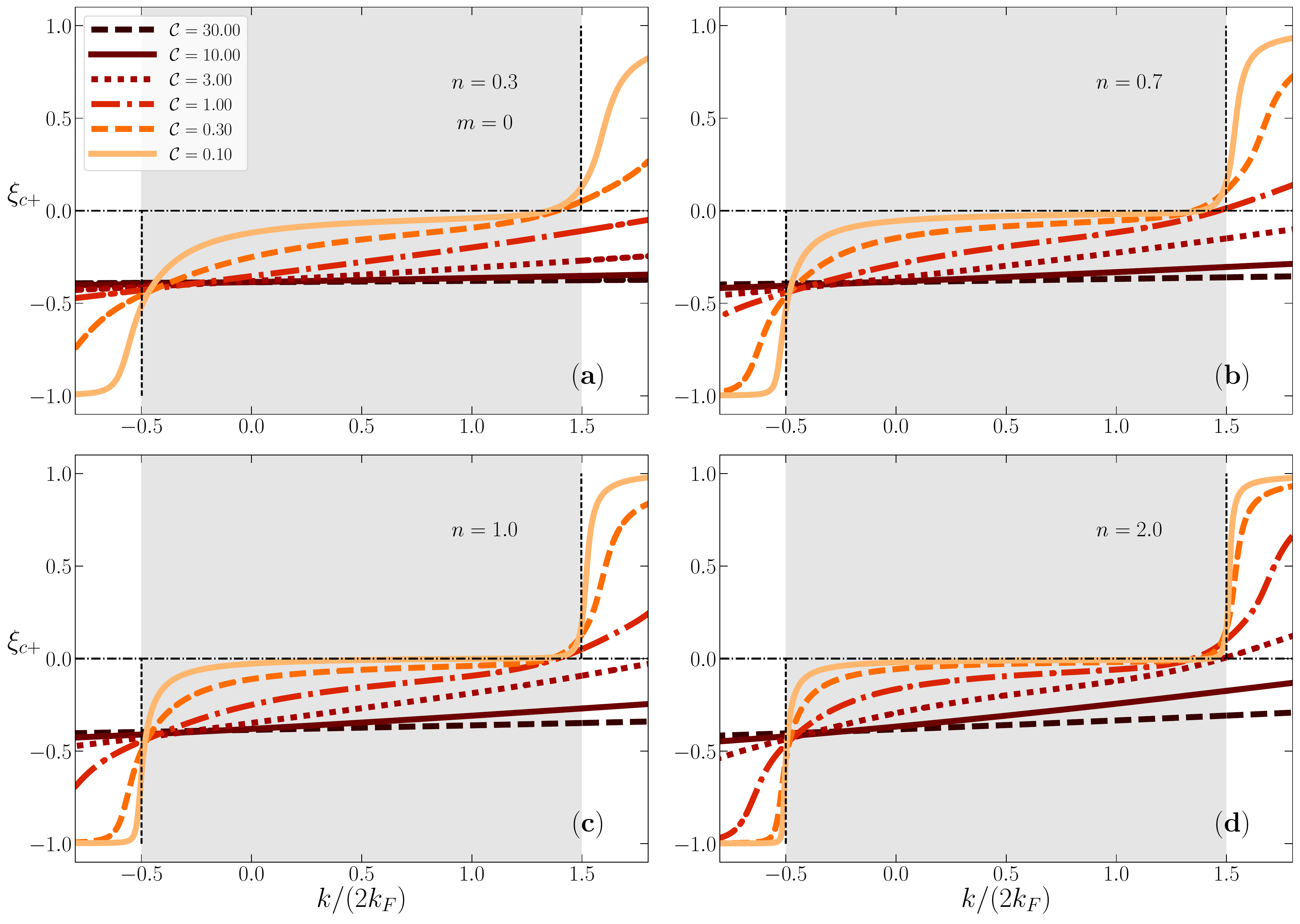}
\caption{The exponent $\xi_{c^{+}} (k)=\xi_{c^{-}} (-k)$, Eq. (\ref{xiRLAcc-m0}),
that controls the singularities in the vicinity of the $c^+$ branch line whose $(k,\omega)$-plane
one-parametric spectrum is defined in Eqs. (\ref{OkudLAcc-m0}) and (\ref{kudLAcc-m0}) 
is plotted for the one-fermion removal and addition spectral function, Eq. (\ref{cpm-branch-m0}), as a function of the momentum
$k/2k_F$. The soft grey region refers to the ground-state occupied Fermi sea.
The curves refer to several ${\cal{C}}=0.10-30.00$ values, fermionic densities 
$n$ (a) $0.3$, (b) $0.7$, (c) $1.0$, (d) $2.0$ and spin density $m= 0$. The type of exponent line associated with
each ${\cal{C}}$ value is for all figures the same. Dashed horizontal and vertical lines denote specific momentum
values between different subbranches and momentum values where the ${\cal{C}}\rightarrow 0$ limiting value
of the exponent changes, respectively.}
\label{figure4}
\end{figure}

The spectrum of general form, Eq. (\ref{dE-dP-bl-m0}), that defines the $(k,\omega)$-plane shape of the 
one-fermion removal and addition $c^+$ branch line is given by,
\begin{eqnarray}
\omega_{c^{+}} (k) & = & \gamma\,\varepsilon_c (q)\hspace{0.20cm}{\rm for}\hspace{0.20cm}\gamma = \pm 1
\hspace{0.20cm}{\rm where}
\nonumber \\
q & \in & [-2k_F,2k_F] \hspace{0.20cm}{\rm for}\hspace{0.20cm}\sigma\hspace{0.20cm}{\rm branch}
\hspace{0.20cm}A\hspace{0.20cm}{\rm fermion}\hspace{0.20cm}\gamma = -1\hspace{0.20cm}{\rm removal} 
\nonumber \\
q & \in & [2k_F,\infty]\hspace{0.20cm}{\rm for}\hspace{0.20cm}\sigma\hspace{0.20cm}{\rm branch}
\hspace{0.20cm}A\hspace{0.20cm}{\rm fermion}\hspace{0.20cm}\gamma = +1\hspace{0.20cm}{\rm addition}
\nonumber \\
q & \in & [-\infty,-2k_F] \hspace{0.20cm}{\rm for}\hspace{0.20cm}\sigma\hspace{0.20cm}{\rm branch}
\hspace{0.20cm}B\hspace{0.20cm}{\rm fermion}\hspace{0.20cm}\gamma = +1\hspace{0.20cm}{\rm addition} \, .
\label{OkudLAcc-m0}
\end{eqnarray}
Here $\varepsilon_c (q)$ is the $c$ band energy dispersion, Eq. (\ref{varepsilon-c-s}) for $\beta =c$.
The excitation momentum $k$ is expressed in terms of the $c$ band momentum $q$ as follows,
\begin{eqnarray}
k & = & \gamma\,q + k_F\hspace{0.20cm}{\rm with}\hspace{0.20cm}{\rm intervals}
\nonumber \\
k &\in & [-k_F,3k_F] \hspace{0.20cm}{\rm for}\hspace{0.20cm}\hspace{0.20cm}{\rm branch}
\hspace{0.20cm}A\hspace{0.20cm}{\rm fermion}\hspace{0.20cm}\gamma = -1\hspace{0.20cm}{\rm removal}
\nonumber \\
k & \in & [3k_F,\infty]\hspace{0.20cm}{\rm for}\hspace{0.20cm}{\rm branch}
\hspace{0.20cm}A\hspace{0.20cm}{\rm fermion}\hspace{0.20cm}\gamma = +1\hspace{0.20cm}{\rm addition}
\nonumber \\
k & \in & [-\infty,-k_F] \hspace{0.20cm}{\rm for}\hspace{0.20cm}{\rm branch}
\hspace{0.20cm}B\hspace{0.20cm}{\rm fermion}\hspace{0.20cm}\gamma = +1\hspace{0.20cm}{\rm addition} \, .
\label{kudLAcc-m0}
\end{eqnarray}
As given in Eq. (\ref{c+-rela-m0}), the corresponding one-fermion 
removal and addition $c^-$ branch line spectrum reads $\omega_{c^{-}} (k) = \omega_{c^{+}} (-k)$.

At excitation momentum $k=k_F$ the removal spectrum is such that,
\begin{equation}
\omega_{c^+} (k_F) = - \varepsilon_c (0)
\hspace{0.2cm}{\rm and}\hspace{0.2cm}
{\partial\omega_{c^+} (k)\over\partial k}\vert_{k=k_F} = 0 \, ,
\nonumber
\end{equation}
where $-\varepsilon_c (0)>0$ is the energy bandwidth of the $c$ band occupied Fermi sea, Eq. (\ref{Wc-expr})
of Appendix \ref{UBAQ} at $m=0$. The limiting behaviors for ${\cal{C}}\rightarrow 0$ and ${\cal{C}}\rightarrow\infty$ of
the spectrum, Eqs. (\ref{OkudLAcc-m0}) and (\ref{kudLAcc-m0}), are given in Eqs.
(\ref{c+BLDCzero-m0}) and (\ref{c+BLDCinfinite-m0}) of Appendix \ref{AIBRL}, respectively.

The use within the PDT of the values of the functional, Eq. (\ref{OESFfunctional-m0}), specific to the excited energy eigenstates that
determine spectral weight distribution near the $c^{\pm}$ branch lines, allows accessing the momentum dependence of the exponents of 
general form, Eq. (\ref{branch-l}), that control such a line shape. The exponent
$\xi_{c^{+}} (k)= \xi_{c^{-}} (-k)$ is found to read,
\begin{equation}
\xi_{c^{+}} (k) = \xi_{c^{-}} (-k) = -{3\over 4} + \sum_{\iota=\pm1}\left({\xi_0\over 4} + \gamma\,\Phi_{c,c}(\iota 2k_F,q)\right)^2 \, ,
\label{xiRLAcc-m0}
\end{equation}
where the parameter $\xi_0$ is defined in Eq. (\ref{xi-0}) of Appendix \ref{UBAQ}. The
phase shift $\Phi_{c,c}(\pm 2k_F,q)$ is defined in Eq. (\ref{PhiqFq}) for $m=0$.
The exponents, Eq. (\ref{xiRLAcc-m0}), are plotted in Fig. \ref{figure4} as a function of the momentum $k$. The curves correspond to 
several ${\cal{C}}=0.10-30.00$ values and fermionic densities $n$ (a) $0.3$, (b) $0.7$, (c) $1.0$, (d) $2.0$.

The specific form of the general PDT expression, Eq. (\ref{branch-l}), of the one-fermion spectral function
$B_{\gamma} (k,\omega)$, Eq. (\ref{Bkomega-m0}), in the vicinity of the present $c^{\pm}$ branch lines is,
\begin{equation}
B_{\gamma} (k,\omega) = C_{\gamma,c^{\pm}} \Bigl(\gamma\omega - \omega_{c^{\pm}} (k)\Bigr)^{\xi_{c^{\pm}} (k)}  
\hspace{0.20cm}{\rm for}\hspace{0.20cm}(\gamma\,\omega - \omega_{c^{\pm}} (k)) \geq 0 
\hspace{0.20cm}{\rm and}\hspace{0.20cm}\gamma = \pm 1 \, .
\label{cpm-branch-m0}
\end{equation}
Here $C_{\gamma,c^{\pm}}$ are constants that have a fixed value 
for the $k$ and $\omega$ ranges corresponding to small values of the energy deviation 
$(\gamma\omega - \omega_{c^{\pm}} (k))$ and the spectra 
$\omega_{c^{+}} (k) = \omega_{c^{-}} (-k)$ in that energy deviation
are given in Eqs. (\ref{OkudLAcc-m0}) and (\ref{kudLAcc-m0}). The exponent $\xi_{c^{+}} (k)=\xi_{c^{-}} (-k)$ is defined in Eq. (\ref{xiRLAcc-m0}).

In the ${\cal{C}}\rightarrow 0$ limit the $c^{+}$ branch line exponent for one-fermion removal ($\gamma=-1$) reads,
\begin{equation}
\lim_{{\cal{C}}\rightarrow 0}\xi_{c^{+}}  (k) = 0 \hspace{0.2cm}{\rm for}\hspace{0.2cm}k \in [-k_F,3k_F]
\hspace{0.2cm}{\rm and}\hspace{0.2cm}\gamma = -1 \, .
\nonumber
\end{equation}
For one-fermion addition ($\gamma=+1$), one finds in that limit,
\begin{eqnarray}
\lim_{{\cal{C}}\rightarrow 0}\xi_{c^{+}} (k) & = & -1\hspace{0.2cm}{\rm for}\hspace{0.2cm}k \in [-\infty,-k_F] 
\hspace{0.2cm}{\rm and}\hspace{0.2cm}\gamma = +1
\nonumber \\
& = 1 & \hspace{0.2cm}{\rm for}\hspace{0.2cm}k \in [3k_F,\infty] 
\hspace{0.2cm}{\rm and}\hspace{0.2cm}\gamma = +1 \, .
\nonumber
\end{eqnarray}
Similar values for the exponent $\xi_{c^{-}} (k)$ are obtained upon exchanging $k$ by $-k$. The important
$c^{-}$ branch line subbranch is that of one-fermion addition for which,
\begin{equation}
\lim_{{\cal{C}}\rightarrow 0}\xi_{c^{-}} (k) = -1 \hspace{0.2cm}{\rm for}\hspace{0.2cm}k \in [k_F,\infty]
\hspace{0.2cm}{\rm and}\hspace{0.2cm}\gamma = +1 \, .
\nonumber
\end{equation}

For the $k$ ranges for which $\lim_{{\cal{C}}\rightarrow 0}\xi_{c^{\pm}} (k) = -1$ for
one-fermion addition, the line shape 
has not the form given in Eq. (\ref{cpm-branch}). It rather is $\delta$-function like,
Eq. (\ref{branch-lexp-1}). In the present case, this gives,
\begin{eqnarray}
\lim_{{\cal{C}}\rightarrow 0} B_{+1} (k,\omega) & = & \delta\Bigl(\omega - \omega_{c^{+}} (k)\Bigr) 
= \delta\Bigl(\omega - (k^2 - k_F^2)\Bigr) \nonumber \\
& & {\rm for}\hspace{0.2cm}k \in [-\infty,-k_F]
\hspace{0.2cm}{\rm and}\hspace{0.2cm}k \in [k_F,\infty] \, ,
\label{cpm-branch-delta-m0}
\end{eqnarray}
where the expression of the $c$ band energy dispersion in the ${\cal{C}}\rightarrow 0$ limit,
Eq. (\ref{varepsilon-c-s-C0}) for $m=0$, has been used.

For the $k$ ranges for which the exponents are for ${\cal{C}}\rightarrow 0$ given by $0$ and/or $1$, the 
one-fermion spectral weight at and near the corresponding branch lines vanishes in the ${\cal{C}}\rightarrow 0$ limit.

In the ${\cal{C}}\rightarrow\infty$ limit the $c^{\pm}$ branch line exponent in Eq. (\ref{xiRLAcc-m0}) has the following values for its
whole $k$ range,
\begin{equation}
\lim_{{\cal{C}}\rightarrow\infty}\xi_{c^{\pm}} (k) = \lim_{{\cal{C}}\rightarrow\infty}\lim_{m\rightarrow 0}\xi_{c^{\pm}}^{\sigma} (k)
= -{3\over 8} \, .
\label{xiudRLAccUim0}
\end{equation}

\subsection{The up-spin and down-spin one-fermion removal and addition $c^{\pm}$ branch lines}
\label{upRcc}

The up-spin and down-spin one-fermion removal and addition $c^{\pm}$ branch lines
are generated by one-parametric processes that correspond to particular cases of the
two-parametric processes that generate the spectra, Eqs. (\ref{SpupElremo})-(\ref{SpdownEladd}).
Hence these lines one-parametric spectra plotted in Figs. \ref{figure2} and \ref{figure3}
are contained within such two-parametric spectra. Those occupy well defined regions in the $(k,\omega)$ plane.

As at zero spin density, Eq. (\ref{c+-rela-m0}), the one-parametric spectra $\omega_{c^{\pm}}^{\sigma} (k)$ 
and the corresponding exponents $\xi_{c^{\pm}}^{\sigma} (k)$
associated with these branch lines are related by the symmetry,
$\omega_{c^{+}}^{\sigma} (k) = \omega_{c^{-}}^{\sigma} (-k)$ and
$\xi_{c^{+}}^{\sigma} (k) = \xi_{c^{-}}^{\sigma} (-k)$.
And again, considering both the $c^{+}$ and $c^{-}$ branch lines for
$k \in [0,\infty]$ or only the $c^{+}$ branch line for $k \in [-\infty,\infty]$ contains exactly the same information.
Here we chose the latter option. 

The up-spin and down-spin one-fermion removal and addition $c^{+}$ branch line refers to excited energy 
eigenstates with the following number deviations relative to those of the initial ground state,
\begin{equation}
\delta N_c^F = 0 \, ; \hspace{0.20cm} \delta J_c^F = \delta_{\sigma,\downarrow}/2 \, ; \hspace{0.20cm} \delta N_c^{NF} = \gamma \, ; \hspace{0.20cm} 
\delta N_s^F = \delta_{\sigma,\downarrow}\,\gamma \, ; \hspace{0.20cm} \delta J_s^F = \gamma_{\sigma}/2 \, .
\nonumber
\end{equation}

The spectrum of general form, Eq. (\ref{dE-dP-bl}), that defines the $(k,\omega)$-plane shape of the 
up-spin and down-spin one-fermion removal and addition $c^+$ branch line reads,
\begin{eqnarray}
\omega_{c^{+}}^{\sigma} (k) & = & \gamma\,\varepsilon_c (q)\hspace{0.20cm}{\rm for}\hspace{0.20cm}\gamma = \pm 1
\hspace{0.20cm}{\rm where} 
\nonumber \\
q & \in & [-2k_F,2k_F] \hspace{0.20cm}{\rm for}\hspace{0.20cm}\sigma\hspace{0.20cm}{\rm branch}
\hspace{0.20cm}A\hspace{0.20cm}{\rm fermion}\hspace{0.20cm}\gamma = -1\hspace{0.20cm}{\rm removal} 
\nonumber \\
q & \in & [2k_F,\infty]\hspace{0.20cm}{\rm for}\hspace{0.20cm}\sigma\hspace{0.20cm}{\rm branch}
\hspace{0.20cm}A\hspace{0.20cm}{\rm fermion}\hspace{0.20cm}\gamma = +1\hspace{0.20cm}{\rm addition}
\nonumber \\
q & \in & [-\infty,-2k_F] \hspace{0.20cm}{\rm for}\hspace{0.20cm}\sigma\hspace{0.20cm}{\rm branch}
\hspace{0.20cm}B\hspace{0.20cm}{\rm fermion}\hspace{0.20cm}\gamma = +1\hspace{0.20cm}{\rm addition} \, .
\label{OkudRLAcc}
\end{eqnarray}
Here $\varepsilon_c (q)$ is the $c$ band energy dispersion, Eq. (\ref{varepsilon-c-s}) for $\beta =c$.
The expression of the excitation momentum $k$ in terms of the $c$ band momentum $q$ is given by,
\begin{eqnarray}
k & = & \gamma\,q + k_{F\bar{\sigma}}\hspace{0.20cm}{\rm where}
\nonumber \\
k &\in & [-k_{F\sigma},(2k_F+k_{F\bar{\sigma}})] \hspace{0.20cm}{\rm for}\hspace{0.20cm}\sigma\hspace{0.20cm}{\rm branch}
\hspace{0.20cm}A\hspace{0.20cm}{\rm fermion}\hspace{0.20cm}\gamma = -1\hspace{0.20cm}{\rm removal}
\nonumber \\
k & \in & [(2k_F+k_{F\bar{\sigma}}),\infty]\hspace{0.20cm}{\rm for}\hspace{0.20cm}\sigma\hspace{0.20cm}{\rm branch}
\hspace{0.20cm}A\hspace{0.20cm}{\rm fermion}\hspace{0.20cm}\gamma = +1\hspace{0.20cm}{\rm addition}
\nonumber \\
k & \in & [-\infty,-k_{F\sigma}] \hspace{0.20cm}{\rm for}\hspace{0.20cm}\sigma\hspace{0.20cm}{\rm branch}
\hspace{0.20cm}B\hspace{0.20cm}{\rm fermion}\hspace{0.20cm}\gamma = +1\hspace{0.20cm}{\rm addition} \, .
\label{OkudLAcc}
\end{eqnarray}

In this equation,
\begin{equation}
{\bar{\uparrow}} = \downarrow \hspace{0.20cm}{\rm and}\hspace{0.20cm} 
{\bar{\downarrow}} = \uparrow \, ,
\label{barsigma}
\end{equation}
so that $k_{F\bar{\uparrow}} = k_{F\downarrow}$ and $k_{F\bar{\downarrow}} = k_{F\uparrow}$.
The two-parametric spectra branches $A$ and $B$ where the $c^+$ branch line is contained
are defined in Eqs. (\ref{SpupElremo})-(\ref{SpdownEladd}).
The corresponding $k$ intervals of the $c^-$ branch line subbranches are obtained from those provided here upon
exchanging $k$ by $-k$.
\begin{figure}
\includegraphics[scale=0.30]{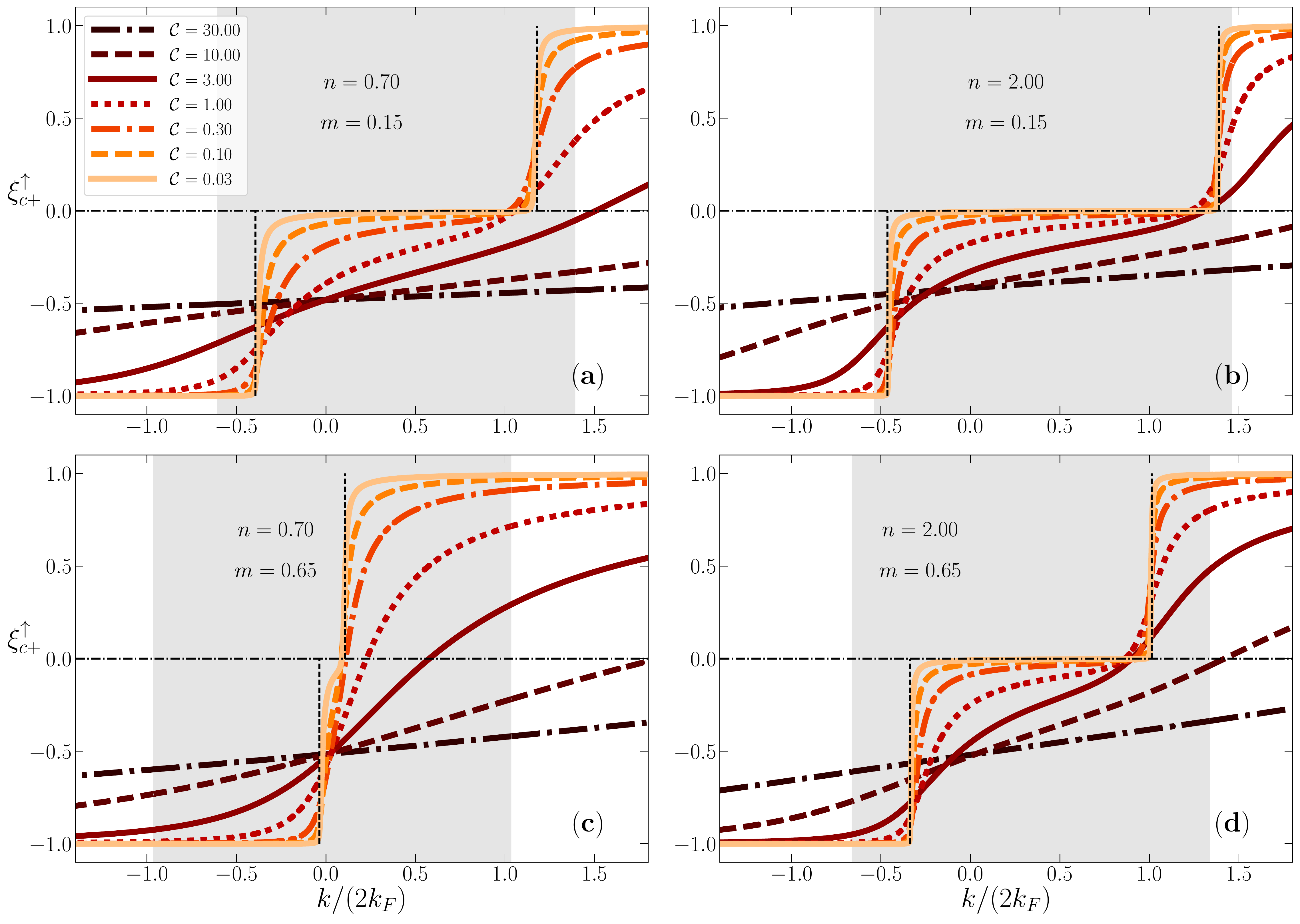}
\caption{The exponent $\xi_{c^{+}}^{\uparrow} (k)=\xi_{c^{-}}^{\uparrow} (-k)$, Eq. (\ref{xiupRLAcc}),
that controls the singularities in the vicinity of the $c^+$ branch line whose $(k,\omega)$-plane
shape is defined by Eqs. (\ref{OkudRLAcc}) and (\ref{OkudLAcc}) for $\sigma =\uparrow$ is plotted 
for the up-spin one-fermion removal and addition spectral function, Eq. (\ref{cpm-branch}) for $\sigma =\uparrow$, 
as a function of the momentum $k/2k_F$. The soft grey region refers to the ground-state occupied Fermi sea.
The curves refer to several ${\cal{C}}=0.03-30.00$ values and spin density $m=0.15$ and
fermionic densities $n$ (a) $0.7$ and (b) $2.0$ and spin density $m=0.65$ and
fermionic densities (c) $0.7$ and (d) $2.0$.}
\label{figure5}
\end{figure}

Combined analysis of the momentum $k$ intervals in Eq. (\ref{OkudLAcc}) with the relation 
$\omega_{c^{+}}^{\sigma} (k) = \omega_{c^{-}}^{\sigma} (-k)$ reveals that the up-spin and down-spin one-fermion
addition $c^{\pm}$ branch lines are the natural continuation of the up-spin and down-spin one-fermion 
removal $c^{\pm}$ branch lines, respectively.

The limiting behaviors for ${\cal{C}}\rightarrow 0$ and ${\cal{C}}\rightarrow\infty$ of
the up-spin and down-spin one-fermion $c^+$ branch-line spectra, Eqs. (\ref{OkudRLAcc}) 
and (\ref{OkudLAcc}), are given in Eqs. (\ref{c+BLDCzero}) and (\ref{c+BLDCinfinite})
of Appendix \ref{AIBRL}, respectively.

We use the values of the functional, Eq. (\ref{OESFfunctional}), specific to the excited energy eigenstates that
determine spectral weight distribution near the $c^{\pm}$ branch lines, to access the momentum dependence of the exponents of 
general form, Eq. (\ref{branch-l}), that control such a line shape. One finds,
\begin{equation}
\xi_{c^{+}}^{\uparrow} (k) = \xi_{c^{-}}^{\uparrow} (-k) = -1 + \sum_{\iota=\pm1}\left({\xi_{c\,s}^1\over 2} + \gamma\,\Phi_{c,c}(\iota 2k_F,q)\right)^2 
+ \sum_{\iota=\pm1}\left({\xi_{s\,s}^1\over 2} + \gamma\,\Phi_{s,c}(\iota k_{F\downarrow},q)\right)^2 \, ,
\label{xiupRLAcc}
\end{equation}
for the up-spin one-fermion $c^{\pm}$ branch lines and,
\begin{eqnarray}
\xi_{c^{+}}^{\downarrow} (k) & = & \xi_{c^{-}}^{\downarrow} (-k) = -1 + 
\sum_{\iota=\pm1}\left({\iota\,\gamma\,\xi_{c\,s}^0\over 2} + {(\xi_{c\,c}^1-\xi_{c\,s}^1)\over 2} 
+ \gamma\,\Phi_{c,c}(\iota 2k_F,q)\right)^2 
\nonumber \\
& + & \sum_{\iota=\pm1}\left({\iota\,\gamma\,\xi_{s\,s}^0\over 2} + {(\xi_{s\,c}^1-\xi_{s\,s}^1)\over 2} 
+ \gamma\,\Phi_{s,c}(\iota k_{F\downarrow},q)\right)^2 \, ,
\label{xidownRLAcc}
\end{eqnarray}
for the down-spin one-fermion $c^{\pm}$ branch lines. The phase shifts $\Phi_{c,c}(\pm 2k_F,q)$ and 
$\Phi_{s,c}(\pm k_{F\downarrow},q)$ in those exponents expressions are defined in Eq. (\ref{PhiqFq}) 
and the $j=0,1$ parameters $\xi^{j}_{\beta\,\beta'}$ are defined in Eq. (\ref{x-aa}).

\begin{figure}
\includegraphics[scale=0.30]{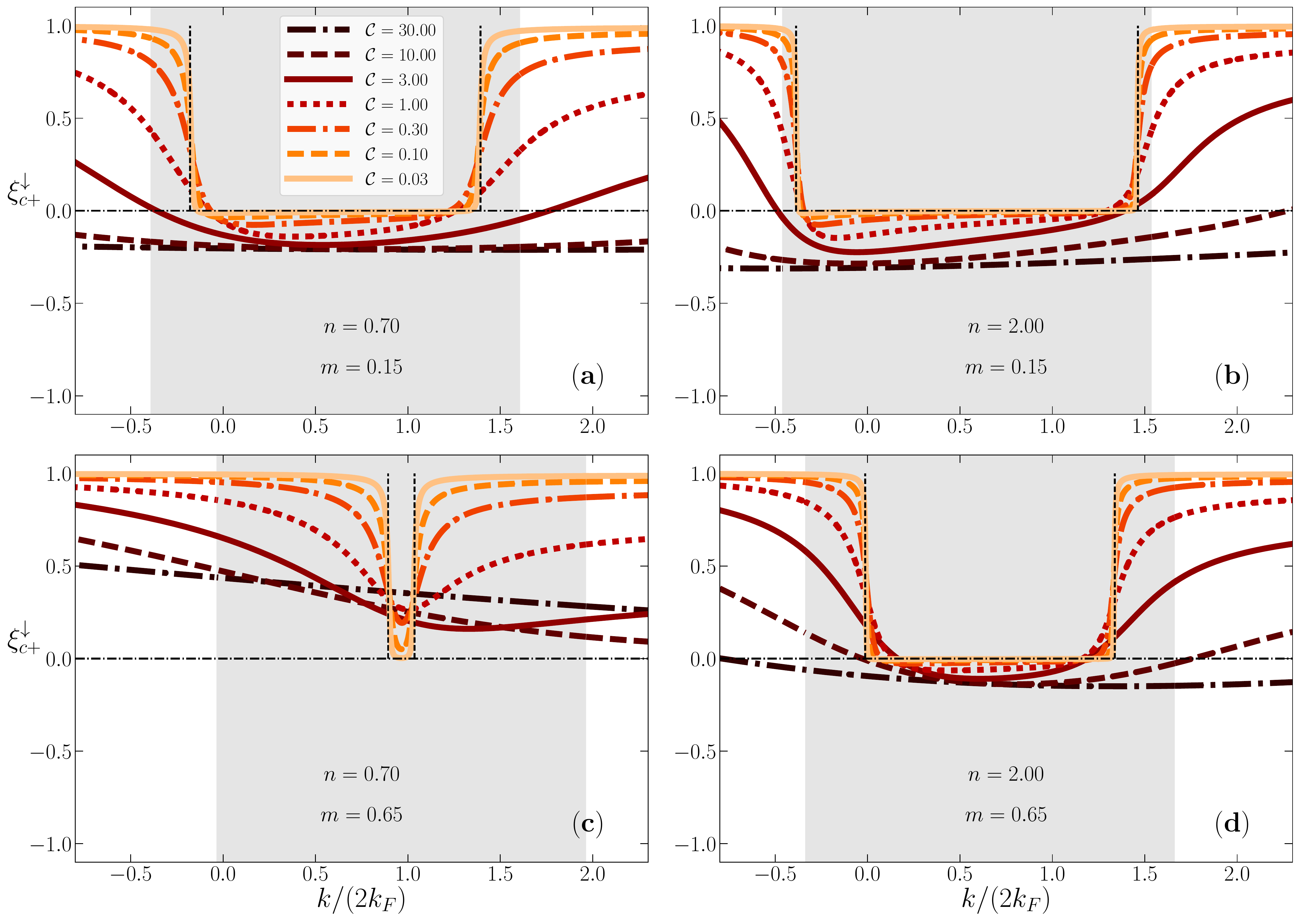}
\caption{The exponent $\xi_{c^{+}}^{\downarrow} (k)=\xi_{c^{-}}^{\downarrow} (-k)$, Eq. (\ref{xidownRLAcc}),
that controls the singularities in the vicinity of the $c^+$ branch line whose $(k,\omega)$-plane
one-parametric spectrum is defined by Eqs. (\ref{OkudRLAcc}) and (\ref{OkudLAcc}) for $\sigma =\downarrow$
is plotted for the one-fermion removal and addition spectral function, Eq. (\ref{cpm-branch}) for $\sigma =\downarrow$, 
as a function of the momentum $k/2k_F$. The soft grey region refers to the ground-state occupied Fermi sea.
The curves refer to several ${\cal{C}}=0.03-30.00$ values and spin density $m=0.15$ and
fermionic densities $n$ (a) $0.7$ and (b) $2.0$ and spin density $m=0.65$ and
fermionic densities (c) $0.7$ and (d) $2.0$. The type of exponent line associated with
each ${\cal{C}}$ value is for all figures the same. Dashed horizontal and vertical lines denote specific momentum
values between different subbranches and momentum values where the ${\cal{C}}\rightarrow 0$ limiting value
of the exponent changes, respectively.}
\label{figure6}
\end{figure}

The up-spin and down-spin one-fermion exponents are plotted in Figs. \ref{figure5} and 
\ref{figure6}, respectively, as a function of the momentum $k$. The curves correspond to several ${\cal{C}}=0.03-30.00$ values, 
fermionic densities $n =0.7$ and $2.0$, and spin densities $m=0.15$ and $m=0.65$.

The specific form of the general expression, Eq. (\ref{branch-l}), of the up-spin and down-spin one-fermion spectral function
$B_{\sigma,\gamma} (k,\omega)$, Eq. (\ref{Bkomega}), in the vicinity of the present $c^{\pm}$ branch lines is,
\begin{equation}
B_{\sigma,\gamma} (k,\omega) = C_{\sigma,\gamma,c^{\pm}} \Bigl(\gamma\omega - \omega_{c^{\pm}}^{\sigma} (k)\Bigr)^{\xi_{c^{\pm}}^{\sigma} (k)}  
\hspace{0.20cm}{\rm for}\hspace{0.20cm}(\gamma\,\omega - \omega_{c^{\pm}}^{\sigma} (k)) \geq 0 
\hspace{0.20cm}{\rm where}\hspace{0.20cm}\gamma = \pm 1 \, ,
\label{cpm-branch}
\end{equation}
and $C_{\sigma,\gamma,c^{\pm}}$ are constants that have a fixed value 
for the $k$ and $\omega$ ranges corresponding to small values of the energy deviation 
$(\gamma\omega - \omega_{c^{\pm}}^{\sigma} (k))$. The spectra $\omega_{c^{+}}^{\sigma} (k) = \omega_{c^{-}}^{\sigma} (-k)$ 
in that energy deviation are given in Eqs. (\ref{OkudRLAcc}) and (\ref{OkudLAcc})
and the exponents $\xi_{c^{+}}^{\sigma} (k)=\xi_{c^{-}}^{\sigma} (-k)$ are defined in Eqs. (\ref{xiupRLAcc}) 
and (\ref{xidownRLAcc}) for $\sigma =\uparrow$ and $\sigma =\downarrow$, respectively.

In ${\cal{C}}\rightarrow 0$ limit the $c^{+}$ branch line exponents for up-spin one-fermion removal ($\gamma=-1$) read,
\begin{eqnarray}
\lim_{{\cal{C}}\rightarrow 0}\xi_{c^{+}}^{\uparrow} (k) & = & -1\hspace{0.2cm}{\rm for}\hspace{0.2cm}k \in [-k_{F\uparrow},-k_{F\downarrow}] 
\hspace{0.2cm}{\rm and}\hspace{0.2cm}\gamma = -1
\nonumber \\
& = & 0 \hspace{0.2cm}{\rm for}\hspace{0.2cm}k \in [-k_{F\downarrow},3k_{F\downarrow}]\hspace{0.2cm}{\rm and}\hspace{0.2cm}\gamma = -1
\nonumber \\
& = & 1 \hspace{0.2cm}{\rm for}\hspace{0.2cm}k \in [3k_{F\downarrow},(2k_F+k_{F\downarrow})]\hspace{0.2cm}{\rm and}\hspace{0.2cm}\gamma = -1 \, .
\label{xiupRc+U0--1}
\end{eqnarray}

For one-fermion addition ($\gamma=+1$), one finds,
\begin{eqnarray}
\lim_{{\cal{C}}\rightarrow 0}\xi_{c^{+}}^{\uparrow} (k) & = & -1\hspace{0.2cm}{\rm for}\hspace{0.2cm}k \in [-\infty,-k_{F\uparrow}] 
\hspace{0.2cm}{\rm and}\hspace{0.2cm}\gamma = +1
\nonumber \\
& = 1 & \hspace{0.2cm}{\rm for}\hspace{0.2cm}k \in [(2k_F+k_{F\bar{\sigma}}),\infty] 
\hspace{0.2cm}{\rm and}\hspace{0.2cm}\gamma = +1 \, .
\label{xiupLAccU0}
\end{eqnarray}

Similar values for the exponent $\xi_{c^{-}}^{\uparrow} (k)$ are obtained upon exchanging $k$ by $-k$. Important
$c^{-}$ branch line subbranches are those for which $\lim_{{\cal{C}}\rightarrow 0}\xi_{c^{-}}^{\uparrow} (k) = -1$. They
refer to the $k$ ranges,
\begin{eqnarray}
\lim_{{\cal{C}}\rightarrow 0}\xi_{c^{-}}^{\uparrow} (k) & = & -1 \hspace{0.2cm}{\rm for}\hspace{0.2cm}k \in [k_{F\downarrow},k_{F\uparrow}] 
\hspace{0.2cm}{\rm and}\hspace{0.2cm}\gamma = -1
\nonumber \\
& = & -1 \hspace{0.2cm}{\rm for}\hspace{0.2cm}k \in [k_{F\uparrow},\infty]
\hspace{0.2cm}{\rm and}\hspace{0.2cm}\gamma = +1 \, .
\nonumber
\end{eqnarray}

As discussed previously, for the $k$ ranges for which $\lim_{{\cal{C}}\rightarrow 0}\xi_{c^{\pm}}^{\uparrow} (k) = -1$, the line shape 
has not the form given in Eq. (\ref{cpm-branch}). It rather is $\delta$-function like,
Eq. (\ref{branch-lexp-1}). In the present case this gives,
\begin{eqnarray}
\lim_{{\cal{C}}\rightarrow 0} B_{\uparrow,-1} (k,\omega) & = & \delta\Bigl(\omega + \omega_{c^{+}}^{\uparrow} (k)\Bigr) 
= \delta\Bigl(\omega - (k^2 - k_{F\uparrow}^2)\Bigr)  
\nonumber \\
& & {\rm for}\hspace{0.2cm}k \in [-k_{F\uparrow},-k_{F\downarrow}]
\nonumber \\
\lim_{{\cal{C}}\rightarrow 0} B_{\uparrow,-1} (k,\omega) & = & \delta\Bigl(\omega + \omega_{c^{-}}^{\uparrow} (k)\Bigr) 
= \delta\Bigl(\omega - (k^2 - k_{F\uparrow}^2)\Bigr)  
\nonumber \\
& & {\rm for}\hspace{0.2cm}k \in [k_{F\downarrow},k_{F\uparrow}]
\nonumber \\
\lim_{{\cal{C}}\rightarrow 0} B_{\uparrow,+1} (k,\omega) & = & \delta\Bigl(\omega - \omega_{c^{+}}^{\uparrow} (k)\Bigr) 
= \delta\Bigl(\omega - (k^2 - k_{F\uparrow}^2)\Bigr) \nonumber \\
& & {\rm for}\hspace{0.2cm}k \in [-\infty,-k_{F\uparrow}]
\nonumber \\
\lim_{{\cal{C}}\rightarrow 0} B_{\uparrow,+1} (k,\omega) & = & \delta\Bigl(\omega - \omega_{c^{-}}^{\uparrow} (k)\Bigr) 
= \delta\Bigl(\omega - (k^2 - k_{F\uparrow}^2)\Bigr) \nonumber \\
& & {\rm for}\hspace{0.2cm}k \in [k_{F\uparrow},\infty] \, ,
\label{cpm-branch-delta}
\end{eqnarray}
where the expression of the $c$ band energy dispersion in the ${\cal{C}}\rightarrow 0$ limit,
Eq. (\ref{varepsilon-c-s-C0}), has been used. The spectra $\omega_{c^{\pm}}^{\uparrow} (k)$ 
indeed become in the ${\cal{C}}\rightarrow 0$ limit the corresponding exact ${\cal{C}}=0$ non-interacting spectra.
In the case of the up-spin one-fermion spectra this applies to all its momentum intervals except for 
$k\in [-k_{F\uparrow},k_{F\uparrow}]$.

For the excitation momentum $k$ intervals for which the exponents are for ${\cal{C}}\rightarrow 0$ given by $0$ and/or $1$, the 
up-spin one-fermion spectral weight at and near the corresponding branch lines vanishes in the ${\cal{C}}\rightarrow 0$ limit.
Specifically, one finds that in the ${\cal{C}}\rightarrow 0$ limit the down-spin one-fermion removal exponent, Eq. (\ref{xidownRLAcc}), 
has the following behaviors,
\begin{eqnarray}
\lim_{{\cal{C}}\rightarrow 0}\xi_{c^{+}}^{\downarrow} (k) & = & 1\hspace{0.2cm}{\rm for}\hspace{0.2cm}
k \in [-k_{F\downarrow},(k_{F\uparrow}-2k_{F\downarrow})]\hspace{0.2cm}{\rm and}\hspace{0.2cm}\gamma = -1
\nonumber \\
\lim_{{\cal{C}}\rightarrow 0}\xi_{c^{+}}^{\downarrow} (k) & = & 0\hspace{0.2cm}{\rm for}\hspace{0.2cm}
k \in [(k_{F\uparrow}-2k_{F\downarrow}),(2k_F+k_{F\downarrow})]\hspace{0.2cm}{\rm and}\hspace{0.2cm}\gamma = -1
\nonumber \\
\lim_{{\cal{C}}\rightarrow 0}\xi_{c^{+}}^{\downarrow} (k) & = & 1\hspace{0.2cm}{\rm for}\hspace{0.2cm}
k \in [(2k_F+k_{F\downarrow}),(2k_F+k_{F\uparrow})]\hspace{0.2cm}{\rm and}\hspace{0.2cm}\gamma = -1 \, .
\label{xidownRc+U0-0}
\end{eqnarray}
The down-spin one-fermion addition exponent is found to behave in that limit as,
\begin{equation}
\lim_{{\cal{C}}\rightarrow 0}\xi_{c^{+}}^{\downarrow} (k) = 1\hspace{0.2cm}{\rm for}\hspace{0.2cm}
k\in [-\infty,-k_{F\downarrow}]\, ; k\in [(2k_F+k_{F\uparrow}),\infty]\hspace{0.2cm}{\rm and}\hspace{0.2cm}\gamma = +1 \, .
\label{xidownRccU0}
\end{equation}
Hence the down-spin one-fermion spectral weight at and near these branch lines vanishes in the ${\cal{C}}\rightarrow 0$ limit both
for one-fermion removal and addition. Similar values for the exponent $\xi_{c^{-}}^{\downarrow} (k)$ are obtained upon exchanging $k$ by $-k$. 

In the ${\cal{C}}\rightarrow\infty$ limit the $c^{\pm}$ branch lines exponents in 
Eqs. (\ref{xiupRLAcc}) and (\ref{xidownRLAcc}) have for $m \rightarrow n$ the following values for their whole $k$ intervals,
\begin{equation}
\lim_{{\cal{C}}\rightarrow\infty}\xi_{c^{\pm}}^{\sigma} (k) = 
- {\gamma_{\sigma}\over 2} \hspace{0.2cm}{\rm for}\hspace{0.2cm} m \rightarrow n \, .
\label{xiudRLAccUimn}
\end{equation}

On the one hand and as shown in Fig. \ref{figure5}, the main effect on the $k$ dependence of the 
up-spin one-fermion removal and addition exponent 
$\xi_{c^{+}}^{\uparrow} (k) = \xi_{c^{-}}^{\uparrow} (-k)$, Eq. (\ref{xiupRLAcc}), of increasing the interaction
${\cal{C}}$ from ${\cal{C}}\ll 1$ to ${\cal{C}}\gg 1$ is to continuously changing its ${\cal{C}}\rightarrow 0$ values $-1$,
$0$, and $1$ for the $k$ ranges given in Eqs. (\ref{xiupRc+U0--1}) and (\ref{xiupLAccU0})
to a $k$ independent value for $k\in [-\infty,\infty]$ as ${\cal{C}}\rightarrow\infty$,
Eq. (\ref{xiudRLAccUim0}) and Eq. (\ref{xiudRLAccUimn}) for $\sigma=\uparrow$. The latter smoothly changes 
from $-3/8$ for $m\rightarrow 0$ to $-1/2$ for $m\rightarrow n$.
The general trend of such an exponent ${\cal{C}}$ dependence is the following: For the momentum
$k$ ranges for which it reads $0$ and $1$ in the ${\cal{C}}\rightarrow 0$ limit, it decreases
upon increasing ${\cal{C}}$; For the $k$ intervals for which it is given by $-1$
in that limit, it rather increases for increasing ${\cal{C}}$ values.

On the other hand, the exponent $\xi_{c^{+}}^{\downarrow} (k)=\xi_{c^{-}}^{\downarrow} (k)$, Eq. (\ref{xidownRLAcc}),
plotted in Fig. \ref{figure6} becomes negative only for large ${\cal{C}}$ and small spin density values.
For ${\cal{C}}\rightarrow 0$ it reads $0$ and $1$ for the $k$ intervals provided in Eqs. (\ref{xidownRc+U0-0}) and (\ref{xidownRccU0}).
As ${\cal{C}}\rightarrow\infty$ it continuously evolves to a $k$ independent value for $k\in [-\infty,\infty]$.
Such a value smoothly changes from $-3/8$ for $m\rightarrow 0$ to $1/2$ for $m\rightarrow n$.
The general trend of that exponent ${\cal{C}}$ dependence is different upon changing the densities.
As shown in Fig. \ref{figure6}, for some densities it always decreases upon increasing ${\cal{C}}$.
For other densities it first decreases upon increasing ${\cal{C}}$ until reaching some minimum
at a finite ${\cal{C}}$ value above which it increases upon further increasing ${\cal{C}}$.

\subsection{The one-fermion removal and addition $s$ branch line at zero magnetic field}
\label{upRs-m0}

The one-parametric spectrum of this branch line is an even function of $k$,
$\omega_s (k) = \omega_s (-k)$. The corresponding
exponent given below is also an even function of $k$, $\xi_s (k) = \xi_s (-k)$.
Hence for simplicity we restrict our following analysis to $k \geq 0$. For
such a momentum range the one-fermion removal and addition parts
of the $s$ branch line refer to excited energy eigenstates with the following number deviations relative 
to those of the initial ground state,
\begin{equation}
\delta N_c^F = \gamma \, ; \hspace{0.20cm} \delta J_c^F = \delta_{\gamma, +1}/2 \, ; \hspace{0.20cm} 
\delta N_s^F =  \delta_{\gamma, +1} \, ; \hspace{0.20cm} \delta J_s^F = 0 \, ; \hspace{0.20cm} 
\delta N_s^{NF} = - 1  \, .
\nonumber
\end{equation}

The spectrum $\omega_s (k)$ of general form, Eq. (\ref{dE-dP-bl-m0}), is for the present branch line at $k>0$ given by,
\begin{eqnarray}
\omega_s (k) & = & - \varepsilon_s (q') \hspace{0.20cm}{\rm where}
\nonumber \\
q' & \in & [-k_F,0]  \hspace{0.20cm}{\rm for}\hspace{0.20cm}{\rm branch}
\hspace{0.20cm}A\hspace{0.20cm}{\rm fermion}\hspace{0.20cm}{\rm removal}
\hspace{0.20cm}\gamma = -1 
\nonumber \\
q' & \in & [-k_F,k_F] \hspace{0.20cm}{\rm for}\hspace{0.20cm}{\rm branch}
\hspace{0.20cm}A\hspace{0.20cm}{\rm fermion}\hspace{0.20cm}{\rm addition}
\hspace{0.20cm}\gamma = +1 \, .
\label{OkudRs-m0}
\end{eqnarray}

The relation of the $s$ band momentum $q'$ to the excitation momentum $k$ is,
\begin{equation}
k = \delta_{\gamma,+1}\,2k_F - q' \geq 0 \, .
\label{kqsup-m0}
\end{equation}
The corresponding intervals of the excitation momentum $k$ are,
\begin{eqnarray}
k & \in & [0,k_F]  \hspace{0.20cm}{\rm for}\hspace{0.20cm}{\rm fermion}\hspace{0.20cm}{\rm removal}
\hspace{0.20cm}\gamma = -1 
\nonumber \\
k & \in & [k_F,3k_F] \hspace{0.20cm}{\rm for}\hspace{0.20cm}{\rm fermion}\hspace{0.20cm}{\rm addition}
\hspace{0.20cm}\gamma = +1 \, .
\label{OkRLAsoth-m0}
\end{eqnarray}
\begin{figure}
\includegraphics[scale=0.30]{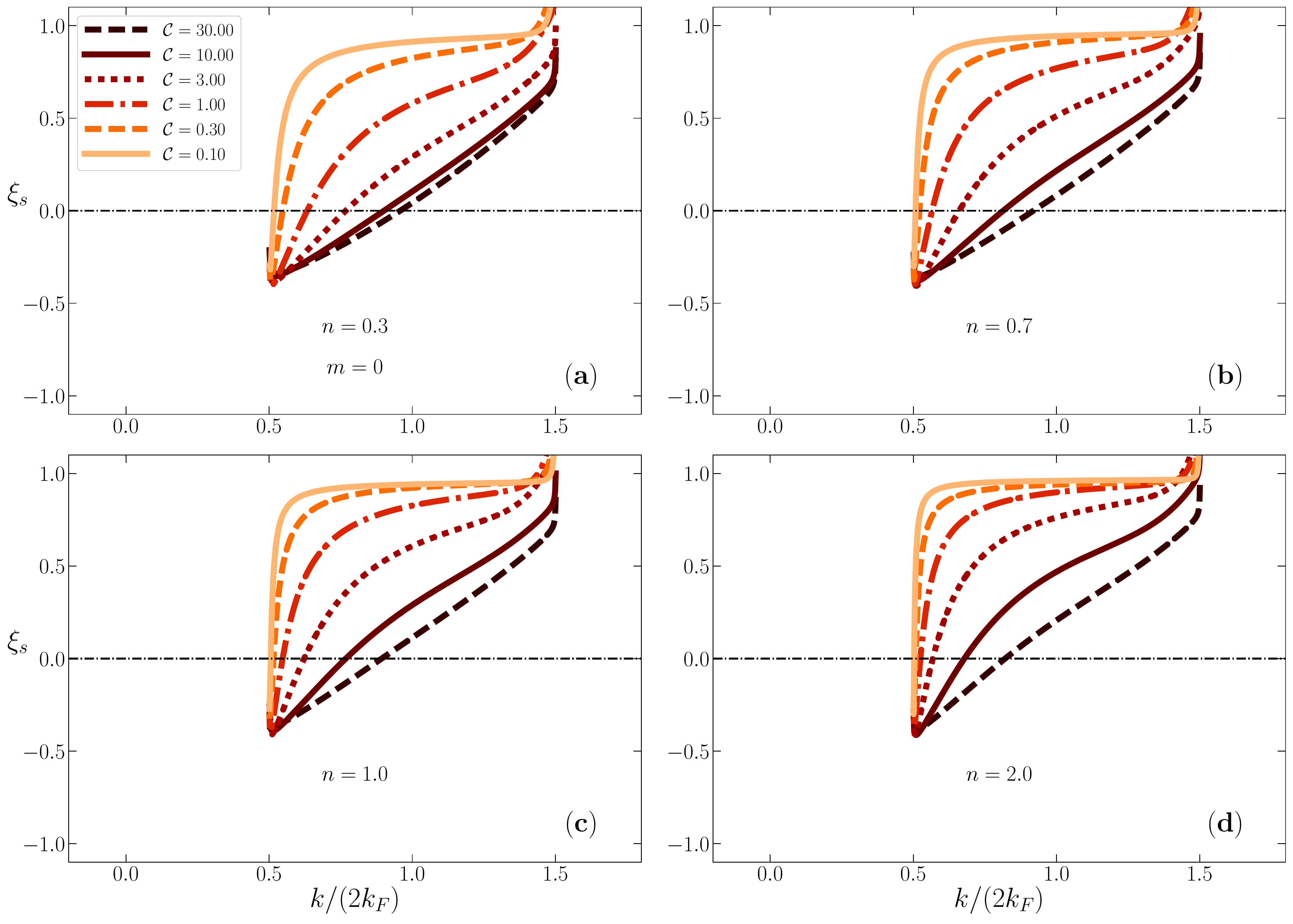}
\caption{The exponent $\xi_{s} (k)$, Eq. (\ref{xiRsUi-m0}) for $\gamma = +1$,
that controls the singularities in the vicinity of the $s$ branch line whose $(k,\omega)$-plane
shape is defined by Eqs. (\ref{OkudRs-m0}) and (\ref{OkRLAsoth-m0}) for $\gamma = +1$ is plotted for the 
one-fermion addition spectral function, Eq. (\ref{s-branch-m0}) for $\gamma = +1$, as a function of the momentum
$k/2k_F$. The curves refer to several ${\cal{C}}=0.10-30.00$ values, fermionic densities 
$n$ (a) $0.3$, (b) $0.7$, (c) $1.0$, (d) $2.0$ and spin density $m = 0$. The type of exponent line associated with
each ${\cal{C}}$ value is for all figures the same.}
\label{figure7}
\end{figure}

The limiting behavior for ${\cal{C}}\rightarrow 0$ of the spectrum, Eqs. (\ref{OkudRs-m0}) and (\ref{kqsup-m0}), is given
in Eq. (\ref{sBLDCzero-m0}) of Appendix \ref{AIBRL}.
For ${\cal{C}}\rightarrow\infty$ it reads $\omega_s (k)=0$ for its whole $k$ intervals.

One finds from inspection of the momentum $k$ intervals in Eq. (\ref{OkRLAsoth-m0}) that the 
one-fermion addition $s$ branch line is the natural continuation of the 
one-fermion removal $s$ branch line. The momentum dependent exponent of general form, Eq. (\ref{branch-l-m0}),
that controls the line shape near the one-fermion removal and addition $s$ branch line is given by,
\begin{equation}
\xi_s (k) = - 1 + \sum_{\iota=\pm1}\left({\iota\over 2\xi_0} + (1+\gamma){\xi_0\over 4}
- \gamma\,\Phi_{c,s}(\iota 2k_F,q')\right)^2 \, .
\label{xiRLAs-m0}
\end{equation}
As reported above, the parameter $\xi_0$ is defined by Eq. (\ref{xi-0}) of Appendix \ref{UBAQ}.
The phase shift $\Phi_{c,s}(\pm 2k_F,q')$ is defined in Eq. (\ref{PhiqFq}) for $m=0$.

\begin{figure}
\includegraphics[scale=0.30]{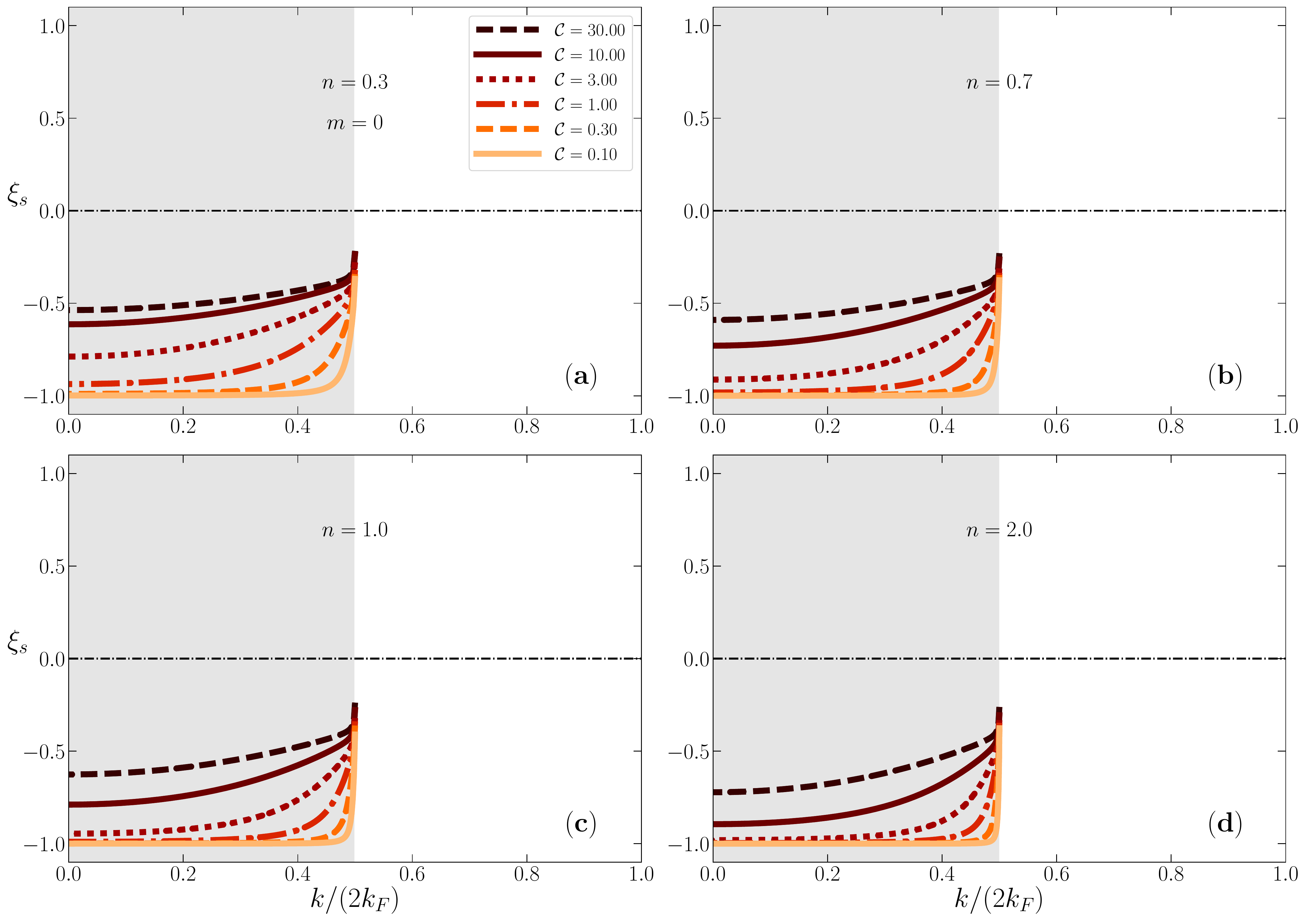}
\caption{The exponent $\xi_{s} (k)$, Eq. (\ref{xiRsUi-m0}) for $\gamma = -1$,
that controls the singularities in the vicinity of the $s$ branch line whose $(k,\omega)$-plane
shape is defined by Eqs. (\ref{OkudRs-m0}) and (\ref{OkRLAsoth-m0}) for $\gamma = -1$ is plotted for the 
one-fermion removal spectral function, Eq. (\ref{s-branch-m0}) for $\gamma = -1$, as a function of the momentum
$k/2k_F$. The soft grey region refers to the ground-state occupied Fermi sea.
The curves refer to several ${\cal{C}}=0.10-30.00$ values, fermionic densities 
$n$ (a) $0.3$, (b) $0.7$, (c) $1.0$, (d) $2.0$ and spin density $m = 0$. The type of exponent line associated with
each ${\cal{C}}$ value is for all figures the same.}
\label{figure8}
\end{figure}

The $s$ branch line one-fermion exponents are plotted as a function of the momentum $k$
in Fig. \ref{figure7} for one-fermion addition and in Fig. \ref{figure8} for one-fermion removal.
The curves correspond to several ${\cal{C}}=0.10-30.00$ values and fermionic densities 
$n$ (a) $0.3$, (b) $0.7$, (c) $1.0$, (d) $2.0$. 

The general expression, Eq. (\ref{branch-l-m0}), of the one-fermion spectral function
$B_{\gamma} (k,\omega)$, Eq. (\ref{Bkomega-m0}), near the present $s$ branch lines reads,
\begin{equation}
B_{\gamma} (k,\omega) = C_{\gamma,s} \Bigl(\gamma\omega - \omega_s (k)\Bigr)^{\xi_s (k)}  
\hspace{0.20cm}{\rm for}\hspace{0.20cm}(\gamma\,\omega - \omega_s (k)) \geq 0 
\hspace{0.20cm}{\rm where}\hspace{0.20cm}\gamma = \pm 1 \, ,
\label{s-branch-m0}
\end{equation}
and $C_{\gamma,s}$ is a constant that has a fixed value 
for the $k$ and $\omega$ ranges corresponding to small values of the energy deviation 
$(\gamma\omega - \omega_s (k))$. The spectrum $\omega_s (k)$ in such an energy deviation
is that in Eq. (\ref{OkudRs-m0}). 

The exponent $\xi_s (k)$, Eq. (\ref{xiRLAs-m0}), in the spectral function expression, Eq. (\ref{s-branch-m0}),
has in the ${\cal{C}}\rightarrow 0$ limit the following behavior for its whole momentum $k$ intervals,
\begin{equation}
\lim_{{\cal{C}}\rightarrow 0}\xi_s (k) = \gamma \, .
\nonumber
\end{equation}
Hence the $\gamma = +1$ addition one-fermion spectral weight at and near these $s$ branch lines 
vanishes in the ${\cal{C}}\rightarrow 0$ limit. 

As given generally in Eq. (\ref{branch-lexp-1}), for one-fermion removal for which
$\lim_{{\cal{C}}\rightarrow 0}\xi_s (k) = -1$ the line shape
near the $s$ branch line is not of the power-law form, Eq. (\ref{s-branch-m0}),
in the ${\cal{C}}\rightarrow 0$ limit. In that limit it rather corresponds to
the following $\delta$-function-like one-fermion removal spectral weight distribution,
\begin{equation}
\lim_{{\cal{C}}\rightarrow 0} B_{-1} (k,\omega) = \delta\Bigl(\omega + \omega_s (k)\Bigr) 
= \delta\Bigl(\omega - (k^2 - k_F^2)\Bigr)
\hspace{0.20cm}{\rm for}\hspace{0.20cm}k \in [-k_F,k_F]  \, ,
\label{s-branch-delta-m0}
\end{equation}
where the expression of the $s$ band energy dispersion for
${\cal{C}}\rightarrow 0$, Eq. (\ref{varepsilon-c-s-C0}) for $m=0$,
has been used.

At zero spin density, $m=0$, the importance of the $s$ and $c^{\pm}$ branch lines is confirmed by in the 
${\cal{C}}\rightarrow 0$ limit they leading to the whole ${\cal{C}}=0$ non-interacting $\delta$-function-like one-fermion 
removal and addition spectrum. Specifically, the $s$ branch line gives rise in the ${\cal{C}}\rightarrow 0$ limit to the
${\cal{C}}=0$ non-interacting one-fermion removal spectrum, Eq. (\ref{s-branch-delta-m0}), for its
whole momentum interval $k \in [-k_F,k_F]$. Furthermore, the $c^{+}$ and $c^{-}$ branch lines lead 
in the ${\cal{C}}\rightarrow 0$ limit to the ${\cal{C}}=0$ non-interacting one-fermion addition spectrum
for its whole momentum intervals $k \in [-\infty,-k_F]$ and $k \in [k_F,\infty]$, respectively, as
given in Eq. (\ref{cpm-branch-delta-m0}).

In the opposite ${\cal{C}}\rightarrow\infty$ limit the exponent, Eq. (\ref{xiRLAs-m0}), in the spectral function 
expression, Eq. (\ref{s-branch-m0}), that controls the line shape near both the one-fermion removal and 
addition $s$ branch lines reads,
\begin{eqnarray}
\xi_{s} (k) & = & - {1\over 2}\left(1- \left({k\over \pi n}\right)^2\right) 
\nonumber \\
& & {\rm for}\hspace{0.2cm}{\rm fermion}\hspace{0.2cm}{\rm removal}\hspace{0.2cm}{\rm at}\hspace{0.2cm}k\in [0,k_F] 
\nonumber \\
& & {\rm for}\hspace{0.2cm}{\rm fermion}\hspace{0.2cm}{\rm addition}\hspace{0.2cm}{\rm at}\hspace{0.2cm}k\in [k_F,3k_F] \, .
\label{xiRsUi-m0}
\end{eqnarray}
This implies that,
\begin{equation}
\lim_{k\rightarrow 0}\xi_{s} (k) = - {1\over 2} \, ; \hspace{0.20cm}
\lim_{k\rightarrow k_F}\xi_{s} (k) = - {3\over 8} \, ; \hspace{0.20cm}
\lim_{k\rightarrow 2k_F}\xi_{s} (k) = 0 \, ; \hspace{0.20cm}
\lim_{k\rightarrow 3k_F}\xi_{s} (3k_F) = {5\over 8}  \, . 
\nonumber
\end{equation}

For ${\cal{C}}\rightarrow\infty$ and $m= 0$ the one-fermion addition 
exponent $\xi_{s} (k)$, Eq. (\ref{xiRLAs-m0}), continuously changes from $\xi_{s} (k) = - 3/8$ for $k\rightarrow k_F$
to $\xi_{s} (k) = 0$ for $k\rightarrow 2k_F$. For its other $k$ ranges it is
positive. In the case of one-fermion removal it
continuously changes in that limit from $\xi_{s} (k) = - 1/2$ for $k\rightarrow 0$
to $\xi_{s} (k) = - 3/8$ for $k\rightarrow k_F$. 

\subsection{The up-spin and down-spin one-fermion removal and addition spectral functions near the $s$ branch line}
\label{upRs}

The up-spin and down-spin fermion removal and addition $s$ branch line
is generated by processes that correspond to particular cases of the
two-parametric processes that generate the spectra, Eqs. (\ref{SpupElremo})-(\ref{SpdownEladd}). 
For the up-spin and down-spin one-fermion spectral 
functions its one-parametric spectrum plotted in Figs. \ref{figure2} and \ref{figure3}
is thus contained within such two-parametric spectra.
(Online, the $s$ branch lines are green in these figures.)

As at zero magnetic field, the one-parametric spectrum of this branch line is an even function of $k$,
$\omega_s^{\sigma} (k) = \omega_s^{\sigma} (-k)$. The corresponding
exponent given below is also an even function of $k$, $\xi_s^{\sigma} (k) = \xi_s^{\sigma} (-k)$.
Hence for simplicity we restrict again our following analysis to $k \geq 0$. For
such a momentum range the up-spin and down-spin fermion removal and addition parts
of the $s$ branch line refer to excited energy eigenstates with the following number deviations relative 
to those of the initial ground state,
\begin{equation}
\delta N_c^F = \gamma \, ; \hspace{0.20cm} \delta J_c^F = \delta_{\sigma,\uparrow}/2 \, ; \hspace{0.20cm} 
\delta N_s^F = \delta_{\sigma,\uparrow}\,\gamma \, ; \hspace{0.20cm} \delta J_s^F = 0 \, ; \hspace{0.20cm} 
\delta N_s^{NF} = - \gamma_{\sigma}\,\gamma  \, .
\nonumber
\end{equation}

The spectrum $\omega_s^{\sigma} (k)$ of general form, Eq. (\ref{dE-dP-bl}), is for the present branch line at $k>0$ given by,
\begin{eqnarray}
\omega_s^{\sigma} (k) & = & 
- \gamma_{\sigma}\,\gamma\, \varepsilon_s (q')\hspace{0.20cm}{\rm where}
\nonumber \\
q' & \in & [-k_{F\uparrow},-k_{F\downarrow}] \hspace{0.20cm}{\rm for}\hspace{0.20cm}\uparrow\hspace{0.20cm}{\rm branch}
\hspace{0.20cm}B\hspace{0.20cm}{\rm fermion}\hspace{0.20cm}\gamma = -1\hspace{0.20cm}{\rm removal}
\nonumber \\
q' & \in & [-k_{F\downarrow},k_{F\downarrow}] \hspace{0.20cm}{\rm for}\hspace{0.20cm}\uparrow\hspace{0.20cm}{\rm branch}
\hspace{0.20cm}A\hspace{0.20cm}{\rm fermion}\hspace{0.20cm}\gamma = +1\hspace{0.20cm}{\rm addition}
\nonumber \\
q' & \in & [-k_{F\downarrow},0]  \hspace{0.20cm}{\rm for}\hspace{0.20cm}\downarrow\hspace{0.20cm}{\rm branches}
\hspace{0.20cm}A\hspace{0.20cm}{\rm and}\hspace{0.20cm}B\hspace{0.20cm}{\rm fermion}\hspace{0.20cm}\gamma = -1\hspace{0.20cm}{\rm removal}
\nonumber \\
q' & \in & [-k_{F\downarrow},k_{F\downarrow}] \hspace{0.20cm}{\rm for}\hspace{0.20cm}
\downarrow\hspace{0.20cm}{\rm branch}\hspace{0.20cm}A\hspace{0.20cm}{\rm fermion}\hspace{0.20cm}
\gamma = +1\hspace{0.20cm}{\rm addition} \, ,
\label{OkudRs}
\end{eqnarray}
and $\varepsilon_s (q')$ is the $s$ band energy dispersion, Eq. (\ref{varepsilon-c-s}) for $\beta =s$. 
In the case of down-spin one-fermion removal both the branches $A$ and $B$ of the spectrum,
Eq. (\ref{SpdownElremo}), contain the $s$ branch line. 
\begin{figure}
\includegraphics[scale=0.30]{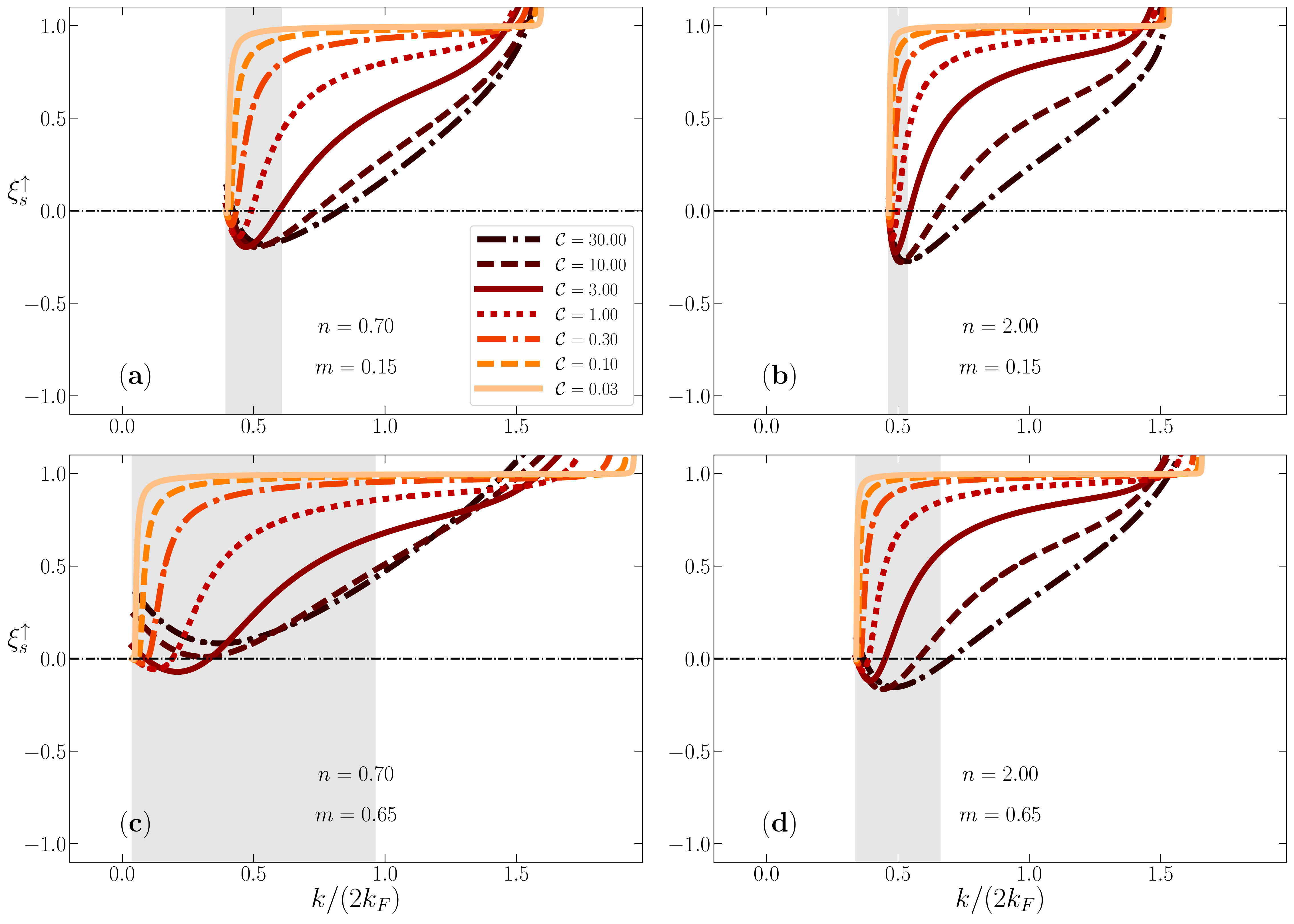}
\caption{The exponent $\xi_{s}^{\uparrow} (k)$, Eq. (\ref{xiupRLAs}),
that controls the singularities in the vicinity of the $s$ branch line whose $(k,\omega)$-plane
shape is defined by Eqs. (\ref{OkudRs}) and (\ref{kqsup}) for $\sigma =\uparrow$ and Eq. (\ref{kupRLAs}) is plotted for the 
one-fermion removal and addition spectral function, Eq. (\ref{s-branch}) for $\sigma =\uparrow$, 
as a function of the momentum $k/2k_F$. The soft grey region refers to the ground-state occupied Fermi sea.
The curves refer to several ${\cal{C}}=0.03-30.00$ values and spin density $m=0.15$ and
fermionic densities $n$ (a) $0.7$ and (b) $2.0$ and spin density $m=0.65$ and
fermionic densities (c) $0.7$ and (d) $2.0$. The type of exponent line associated with
each ${\cal{C}}$ value is for all figures the same.}
\label{figure9}
\end{figure}

The relation of the $s$ band momentum $q$ to the excitation momentum $k$ is,
\begin{equation}
k = \delta_{\sigma,\uparrow}\,2k_F - \gamma_{\sigma}\,\gamma\,q' \geq 0 \, .
\label{kqsup}
\end{equation}
This gives,
\begin{eqnarray}
k & \in & [k_{F\downarrow},k_{F\uparrow}]  
\hspace{0.20cm}{\rm for}\hspace{0.20cm}\uparrow\hspace{0.05cm}{\rm fermion}\hspace{0.20cm}{\rm removal}
\hspace{0.20cm}\gamma = -1 \, ,
\nonumber \\
k & \in &  [k_{F\uparrow},(2k_F+k_{F\downarrow})] 
\hspace{0.20cm}{\rm for}\hspace{0.20cm}\uparrow\hspace{0.05cm}{\rm fermion}\hspace{0.20cm}{\rm addition}
\hspace{0.20cm}\gamma = +1 \, ,
\label{kupRLAs}
\end{eqnarray}
and
\begin{eqnarray}
k & \in & [0,k_{F\downarrow}]  
\hspace{0.20cm}{\rm for}\hspace{0.20cm}\downarrow\hspace{0.05cm}{\rm fermion}\hspace{0.20cm}{\rm removal}
\hspace{0.20cm}\gamma = -1 \, ,
\nonumber \\
k & \in & [k_{F\downarrow},k_{F\uparrow}] 
\hspace{0.20cm}{\rm for}\hspace{0.20cm}\downarrow\hspace{0.05cm}{\rm fermion}\hspace{0.20cm}{\rm addition}
\hspace{0.20cm}\gamma = +1 \, .
\label{OkdownRLAsoth}
\end{eqnarray}

The limiting behavior for ${\cal{C}}\rightarrow 0$ of the spectrum, Eqs. (\ref{OkudRs}) and (\ref{kqsup}), is given
Eq. (\ref{sBLDCzero}) of Appendix \ref{AIBRL} where the subdomains of the $k$ intervals that correspond 
to one-fermion addition $(\gamma = +1)$ and removal $(\gamma = -1)$ are provided in Eqs. (\ref{kupRLAs}) and (\ref{OkdownRLAsoth}).
For ${\cal{C}}\rightarrow\infty$ this spectrum reads $\omega_s (k)=0$ for its whole $k$ intervals.

As for zero spin density, the momentum $k$ intervals in Eq. (\ref{OkdownRLAsoth}) reveal that the 
up-spin and down-spin one-fermion addition $s$ branch line is the natural continuation of the up-spin and down-spin
one-fermion removal $s$ branch line. The momentum dependent exponent of general form, Eq. (\ref{branch-l}),
that controls the line shape near the up-spin one-fermion removal and addition $s$ branch lines is given by,
\begin{eqnarray}
\xi_s^{\uparrow} (k) & = & -1 + \sum_{\iota=\pm1}\left({\iota \,\gamma(\xi_{c\,c}^0+\xi_{c\,s}^0)\over 2} 
+ {\xi_{c\,c}^1\over 2} - \gamma\,\Phi_{c,s}(\iota 2k_F,q')\right)^2 
\nonumber \\
& + & \sum_{\iota=\pm1}\left({\iota \,\gamma(\xi_{s\,c}^0+\xi_{s\,s}^0)\over 2} + 
{\xi_{s\,c}^1\over 2} - \gamma\,\Phi_{s,s}(\iota k_{F\downarrow},q')\right)^2  \, .
\label{xiupRLAs}
\end{eqnarray}
The exponent that controls it in the vicinity of the down-spin one-fermion removal and addition $s$ branch line reads,
\begin{equation}
\xi_s^{\downarrow} (k) = -1 + \sum_{\iota=\pm1}\left({\iota\,\xi_{c\,c}^0\over 2} + \Phi_{c,s}(\iota 2k_F,q')\right)^2 
+ \sum_{\iota=\pm1}\left({\iota\,\xi_{s\,c}^0\over 2} + \Phi_{s,s}(\iota k_{F\downarrow},q')\right)^2  \, .
\label{xidownRLAsoth}
\end{equation}
This latter exponent has the same formal expression for $\gamma =-1$ and $\gamma =+1$, respectively. The corresponding $q'$
ranges are though different, as given in Eq. (\ref{OkudRs}). The phase shifts $\Phi_{c,s}(\pm 2k_F,q')$ and 
$\Phi_{s,s}(\pm k_{F\downarrow},q')$ in those exponents expressions are
defined in Eq. (\ref{PhiqFq}) and the $j=0,1$ parameters $\xi^{j}_{\beta\,\beta'}$ are defined in Eq. (\ref{x-aa}).
\begin{figure}
\includegraphics[scale=0.30]{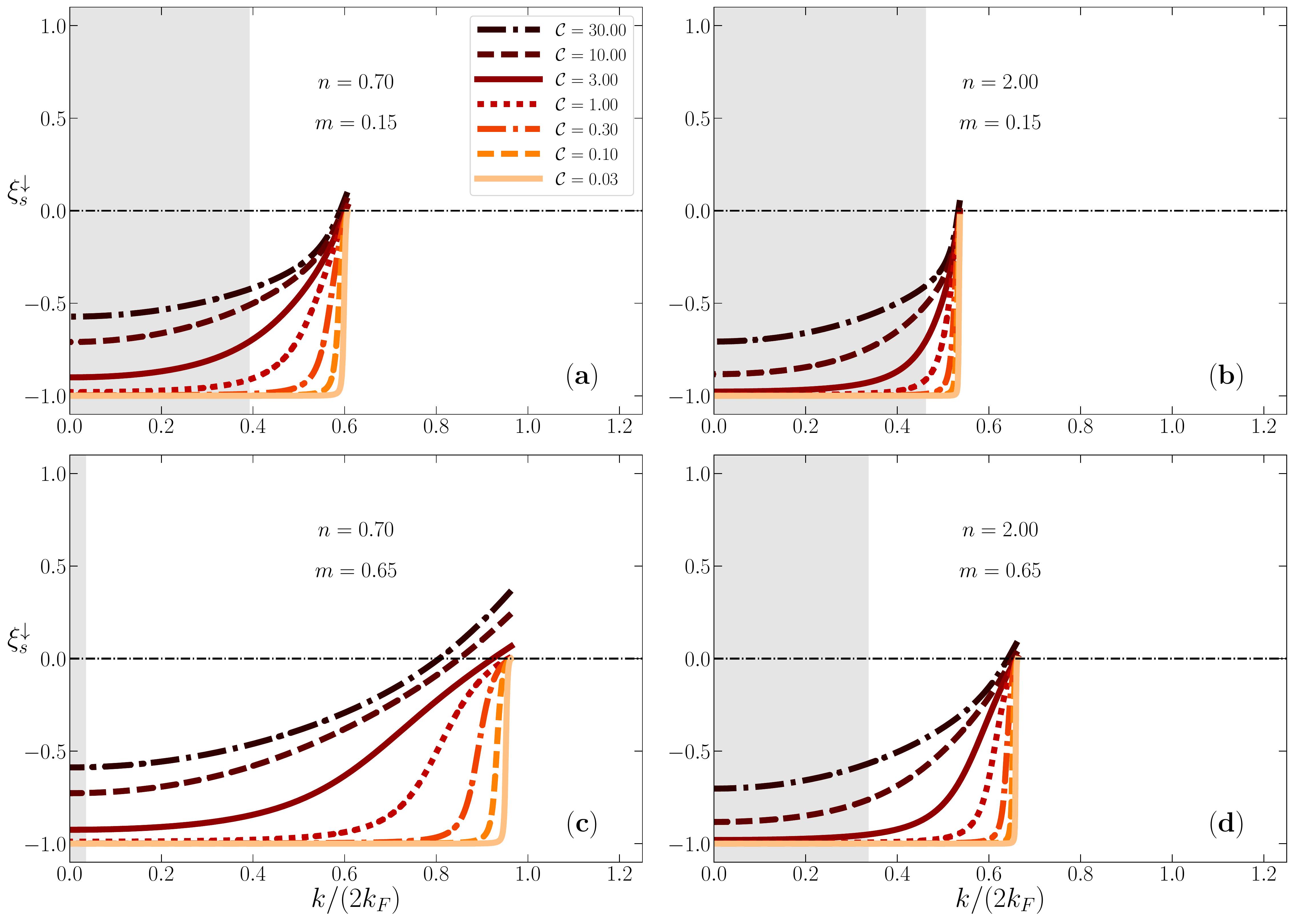}
\caption{The exponent $\xi_{s}^{\downarrow} (k)$, Eq. (\ref{xidownRLAsoth}),
that controls the singularities in the vicinity of the $s$ branch line whose $(k,\omega)$-plane
one-parametric spectrum is defined by Eqs. (\ref{OkudRs}) and (\ref{kqsup}) for $\sigma =\downarrow$ and Eq.
(\ref{OkdownRLAsoth}) is plotted for the down-spin
one-fermion removal and addition spectral function, Eq. (\ref{s-branch}) for $\sigma =\downarrow$, as a function of the momentum
$k/2k_F$. The soft grey region refers to the ground-state occupied Fermi sea.
The curves refer to several ${\cal{C}}=0.03-30.00$ values and spin density $m=0.15$ and
fermionic densities $n$ (a) $0.7$ and (b) $2.0$ and spin density $m=0.65$ and
fermionic densities (c) $0.7$ and (d) $2.0$. The type of exponent line associated with
each ${\cal{C}}$ value is for all figures the same.}
\label{figure10}
\end{figure}

The $s$ branch line one-fermion exponents are plotted as a function of the momentum $k$
in Fig. \ref{figure9} for up-spin one-fermion removal and addition and
in Fig. \ref{figure10} for down-spin one-fermion removal and addition.
The curves correspond to several ${\cal{C}}=0.03-30.00$ values,
fermionic densities $n =0.7$ and $2.0$, and spin densities $m=0.15$ and $m=0.65$.

The general expression, Eq. (\ref{branch-l}), of the up-spin and down-spin one-fermion spectral function
$B_{\sigma,\gamma} (k,\omega)$, Eq. (\ref{Bkomega}), is near the present $s$ branch line given by, 
\begin{equation}
B_{\sigma,\gamma} (k,\omega) = C_{\sigma,\gamma,s} \Bigl(\gamma\omega - \omega_s^{\sigma} (k)\Bigr)^{\xi_s^{\sigma} (k)}  
\hspace{0.20cm}{\rm for}\hspace{0.20cm}(\gamma\,\omega - \omega_s^{\sigma} (k)) \geq 0 
\hspace{0.20cm}{\rm where}\hspace{0.20cm}\gamma = \pm 1 \, ,
\label{s-branch}
\end{equation}
and $C_{\sigma,\gamma,s}$ is a constant that has a fixed value 
for the $k$ and $\omega$ ranges corresponding to small values of the energy deviation 
$(\gamma\omega - \omega_s^{\sigma} (k))$. The spectrum $\omega_s^{\sigma} (k)$ in such an energy deviation
is that in Eq. (\ref{OkudRs}). The exponent $\xi_s^{\sigma} (k)$
is given in Eqs. (\ref{xiupRLAs}) and (\ref{xidownRLAsoth}).

The exponent $\xi_s^{\sigma} (k)$ has the following related behavior in the ${\cal{C}}\rightarrow 0$ limit
for its whole $k$ intervals,
\begin{equation}
\lim_{{\cal{C}}\rightarrow 0}\xi_s^{\sigma} (k) = \gamma_{\sigma} \, .
\nonumber
\end{equation}
Hence the up-spin one-fermion spectral weight at and near these $s$ branch lines vanishes in the ${\cal{C}}\rightarrow 0$ limit
both for one-fermion removal and addition. 

As given generally in Eq. (\ref{branch-lexp-1}), for the $n$, $m$, and $k$ ranges for which 
$\lim_{{\cal{C}}\rightarrow 0}\xi_s^{\downarrow} (k) = -1$ the line shape
near the branch line is not of the power-law form, Eq. (\ref{s-branch}). As for zero spin density, in that limit it rather corresponds to
the following $\delta$-function-like down-spin one-fermion spectral weight distribution,
\begin{eqnarray}
\lim_{{\cal{C}}\rightarrow 0} B_{\downarrow,-1} (k,\omega) & = & \delta\Bigl(\omega + \omega_s^{\downarrow} (k)\Bigr) 
= \delta\Bigl(\omega - (k^2 - k_{F\downarrow}^2)\Bigr)
\nonumber \\
& & {\rm for}\hspace{0.2cm}k \in [-k_{F\downarrow},k_{F\downarrow}]  \, ,
\nonumber \\
\lim_{{\cal{C}}\rightarrow 0} B_{\downarrow,+1} (k,\omega) & = & \delta\Bigl(\omega - \omega_s^{\downarrow} (k)\Bigr) 
= \delta\Bigl(\omega - (k^2 - k_{F\downarrow}^2)\Bigr)
\nonumber \\
& & {\rm for}\hspace{0.2cm}k \in [-k_{F\uparrow},-k_{F\downarrow}]\hspace{0.2cm}{\rm and}\hspace{0.2cm}
k \in [k_{F\downarrow},k_{F\uparrow}] \, ,
\label{s-branch-delta}
\end{eqnarray}
where the expression of the $s$ band energy dispersion for
${\cal{C}}\rightarrow 0$, Eq. (\ref{varepsilon-c-s-C0}), has been used.
The ${\cal{C}}\rightarrow 0$ limiting behavior reported in the latter equation
for that energy dispersion appearing in the spectrum $\omega_s^{\downarrow} (k)$, 
Eq. (\ref{OkudRs}), confirms that the latter spectrum becomes in the ${\cal{C}}\rightarrow 0$ limit the corresponding ${\cal{C}}=0$ 
non-interacting spin-down fermionic spectrum, as given in Eq. (\ref{s-branch-delta}). This applies
to its whole $k$ interval except for down-spin fermion addition for
$k \in [-\infty,-k_{F\uparrow}]$ and $k \in [k_{F\uparrow},\infty]$, as further discussed in Section \ref{upRsl}.

For the $k$ interval for which $\lim_{{\cal{C}}\rightarrow 0}\xi_s^{\downarrow} (k) = 0$, the down-spin one-fermion addition spectral 
weight at and near the present $s$ branch line vanishes in the ${\cal{C}}\rightarrow 0$ limit.

For ${\cal{C}}\rightarrow\infty$ and $m\rightarrow n$ we find the following
exponent expressions for the up-spin one-fermion removal
and down-spin one-fermion addition $s$ branch line,
\begin{eqnarray}
\xi_{s}^{\uparrow} (k) & = & {1\over 2}\left({k\over \pi n}\right)^2 
+ {2\over\pi^2}\left\{\arctan\left({1\over 2}\cot \left({k\over 2n}\right)\right)\right\}^2 
\nonumber \\
& & \uparrow {\rm fermion}\hspace{0.2cm}{\rm removal}\hspace{0.2cm}{\rm for}\hspace{0.2cm}k\in [0,2k_F] \, ,
\nonumber \\
\xi_{s}^{\downarrow} (k) & = & - {1\over 2}\left(1-\left({k\over \pi n}\right)^2 \right)
+ {2\over\pi^2}\left\{\arctan\left({1\over 2}\tan \left({k\over 2n}\right)\right)\right\}^2 
\nonumber \\
& & \downarrow {\rm fermion}\hspace{0.2cm}{\rm addition}\hspace{0.2cm}{\rm for}\hspace{0.2cm}k\in [0,2k_F] \, .
\nonumber
\end{eqnarray}
This then implies that,
\begin{eqnarray}
\lim_{k\rightarrow 0}\xi_{s}^{\uparrow} (k) & = & {1\over 2} \, ; \hspace{0.20cm}
\lim_{k\rightarrow k_F}\xi_{s}^{\uparrow} (k) = {1\over 8} + 2\left({1\over\pi}\arctan\left({1\over 2}\right)\right)^2 
\approx 0.16856 \, ; \hspace{0.20cm}
\lim_{k\rightarrow 2k_F}\xi_{s}^{\uparrow} (k) = {1\over 2} 
\nonumber \\
\lim_{k\rightarrow 0}\xi_{s}^{\downarrow} (k) & = & - {1\over 2} \, ; \hspace{0.20cm}
\lim_{k\rightarrow k_F}\xi_{s}^{\downarrow} (k) =  - {3\over 8}
+ 2\left({1\over\pi}\arctan\left({1\over 2}\right)\right)^2 
\approx - 0.33144 \, ; \hspace{0.20cm}
\lim_{k\rightarrow 2k_F}\xi_{s}^{\downarrow} (k) = {1\over 2} \, .
\nonumber
\end{eqnarray}

Analysis of these expressions and values reveals that in the ${\cal{C}}\rightarrow\infty$ and $m\rightarrow n$ limits
the up-spin one-fermion removal exponent $\xi_{s}^{\uparrow} (k)$
smoothly decreases from $\xi_{s}^{\uparrow} (k) = 1/2$ for $k\rightarrow 0$
until it reaches a minimum value at $k=k_F$. For $k>k_F$ it continuously increases
to $\xi_{s}^{\uparrow} (k) = 1/2$ as $k\rightarrow 2k_F$.
In the same limits, the down-spin one-fermion addition exponent $\xi_{s}^{\downarrow} (k)$
smoothly varies from $\xi_{s}^{\downarrow} (k) = - 1/2$ for $k\rightarrow 0$
to $\xi_{s}^{\downarrow} (k) = 1/2$ for $k\rightarrow 2k_F$. 

Moreover, analysis of Fig. \ref{figure9} for other spin densities
shows that the exponent $\xi_s^{\uparrow} (k)$ 
only becomes negative for a part of the $s$ branch line $k$ interval. It
starts at $k=k_{F\downarrow}$ and ends at a $k$ momentum that 
for smaller and larger spin density values refers to one-fermion
addition and removal, respectively. The ${\cal{C}}$ values for which it
is negative depend on the densities.

For ${\cal{C}}>0$ the exponent $\xi_s^{\downarrow} (k)$, whose $k$
dependence is plotted in Fig. \ref{figure10}, is in general negative. The exception
refers to a small $k$ region. It corresponds to the larger $k$ values of its range.
For the $k$ ranges for which it reads $-1$ for ${\cal{C}}\rightarrow 0$, it remains being an
increasing function of ${\cal{C}}$ for the whole ${\cal{C}}$ interval. However, for the $k$ 
domains for which it is given by $0$ in the ${\cal{C}}\rightarrow 0$ limit, upon increasing
${\cal{C}}$ it first decreases, goes through a minimum value, and then becomes an
increasing function of ${\cal{C}}$, until reaching its ${\cal{C}}\rightarrow\infty$ limit 
$k$ dependent values.

\subsection{The up-spin one-fermion removal and down-spin one-fermion addition ``non-branch lines''}
\label{upRsl}

On the one hand and as discussed in Section \ref{upRs-m0}, at zero magnetic field the $c^{\pm}$ and $s$ 
branch lines lead in the ${\cal{C}}\rightarrow 0$ limit to the ${\cal{C}}=0$ non-interaction $\delta$-function-like 
one-fermion addition and removal spectra for their whole $k$ intervals, as given in Eqs. (\ref{cpm-branch-delta-m0}) 
and (\ref{s-branch-delta-m0}).

On the other hand, at finite magnetic field and finite spin density the $c^{\pm}$ and $s$ branch lines lead in the 
${\cal{C}}\rightarrow 0$ limit to most $k$ intervals of the ${\cal{C}}=0$ non-interaction $\delta$-function-like 
up-spin and down-spin one-fermion addition and removal spectra, respectively. This is confirmed from
analysis of the up-spin one-fermion spectral function expressions obtained from the $c^{\pm}$ branch lines
in Eq. (\ref{cpm-branch-delta}) and of the expressions of the down-spin one-fermion spectral function obtained 
from the $s$ branch lines in Eq. (\ref{s-branch-delta}).

However and as mentioned above, at finite magnetic field some $k$ subintervals of the ${\cal{C}}=0$ 
non-interacting removal up-spin and addition down-spin one-fermion spectra do not stem from
branch lines. This refers to the momentum interval $k\in [-k_{F\downarrow},k_{F\downarrow}]$
for up-spin one-fermion removal and to the momentum intervals $k\in [-\infty,-k_{F\uparrow}]$ and
$k\in [k_{F\uparrow},\infty]$ for down-spin one-fermion addition.

The ${\cal{C}}=0$ non-interacting up-spin one-fermion removal spectral weight missing 
for $k\in [-k_{F\downarrow},k_{F\downarrow}]$ stems in the ${\cal{C}}\rightarrow 0$ limit 
from a ${\cal{C}}>0$ $s'$ spectral feature that is generated by transitions to excited energy 
eigenstates whose number deviations relative to those of the initial ground state are given by,
\begin{equation}
\delta N_c^F = \delta J_c^F = 0 \, ; \hspace{0.20cm} \delta N_c^{NF} = - 1  \, ; \hspace{0.20cm} 
\delta N_s^F = 1 \, ; \hspace{0.20cm} \delta J_s^F = \pm 1 \, ; \hspace{0.20cm} \delta N_s^{NF} = - 1 \, .
\nonumber
\end{equation}

The one-parametric spectrum of this line reads,
\begin{eqnarray}
\omega_{s'}^{\uparrow} (k) & = & - \varepsilon_s (-k) - \varepsilon_c (\pm 2k_{F\downarrow}) = - \varepsilon_s (q') 
- \varepsilon_c (\pm 2k_{F\downarrow}) \hspace{0.20cm}{\rm where}\hspace{0.20cm} q' \in [-k_{F\downarrow},k_{F\downarrow}] 
\hspace{0.20cm}{\rm and}
\nonumber \\
k & = & - q' \in [-k_{F\downarrow},k_{F\downarrow}] \, .
\nonumber
\end{eqnarray}
Here $\varepsilon_c (q)$ and $\varepsilon_s (q')$ are the $c$ and $s$ band energy dispersions, Eq. (\ref{varepsilon-c-s}).
While the line shape expression near the present $s'$ line involves within the PDT state summations difficult to
be performed analytically for ${\cal{C}}>0$, in the ${\cal{C}}\rightarrow 0$ limit
its exact line shape becomes $\delta$-function-like,
\begin{equation}
\lim_{{\cal{C}}\rightarrow 0} B_{\uparrow,-1} (k,\omega) = \delta\Bigl(\omega + \omega_{s'}^{\uparrow} (k)\Bigr) 
= \delta\Bigl(\omega - (k^2 - k_{F\uparrow}^2)\Bigr) \hspace{0.20cm}{\rm for}\hspace{0.20cm}k \in [-k_{F\downarrow},k_{F\downarrow}] \, .
\nonumber
\end{equation}
Here the expression of the $c$ and $s$ energy dispersions for ${\cal{C}}\rightarrow 0$, Eq. (\ref{varepsilon-c-s-C0}), 
have been used.

The ${\cal{C}}=0$ non-interacting down-spin one-fermion addition spectral weight missing 
for $k\in [-\infty,-k_{F\downarrow}]$ and $k\in [k_{F\downarrow},\infty]$ stems in the ${\cal{C}}\rightarrow 0$ limit
from a ${\cal{C}}>0$ $c'$ spectral feature that 
is generated by transitions to excited energy eigenstates whose number deviations relative to those of the
initial ground state read,
\begin{equation}
\delta N_c^F = 0 \, ; \hspace{0.20cm} \delta J_c^F = \pm 1/2 \, ; \hspace{0.20cm} \delta N_c^{NF} = 1  \, ; \hspace{0.20cm} 
\delta N_s^F = \delta J_s^F = 0 \, ; \hspace{0.20cm} \delta N_s^{NF} = 1 \, .
\nonumber
\end{equation}

The one-parametric spectrum of this line is given by,
\begin{eqnarray}
\omega_{c'}^{\downarrow} (k) & = & \varepsilon_c (k -k_{F\downarrow}) + \varepsilon_s (-k_{F\uparrow}) 
\hspace{0.20cm}{\rm with}\hspace{0.20cm}{\rm interval}
\nonumber \\
k & = & q + k_{F\downarrow} \in [-\infty,-k_{F\uparrow}]\hspace{0.2cm}{\rm for}\hspace{0.2cm}q \in [-\infty,-2k_F]
\nonumber \\
\omega_{c'}^{\downarrow} (k) & = & \varepsilon_c (k + k_{F\downarrow}) + \varepsilon_s (k_{F\uparrow})
\hspace{0.20cm}{\rm with}\hspace{0.20cm}{\rm interval}
\nonumber \\
k & = & q - k_{F\downarrow} \in [k_{F\uparrow},\infty]\hspace{0.2cm}{\rm for}\hspace{0.2cm}q \in [2k_F,\infty] \, .
\nonumber
\end{eqnarray}

Again, the line shape expression in the vicinity of this $c'$ line involves within the PDT state summations difficult to
be performed analytically for ${\cal{C}}>0$. In the ${\cal{C}}\rightarrow 0$ limit
that line shape becomes $\delta$-function-like,
\begin{equation}
\lim_{{\cal{C}}\rightarrow 0} B_{\downarrow,+1} (k,\omega) = \delta\Bigl(\omega - \omega_{c'}^{\uparrow} (k)\Bigr) 
= \delta\Bigl(\omega - (k^2 - k_{F\downarrow}^2)\Bigr) \hspace{0.20cm}{\rm for}\hspace{0.20cm}
k \in [-\infty,-k_{F\uparrow}]\hspace{0.20cm}{\rm and}\hspace{0.20cm}[k_{F\uparrow},\infty] \, ,
\nonumber
\end{equation}
where the expressions of the $c$ and $s$ energy dispersions for ${\cal{C}}\rightarrow 0$, Eq. (\ref{varepsilon-c-s-C0}), 
have again been used.

\subsection{The one-fermion removal boundary line at zero magnetic field}
\label{OneFBLs-m0}

As given in Eq. (\ref{B-bol-m0}), the line shape near a one-fermion removal boundary line singularity is power law like,
$B_{\gamma} (k,\omega) \propto (\omega +\omega_{BL} (k))^{-1/2}$,
with a negative momentum independent exponent, $-1/2$, and an energy spectrum $\omega_{BL} (k)$ whose details 
at zero spin density are further studied in this section. Such one-fermion removal boundary lines emerge from the 
$c^{\pm}$ branch lines of energy spectrum $\omega_{c^{\pm}} (k)$, Eqs. (\ref{OkudLAcc-m0}) and (\ref{kudLAcc-m0}) for $\gamma = -1$, at 
a well-defined momentum $k_{min}$. 

For simplicity, here we consider the 
boundary line emerging from the $c^{+}$ branch line. The spectrum of that emerging
from the $c^{-}$ branch line is generated from that considered here by interchanging $k$ and $-k$.
We call $q=q_c^{BL}$ and $q'=q_s^{BL}$ a pair of $c$ band and $s$ band momenta such that, $v_c (q_c^{BL}) = v_{s} (q_s^{BL})$,
which as given in Eq. (\ref{dE-dP-c-s-m0}) are those that contribute to a boundary line.
For the branch A of the two-parametric spectrum $\omega_{R} (k)$, Eq. (\ref{SpElremo-m0}), the
one-fermion removal boundary line spectrum reads,
\begin{eqnarray}
\omega_{BL} (k) & = & - \varepsilon_c (2k_F - k - q_s^{BL}) 
- \varepsilon_{s} (q_s^{BL}) = - \varepsilon_c (q_c^{BL}) - \varepsilon_{s} (2k_F - k + q_c^{BL}])
\nonumber \\ 
& & {\rm for}\hspace{0.2cm}
v_c (2k_F - k - q_s^{BL}) = v_{s} (q_s^{BL})\hspace{0.2cm}
{\rm and}\hspace{0.2cm}{\rm equivalently}\hspace{0.2cm}
v_c (q_c^{BL}) = v_{s} (2k_F - k + q_c^{BL}]) 
\nonumber \\ 
& & {\rm with}\hspace{0.2cm} {\rm sgn}\{q_c^{BL}\} = {\rm sgn}\{q_s^{BL}\}
\hspace{0.20cm}{\rm and}\hspace{0.20cm}k\hspace{0.20cm}{\rm interval}
\nonumber \\ 
k & = & 2k_F - q_c^{BL} - q_s^{BL} \in [k_{min},k_{max}]  
\hspace{0.20cm}{\rm where}\hspace{0.2cm}k_{min} = k_F - q_c^0  \in [-k_F,k_F]
\hspace{0.20cm}{\rm and}\hspace{0.20cm}
q_c^0 \in [0, 2k_F] \, .
\label{BLSPD-m0}
\end{eqnarray}

The excitation momentum interval, $k\in [k_{min},k_{max}]$, is that for which the boundary line exists. 
Only for that interval is it a limiting line of the two-dimensional $(k,\omega)$-plane 
domain associated with the branch A of the two-parametric spectrum $\omega_{R} (k)$, Eq. 
(\ref{SpElremo-m0}). The limiting values of the interval $k\in [k_{min},k_{max}]$ read,
\begin{eqnarray}
k_{min} & = & k_F - q_c^0 \in [-k_F,k_F]
\nonumber \\
k_{max} & = & 4k_F-k_{min} = 3k_F + q_c^0 \in [3k_F,5k_F] \, .
\label{kminkmax-m0}
\end{eqnarray}
The reference $c$ band momentum value $q_c^0$ appearing here is
defined by the following velocities relation,
\begin{equation}
v_c (q_c^0) = v_s (k_F) \, .
\label{vcvs-m0}
\end{equation}
Useful limiting values of the reference $c$ band momentum $q_c^0$ and of the momenta $k_{min}$ and $k_{max}$ 
defined in Eq. (\ref{kminkmax-m0}) are given in Eq. (\ref{qkklimits-m0}) of Appendix \ref{AIRBL}.

The present one-fermion removal singular boundary line is at zero spin density represented in 
the spectrum plotted in Fig. \ref{figure1} by a dashed-dotted line. For momentum 
intervals different from $k\in [k_{min},k_{max}]$ of the two-parametric spectrum 
$\omega_{R} (k)$, Eq.  (\ref{SpElremo-m0}), its two-dimensional $(k,\omega)$-plane domain is not limited by a boundary line
as defined in Eq. (\ref{dE-dP-c-s-m0}). Indeed, for such $k$ values one has that 
$\vert v_c (q)\vert > v_s (k_F)$ and the condition $v_c (2k_F - k - q') = v_{s} (q')$ cannot be met, 
as $v_s (k_F)$ is the maximum absolute value of the $s$ band velocity
for $q' \in [-k_F,k_F]$.

Further information on the boundary line spectrum, Eq. (\ref{BLSPD-m0}), is given in Appendix \ref{AIRBL}.

\subsection{The up-spin and down-spin one-fermion removal boundary lines}
\label{UpDownOneFBLs}
  
As at zero spin density, the line shape near a up-spin and down-spin one-fermion boundary line 
singularity given in Eq. (\ref{B-bol}) is power law like, $B_{\sigma,\gamma} (k,\omega) \propto (\omega +\omega_{BL}^{\sigma} (k))^{-1/2}$,
again with a negative momentum independent exponent, $-1/2$, and an energy spectrum $\omega_{BL}^{\sigma} (k)$ 
whose details are further studied in this section. Such up-spin and down-spin one-fermion removal boundary lines
emerge from the $c^{\pm}$ branch line of energy spectrum $\omega_{c^{\pm}}^{\sigma} (k)$, 
Eqs. (\ref{OkudRLAcc}) and (\ref{OkudLAcc}) for $\gamma=-1$, at momentum $k_{min}^{\sigma}$. 

As for zero spin density and for simplicity, here we consider the 
boundary lines emerging from the $c^{+}$ branch line. The spectrum of those emerging
from the $c^{-}$ branch line is generated from that considered here by replacing $k$ by $-k$.
For branches A of the two-parametric spectra $\omega_{R}^{\sigma} (k)$, Eqs. 
(\ref{SpupElremo}) and (\ref{SpdownElremo}), the up-spin $(\sigma=\uparrow)$ and down-spin 
$(\sigma=\downarrow)$ one-fermion removal boundary lines spectrum reads,
\begin{eqnarray}
\omega_{BL}^{\uparrow} (k) & = & - \varepsilon_c (k_F - k + q_s^{BL}) + \varepsilon_{s} (q_s^{BL}) 
= - \varepsilon_c (q_c^{BL}) + \varepsilon_{s} (k-q_c^{BL})
\nonumber \\ 
{\rm for} & & v_c (- k + q_s^{BL}) = v_{s} (q_s^{BL})\hspace{0.2cm}
{\rm and}\hspace{0.2cm}{\rm equivalently}\hspace{0.2cm}
v_c (q_c^{BL}) = v_{s} (k-q_c^{BL}) \hspace{0.20cm}{\rm with}
\nonumber \\ 
k & = & - q_c^{BL} + q_s^{BL} \in [k_{min}^{\uparrow},k_{max}^{\uparrow}]  
\nonumber \\ 
\omega_{BL}^{\downarrow} (k) & = & - \varepsilon_c (2k_F - k - q_s^{BL}) - \varepsilon_{s} (q_s^{BL}) 
= - \varepsilon_c (q_c^{BL}) - \varepsilon_{s} (2k_F - k + q_c^{BL})
\nonumber \\ 
{\rm for} && v_c (2k_F - k - q_s^{BL}) = v_{s} (q_s^{BL})\hspace{0.2cm}
{\rm and}\hspace{0.2cm}{\rm equivalently}\hspace{0.2cm}
v_c (q_c^{BL}) = v_{s} (2k_F - k + q_c^{BL}])  \hspace{0.20cm}{\rm with}
\nonumber \\ 
k & = & 2k_F - q_c^{BL} - q_s^{BL} \in [k_{min}^{\downarrow},k_{max}^{\downarrow}]  
\nonumber \\ 
& & {\rm where}\hspace{0.20cm}{\rm sgn}\{q_c^{BL}\} = {\rm sgn}\{q_s^{BL}\}
\hspace{0.20cm}{\rm for}\hspace{0.20cm}{\rm both}\hspace{0.20cm}
\sigma = \uparrow,\downarrow \, .
\label{BLSPD}
\end{eqnarray}

The $\sigma=\uparrow,\downarrow$ excitation momentum intervals 
$k\in [k_{min}^{\sigma},k_{max}^{\sigma}]$ are those for which the corresponding boundary lines exist.
For $k\in [k_{min}^{\sigma},k_{max}^{\sigma}]$ the boundary lines are limiting lines of the corresponding 
two-dimensional $(k,\omega)$-plane domains associated 
with the branches A of the two-parametric spectra $\omega_{R}^{\sigma} (k)$, Eqs. 
(\ref{SpupElremo}) and (\ref{SpdownElremo}), respectively. 

The limiting values of the $\sigma=\uparrow,\downarrow$ intervals $k\in [k_{min}^{\sigma},k_{max}^{\sigma}]$
are given in Eq. (\ref{kminkmax}) of Appendix \ref{AIRBL}. Also further information on the up-spin and down-spin 
boundary line spectra, Eq. (\ref{BLSPD}), is provided in that Appendix.

The up-spin and down-spin one-fermion removal singular boundary lines are at finite spin density 
represented in the spectra plotted in Figs. \ref{figure2} and \ref{figure3} by dashed-dotted lines.
As at zero magnetic field, for momentum intervals different from $k\in [k_{min}^{\sigma},k_{max}^{\sigma}]$ of the two-parametric spectra 
$\omega_{R}^{\sigma} (k)$, Eqs. (\ref{SpupElremo}) and (\ref{SpdownElremo}), 
such spectra two-dimensional $(k,\omega)$-plane domain is not limited by a boundary line
as defined in Eq. (\ref{dE-dP-c-s}). 

\section{The spectral function power-law behaviors in the low-energy TLL regime}
\label{LESF}

The expression of the one-fermion spectral functions near the branch lines, Eqs. (\ref{branch-l-m0}) and (\ref{branch-l}), 
is valid for energy scales beyond the reach of the low-energy TLL \cite{Tomonaga-50,Luttinger-63,Solyom-79,Voit}.
However, as for the related 1D Hubbard model \cite{LE},
the PDT also applies to the TLL low-energy regime whose spectral-function 
exponents near the $c^{\pm}$ and $s$ branch lines are different from those of the high-energy
regime.

The processes that generate a branch line involve creation of a single $c$ or $s$ particle or hole
away from the corresponding Fermi points. They also involve creation of a single $s$ or $c$ particle or hole, respectively, 
at the corresponding Fermi points. Finally, such processes are dressed by low-energy and small-momentum 
multiple particle-hole processes around the two branches Fermi points. 

As reported in Section \ref{PRPS}, the single $c$ or $s$ particle or hole created
away from the corresponding Fermi points within the high-energy regime is in the TLL
regime and cross-over to it rather created at bare $c$ and $s$ band momenta
with absolute values $\vert q\vert\in [2k_F- k_{Fc}^0,2k_F+k_{Fc}^0]$ and 
$\vert q'\vert\in [k_{F\downarrow}-k_{Fs}^0,k_{F\downarrow}+k_{Fs}^0]$, respectively. 
In the high energy regime, the group velocity of the $c$ or $s$ particle or hole created away from its Fermi points 
is different from the $c$ or $s$ band Fermi velocity at any of the $c$ and $s$ band Fermi points, respectively.
In contrast, in the low-energy TLL regime that $c$ or $s$ particle or hole velocity becomes the $c$ or $s$ band 
Fermi velocity, Eq. (\ref{vFcvFs}), of the low-energy particle-hole excitations near one
of the $c$ or $s$ Fermi points. Hence the $c$ or $s$ particle or hole under consideration
loses its identity, in that it cannot be distinguished from the $c$ or $s$ particles or holes in the particle-hole excitations. 

It turns out that, as a result, in the TLL regime the $\gamma =c,c',s$ branch line exponents expression 
$\xi_{\beta} = -1 + 2\Delta_c^{+1} + 2\Delta_c^{-1} + 2\Delta_s^{+1} + 2\Delta_s^{-1}$, Eq. (\ref{branch-l-m0}),
at $m=0$ or $\xi_{\beta}^{\sigma} = -1 + 2\Delta_c^{+1} + 2\Delta_c^{-1} + 
2\Delta_s^{+1} + 2\Delta_s^{-1}$, Eq. (\ref{branch-l}), for $m>0$ loses one
of its four $2\Delta$s. Specifically, in the case of the $c$ or $s$ particle or hole it loses the 
$2\Delta_c^{\iota}$ and $2\Delta_s^{\iota}$ term, respectively, whose sign $\iota = \pm 1$ 
is that of the Fermi point whose velocity is the same as its own velocity. The corresponding 
expressions of the exponents in the high-energy 
spectral function expressions are thus different from those of the TLL regime.

Specifically, in the low-energy TLL limit a branch-line energy reads $\pm \gamma\,v_{\beta}\,(k-k_0)$ where $\beta =c$ and $\beta =s$
for the $c^{\pm}$ and $s$ branch lines, respectively, and $\pm v_{\beta}$ is the corresponding Fermi velocity, Eq. (\ref{vFcvFs}).
Indeed, its group velocity equals in that limit the $\beta$ band Fermi velocity. For small excitation energy 
$\omega \approx \pm \gamma\,v_{\beta}\,(k-k_0)$
the behavior of the one-fermion spectral function
$B_{\gamma} (k,\omega)$, Eq. (\ref{Bkomega-m0}) at $m=0$
and $B_{\sigma,\gamma} (k,\omega)$, Eq. (\ref{Bkomega}), for $m>0$ near such a branch line remain power-law like.
It reads,
\begin{eqnarray}
B_{\gamma} (k,\omega) & \propto & \Bigl(\gamma\,\omega \mp v_{\beta}\,(k-k_0)\Bigr)^{\zeta_{\pm}}
\hspace{0.20cm}{\rm for}\hspace{0.20cm}(\gamma\,\omega \mp v_{\beta}\,(k-k_0)) \geq 0\hspace{0.20cm}{\rm where}
\nonumber \\
\zeta_{\pm} & = & -1- 2\Delta_{\beta}^{\mp 1} +\sum_{\beta' = c,s}\sum_{\iota =\pm 1}2\Delta_{\beta'}^{\iota}
\hspace{0.20cm}{\rm and}\hspace{0.20cm}(\gamma\,\omega) \approx \pm v_{\beta}\,(k-k_0) 
\hspace{0.20cm}{\rm for}\hspace{0.20cm}\beta =c,s \, ,
\label{branchline-m0}
\end{eqnarray}
at $m=0$ and,
\begin{eqnarray}
B_{\sigma,\gamma} (k,\omega) & \propto & \Bigl(\gamma\,\omega \mp v_{\beta}\,(k-k_0)\Bigr)^{\zeta_{\pm}^{\sigma}}  
\hspace{0.20cm}{\rm for}\hspace{0.20cm}(\gamma\,\omega \mp v_{\beta}\,(k-k_0)) \geq 0\hspace{0.20cm}{\rm where}
\nonumber \\
\zeta_{\pm}^{\sigma} & = & -1- 2\Delta_{\beta}^{\mp 1} +\sum_{\beta' = c,s}\sum_{\iota =\pm 1}2\Delta_{\beta'}^{\iota} 
\hspace{0.20cm}{\rm and}\hspace{0.20cm}(\gamma\,\omega) \approx \pm v_{\beta}\,(k-k_0)
\hspace{0.20cm}{\rm for}\hspace{0.20cm}\beta =c,s \, ,
\label{branchline}
\end{eqnarray}
for $m>0$.

The expression of the exponents $\zeta_{\pm}$ and $\zeta_{\pm}^{\sigma}$ now only involves three $2\Delta$s, as
reported above. Moreover, such $\beta =c,s$ and
$\iota =\pm 1$ functionals $2\Delta_{\beta}^{\iota}$, Eqs. (\ref{OESFfunctional-m0}) and (\ref{OESFfunctional}), now do not involve 
high-energy deviations away from the Fermi points. They read,
\begin{equation}
2\Delta^{\iota}_{c} = \left({\iota\over\xi_0}{\delta N^F_c\over 2} 
+ \xi_0\left(\delta J^F_c + {\delta J^F_s\over 2}\right)\right)^2  \hspace{0.20cm}{\rm and}\hspace{0.2cm}
2\Delta^{\iota}_{s} = {1\over 2}\left(\iota\left(\delta N^F_s - {\delta N^F_c\over 2}\right) + \delta J^F_s\right)^2 
\hspace{0.20cm}{\rm for}\hspace{0.20cm}\iota = \pm 1 \, ,
\label{pointfunctional-m0}
\end{equation}
at $m=0$ where $\xi_0$ is the parameter defined in Eq. (\ref{xi-0}) of Appendix \ref{UBAQ} and,
\begin{equation}
2\Delta^{\iota}_{\beta} = \left(\sum_{\beta'=c,s}\left(\iota\, \xi^0_{\beta\,\beta'}\,{\delta N^F_{\beta'}\over 2} 
+ \xi^1_{\beta\,\beta'}\,\delta J^F_{\beta'}\right)\right)^2 
\hspace{0.2cm}{\rm for}\hspace{0.2cm} \beta = c,s\hspace{0.20cm}{\rm and}\hspace{0.20cm}\iota = \pm 1 \, ,
\label{pointfunctional}
\end{equation}
at $m>0$ where the $j=0,1$ parameters $\xi^{j}_{\beta\,\beta'}$ are defined in Eq. (\ref{x-aa}).
The spectral function expressions, Eqs. (\ref{branchline-m0}) and (\ref{branchline}), are valid at small energy 
$(\gamma\,\omega)$ and for small energy deviations $(\gamma\,\omega \mp v_{\beta}\,(k-k_0))$.

In the case of a large finite system, there is a cross-over regime between the low-energy TLL regime and the 
high energy regime within which the above quantity $2\Delta_c^{\iota}$ or $2\Delta_s^{\iota}$ 
gradually vanishes. Such a cross-over regime momentum and energy widths are very small or vanish in the 
thermodynamic limit. Since our studies refer to the thermodynamic limit, such a cross-over regime
is not among the goals of this paper.

Finally, within the TLL regime at finite spectral-weight $(k,\omega)$-plane regions near a point $(k_0,0)$ in 
directions other than a branch line the spectral functions behave at small excitation energy $\omega$ as,
\begin{eqnarray}
B_{\gamma} (k,\omega) & \propto & \Bigl(\gamma\,\omega\Bigr)^{\zeta} 
\hspace{0.20cm}{\rm for}\hspace{0.20cm}(\gamma\,\omega) \geq 0\hspace{0.20cm}{\rm where}
\nonumber \\
\zeta & = & -2+\sum_{\beta' = c,s}\sum_{\iota =\pm 1}2\Delta_{\beta'}^{\iota} 
\hspace{0.20cm}{\rm and}\hspace{0.20cm}(\gamma\,\omega) \neq \pm v_{\beta}\,(k-k_0) \hspace{0.20cm}{\rm for}\hspace{0.20cm}\beta =c,s \, ,
\label{point-m0}
\end{eqnarray}
at $m=0$ and,
\begin{eqnarray}
B_{\sigma,\gamma} (k,\omega) & \propto & \Bigl(\gamma\,\omega\Bigr)^{\zeta^{\sigma}} 
\hspace{0.20cm}{\rm for}\hspace{0.20cm}(\gamma\,\omega) \geq 0\hspace{0.20cm}{\rm where}
\nonumber \\
\zeta^{\sigma} & = & -2+\sum_{\beta' = c,s}\sum_{\iota =\pm 1}2\Delta_{\beta'}^{\iota}
\hspace{0.20cm}{\rm and}\hspace{0.20cm}(\gamma\,\omega) \neq \pm v_{\beta}\,(k-k_0) 
\hspace{0.20cm}{\rm for}\hspace{0.20cm}\beta =c,s \, ,
\label{point}
\end{eqnarray}
for $m>0$ where the $\iota =\pm 1$ functionals $2\Delta_{\beta}^{\iota}$ in the exponent
expression in Eqs. (\ref{point-m0}) and (\ref{point}) are those
given in Eqs. (\ref{pointfunctional-m0}) and (\ref{pointfunctional}), respectively.

\section{Discussion and concluding remarks}
\label{Disconclu}

In this paper we have studied the high-energy one-fermion spectral properties
of the 1D repulsive fermion model, Eq. (\ref{H}), and specifically
the momentum and energy dependence of the 
exponents and energy spectra that control the line shape of the
one-fermion spectral function, Eq. (\ref{Bkomega-m0}), at zero magnetic field and of the
up-spin and down-spin one-fermion spectral functions, Eq. (\ref{Bkomega}), at 
finite magnetic field near those functions singularities.

That fermionic model is an interacting system characterized by a breakdown 
of the basic Fermi liquid quasiparticle picture. Indeed, no quasiparticles and no quasi-holes
with the same quantum numbers as the corresponding free fermions occur when the interacting fermion range of motion is 
restricted to a single spatial dimension \cite{Sutherland-04,Voit}. In 1D, correlated fermions rather split into the basic 
fractionalized charge-only and spin-only particles whose representation is used in our study. That for finite repulsive interaction the 
generators of the exact energy eigenstates onto the fermion vacuum are naturally expressed in terms 
of creation onto it of such fractionalized particles renders it the most suitable representation to study the 
up-spin and down-spin one-fermion spectral functions. 

The many-fermion system non-perturbative character is thus the reason why in this paper 
we have used a {\it language} other than that of a Fermi liquid. 
Our analysis of the problem focused on the vicinity of two types
of singular features: The one-fermion removal and addition branch lines whose $(k,\omega)$-plane spectra
general form is given in Eq. (\ref{dE-dP-bl-m0}) for zero spin density and in Eq. (\ref{dE-dP-bl}) 
for finite spin density and the one-fermion removal boundary lines whose $(k,\omega)$-plane spectra
general form is provided in Eq. (\ref{dE-dP-c-s-m0}) for $m=0$ and in Eq. (\ref{dE-dP-c-s}) for $m>0$. 
The $\beta = c^+,c^-,s$ branch lines are represented in Figs. \ref{figure1}-\ref{figure3}
by solid lines and dashed lines for the $k$ ranges for which the 
corresponding exponent $\xi_{\beta} (k)$, Eq. (\ref{branch-l-m0}), for $m=0$ in 
Fig. \ref{figure1} and $\xi_{\beta}^{\sigma} (k)$, Eq. (\ref{branch-l}), 
for $m>0$ in Figs. \ref{figure2} and \ref{figure3} is negative and positive,
respectively. The one-fermion removal boundary lines are 
in these figures represented by dashed-dotted lines. 

Which is the physics behind the occurrence of separate charge and branch lines beyond the low-energy
TLL regime where the one-fermion spectral singularities are located in the $(k,\omega)$ plane? This 
follows from at all energy scales the exotic charge and spin fractionalized particles 
or holes moving generally with different speeds and in different directions in the 1D many-fermion system. 
The fermions degrees of freedom in the system have this ability because 
they behave like separate waves. When excited upon fermion removal or addition, 
such waves can split into multiple waves, each carrying different characteristics of the fermion. 

This occurs because collective modes take over, so that
the one-fermion removal and addition excitations studied in this paper indeed do not create single Fermi-liquid 
quasiparticles or quasi-holes with the same quantum numbers as the free fermions. 
Such one-fermion excitations rather originate an energy 
continuum of excitations that display non-Fermi-liquid singularities on the charge and spin fractionalized particles branch 
lines and boundary lines. Consistent, the $(k,\omega)$-plane line shape near such spectral features is not $\delta$-function 
like as in a Fermi liquid. It rather is power-law like, controlled by negative exponents that for the charge and spin branch lines are 
momentum, interaction, and fermionic and spin densities dependent. 

On the one hand, the one-fermion removal excitations boundary lines refer to charge 
and spin fractionalized holes moving with the same velocity whose energy spectrum has 
thus contributions from both their charge and spin energy dispersions.
On the other hand, the energy dispersions of the charge and spin fractionalized particles and holes 
fully control the shape of the corresponding charge and spin branch lines spectra, respectively, 
whose momentum slope corresponds to their generally different velocities. 
The fractionalized charge-only and spin-only particles 
and holes associated with such spectral features emerge within the 1D many-fermion system. 
They cannot though exist independently, outside such a system. Moreover, they are not adiabatically connected to free fermions. 

To access the expressions of the one-fermion spectral functions near
the branch lines and boundary lines singularities, we have used the PDT, which applies to the present model and other
integrable models \cite{Carmelo_05,Carmelo_17,Carmelo_18,Carmelo_16,Carmelo_15}. 
For the $k$ ranges for which the (i) branch lines exponents $\xi_{\beta} (k)$ at $m=0$ 
(ii) and $\sigma=\uparrow,\downarrow$ branch lines exponents $\xi_{\beta}^{\sigma} (k)$ for $m>0$ (which are plotted in (i) 
Figs. \ref{figure4}, \ref{figure7}, and \ref{figure8} and (ii) in Figs. \ref{figure5}, \ref{figure6}, \ref{figure9}, and \ref{figure10}, respectively) 
are negative, there are singularity cusps in the corresponding one-fermion spectral functions, 
Eqs. (\ref{Bkomega-m0}) and (\ref{Bkomega}). The same occurs in the $(k,\omega)$-plane vicinity 
of the one-fermion removal boundary lines.

The ${\cal{C}}>0$ branch lines singularity cusps play an important role in the model physics. For instance, at zero
spin density the $c^{\pm}$ and $s$ branch lines lead in the ${\cal{C}}\rightarrow 0$ limit to the 
${\cal{C}}=0$ non-interacting $\delta$-function-like one-fermion addition and removal spectrum for their 
whole $k$ intervals, as given in Eqs. (\ref{cpm-branch-delta-m0}) and (\ref{s-branch-delta-m0}).
At finite spin densities this applies to most of the momentum $k$ ranges of the up-spin and down-spin one-fermion spectrum. 
This can be confirmed from a combined analysis of the up-spin one-fermion spectral function expressions obtained 
from the $c^{\pm}$ branch lines in Eq. (\ref{cpm-branch-delta}) and of the expressions of the down-spin one-fermion 
spectral function obtained from the $s$ branch lines in Eq. (\ref{s-branch-delta}).

The momentum subranges for which at $m> 0$ the ${\cal{C}}=0$ non-interacting $\delta$-function-like one-fermion spectrum
does not stem from branch lines are $k\in [-k_{F\downarrow},k_{F\downarrow}]$
for up-spin one-fermion removal and $k\in [-\infty,-k_{F\uparrow}]$ and
$k\in [k_{F\uparrow},\infty]$ for down-spin one-fermion addition. The PDT also accounts for
the processes that give rise in the ${\cal{C}}\rightarrow 0$ limit to the ${\cal{C}}=0$ 
one-fermion spectrum at such $k$ intervals. However, the expression of the corresponding
one-fermion spectral functions near the spectral line features under consideration involve
state summations difficult to analytically compute for ${\cal{C}}>0$. 
(Those ${\cal{C}}>0$ line features are represented in Figs. \ref{figure2} and \ref{figure3}
by sets of diamond symbols.) 

The interacting spin-$1/2$ fermions described by 1D repulsive fermion model, Eq. (\ref{H}), can either be electrons
or atoms. Which is the relevance and consequences for actual physical systems of the theoretical results of this 
paper on the up-spin and down-spin one-fermion removal and addition spectral functions, Eqs. (\ref{Bkomega-m0}) 
and (\ref{Bkomega})? Can their spectral singularities be observed in actual systems both at zero and finite magnetic fields?

In condensed matter materials at zero magnetic field, angle resolved photoemission spectroscopy (ARPES) directly 
measures the spectral function of the electrons \cite{Damascelli_03}. ARPES removes electrons via the photoelectric effect. 
This technique does not apply at finite magnetic fields and can only measure occupied one-electron states. 
To measure the unoccupied states, there is inverse photo-emission spectroscopy or as well as tunneling 
experiments \cite{Damascelli_03}. Quasi-1D and 1D condensed matter systems are in general rather described
by toy lattice correlated electronic models the simplest of which is the 1D Hubbard model \cite{spectral0,Dionys-87}.

Concerning the relation of our theoretical results on the spectral functions of the 1D continuous fermionic gas with repulsive 
delta-function interaction to actual physical systems, that model can be implemented with ultra-cold atoms \cite{Batchelor_16,Guan_13,Zinner_16,Dao_2007,Stewart_08,Clement_09,Febbri_12}. 
The ability to study ultra-cold atomic Fermi gases actually holds the promise of significant advances in testing fundamental theories 
of the corresponding many-fermion quantum physics. This applies to the results presented in this paper. 

Momentum-resolved radio-frequency (RF) spectroscopy \cite{Dao_2007,Stewart_08} and Bragg spectroscopy 
\cite{Clement_09,Febbri_12} are techniques to measure the spectral functions in ultra-cold atomic gases. 
In particular, momentum-resolved RF is a tool to achieve an analogue of the photoemission spectroscopy measurement 
for ultra-cold atomic gases. The spectroscopy takes advantage of the many spin states of the atoms in these cold gases. 
Can our results about one-fermion singular spectral features at magnetic field be used as theoretical predictions of
ultra-cold spin-$1/2$ atomic gases experiments?

In a magnetic field, the degeneracy of the spin states of the atoms is split by the Zeeman interaction and at magnetic field 
strengths around the Feshbach resonance. This Zeeman splitting is much larger than other energy scales in the system. 
Fortunately, all the spin-relaxation mechanisms available to atoms in some of their spin-states are either forbidden 
or strongly suppressed, so a gas of atoms in those spin states stays that way without relaxing.

Momentum-resolved RF spectroscopy, takes advantage of the fact that the RF photon has a negligible momentum compared to the 
momentum of the atom. As a result, the spin-flip transition does not change its momentum state. In the language of photoemission 
spectroscopy this is a vertical transition. The momentum of the spin-flipped atom, and thus the momentum of the atom inside 
the interacting system, can be measured in a time-of-flight experiment. 
Importantly, with this information, one {\it can indeed} reconstruct the one-fermion spectral function and thus use the present
results as a theoretical prediction and check their relevance and consequences for actual physical systems \cite{Dao_2007,Stewart_08}. 

Finally, it is interesting to briefly discuss here some complications that momentum-resolved RF spectroscopy removes
compared to ARPES for condensed-matter systems as well as other new complications that it though introduces.
For example, ARPES suffers from ejected electrons colliding with other electrons on 
their way out of the material \cite{Damascelli_03}. This introduces a background signal and limits ARPES to probing 
near the material surface. With ultra-cold atoms, the interactions 
between the atoms in the out-coupled spin state and the rest of the atoms is so weak that this is not a problem. 
Indeed, the mean free path of the out-coupled atom is much larger than the system size \cite{Stewart_08}. 
Furthermore, in ARPES there is a matrix element for the process of removing an electron with a photon that depends on the angles and 
wave-vectors of the photon and electron \cite{Damascelli_03}. This matrix element is not always known well enough to divide out. 
In the ultra-cold atom case, the emission process is simply a Zeeman spin-flip transition and the matrix element is 
constant \cite{Chen_09}. Experimentally, the photon source in the atom case is a RF field. It is far easier to produce than 
a synchrotron x-ray radiation ARPES source or even then a laser based ultra-violet ARPES source. 
Moreover, the detection of atoms by time-of-flight absorption imaging is simpler than the detection of electrons with an electron spectrometer. 
It requires only a low power laser, charge-coupled device camera, and imaging optics. Another general advantage is the ability of ultra-cold atom gas 
experiments to reproduce identical samples repeatedly. In contrast, this can be a real challenge in ARPES
for condensed-matter materials.

To acquire the full spectral function, in the case of the ultra-cold atoms a range of frequencies are needed.
On the contrary, with ARPES the entire spectral function can be measured with a single photon frequency.
This follows the non-conservation of momentum and energy in the direction perpendicular to the material surface,
as discussed in Ref. \onlinecite{Damascelli_03}. In ARPES, the 
energy resolution of the measurements is typically set by the detector resolution. It can be made very small compared 
to the Fermi energy, specifically, meV compared to eV \cite{Damascelli_03}. In the ultra-cold atom case, the energy resolution is Fourier limited 
by the duration of the RF pulse. It must be kept much shorter than the oscillation period of the confinement potential.
This is needed in order to keep the momentum of the out-coupled atoms from changing.

In conclusion, the relevance and consequences for actual physical systems of our detailed theoretical study on the up-spin and 
down-spin one-fermion removal and addition spectral functions, Eqs. (\ref{Bkomega-m0}) and (\ref{Bkomega}), of the
1D repulsive fermion model, Eq. (\ref{H}), at and near such functions singularities where most of the spectral weight is located 
in the $(k,\omega)$ plane, can be checked by experimental studies of ultra-cold spin-$1/2$ atomic systems. Those can rely on
momentum-resolved RF spectroscopy \cite{Dao_2007,Stewart_08} or Bragg spectroscopy \cite{Clement_09,Febbri_12}.
An interesting program would thus be the observation of the one-atom spectral weight distributions over the $(k,\omega)$ plane associated with the 
spectral functions studied in this paper in systems of spin $1/2$ ultra-cold fermionic atoms on optical lattices,
simulating both a vanishing and finite magnetic fields.

\acknowledgements

T. C. and J. M. P. C. would like to thank P. D. Sacramento for fruitful discussions. We acknowledge the support from 
NSAF Grant U1530401, NSFC Grant 11650110443, computational resources from CSRC (Beijing), and the FEDER 
through the COMPETE Program and the Portuguese FCT in the framework of the Strategic Project UID/FIS/04650/2013. 
S. N. and J. M. P. C. thank the support of the FCT Grant PTDC/FIS-MAC/29291/2017. J. M. P. C. would like to thank Boston 
University's Condensed Matter Theory Visitors Program for support, the hospitality of MIT, and acknowledges the support from
FCT Grant SFRH/BSAB/142925/2018.

\appendix

\section{Useful quantities and limiting behaviors and values}
\label{UBAQ}

The 1D repulsive fermion model, Eq. (\ref{H}), BA solution ground-state charge momentum rapidity function 
$k = k_0 (q) \in [-\infty,\infty]$ where $q \in [-\infty,\infty]$,
whose inverse function $q = q (k)$ has thus also domain $k \in [-\infty,\infty]$, is defined by the equation,
\begin{eqnarray}
q (k) & = & k + \frac{1}{\pi} \int_{-B}^{B}d\Lambda\,2\pi\sigma (\Lambda)\, \arctan \left({2k - 2\Lambda\over {\cal{C}}}\right) 
\hspace{0.20cm}{\rm for}\hspace{0.20cm}k \in [-\infty,\infty] 
\nonumber \\
q (\pm Q) & = & \pm 2k_F \hspace{0.20cm}{\rm where}\hspace{0.20cm}\pm Q = k_0 (\pm 2k_F) 
\nonumber \\
q (\pm\infty) & = & \pm\infty \hspace{0.20cm}{\rm where}\hspace{0.20cm}\pm\infty = k_0 (\pm\infty) \, ,
\label{eq-cont1}
\end{eqnarray}
where ${\cal{C}}>0$. Moreover, 
the ground-state spin rapidity function $\Lambda = \Lambda_0 (q')$ domain is $q' \in [-k_{F\uparrow},k_{F\uparrow}]$,
whose inverse function $q = q (\Lambda)$ domain is $\Lambda \in [-\infty,\infty]$,
is defined by the equation,
\begin{eqnarray}
q (\Lambda) & = & {1\over\pi}\int_{-Q}^{Q}dk\,2\pi\rho (k)\, \arctan \left({2\Lambda-2k\over {\cal{C}}}\right) 
-  \frac{1}{\pi} \int_{-B}^{B}d\Lambda^{\prime}\,2\pi\sigma (\Lambda^{\prime})\, \arctan \left({\Lambda - \Lambda^{\prime}\over {\cal{C}}}\right) 
\hspace{0.20cm}{\rm for}\hspace{0.20cm}\Lambda \in [-\infty,\infty] 
\nonumber \\
q (\pm B) & = & \pm k_{F\downarrow} \hspace{0.20cm}{\rm where}\hspace{0.20cm} \pm B = \Lambda_0 (\pm k_{F\downarrow})
\nonumber \\
q (\pm\infty) & = & \pm k_{F\uparrow} \hspace{0.20cm}{\rm where}\hspace{0.20cm} \pm\infty = \Lambda_0 (\pm k_{F\uparrow}) 
\hspace{0.20cm}{\rm and}\hspace{0.20cm}k_{F\uparrow} = 2k_F-k_{F\downarrow} = \pi\,n_{\uparrow} \, .
\label{eq-cont}
\end{eqnarray}
(Here we distinguish the functions $q = q (k) \in [-\infty,\infty]$ and 
$q = q (\Lambda)\in [-k_{F\uparrow},k_{F\uparrow}]$ by their variables reading $k$ and $\Lambda$, respectively.)

The charge distribution $2\pi\rho (k)$ and the spin distribution $2\pi\sigma (\Lambda)$ appearing in the above
equations are such that,
\begin{equation}
2\pi\rho (k) = {\partial q (k)\over\partial k} \, ; \hspace{0.20cm} {\partial k (q)\over\partial q} = {1\over 2\pi\rho (k(q))} \, ;
\hspace{0.20cm}
2\pi\sigma (\Lambda) = {\partial q (\Lambda)\over\partial \Lambda} 
\, ; \hspace{0.20cm} {\partial \Lambda (q)\over\partial q} = {1\over 2\pi\sigma (\Lambda(q))} \, .
\label{defrhosima}
\end{equation}
They are solutions of the coupled integral equations,
\begin{equation}
2\pi\rho (k) = 1 + \frac{2}{\pi\,{\cal{C}}} \int_{-B}^{B}d\Lambda\,{2\pi\sigma (\Lambda)\over 1 +  \left({2k - 2\Lambda\over {\cal{C}}}\right)^2} \, ,
\label{rho}
\end{equation}
and
\begin{equation}
2\pi\sigma (\Lambda) = {2\over\pi\,{\cal{C}}}\int_{-Q}^{Q}dk\,{2\pi\rho (k)\over 1 +  \left({2\Lambda-2k\over {\cal{C}}}\right) ^2} 
- \frac{1}{\pi\,{\cal{C}}} \int_{-B}^{B}d\Lambda^{\prime}\,{2\pi\sigma (\Lambda^{\prime})\over 1 +  \left({\Lambda - \Lambda^{\prime}\over {\cal{C}}})\right)^2} \, ,
\label{sigma}
\end{equation}
respectively. From the use of Eq. (\ref{rho}) in Eq. (\ref{sigma}) one obtains the following single integral equation for  
the spin distribution $2\pi\sigma (\Lambda)$,
\begin{equation}
2\pi\sigma (\Lambda) = {2\over\pi\,{\cal{C}}}\int_{-Q}^{Q}dk\,{1\over 1 +  \left({2\Lambda-2k\over {\cal{C}}}\right) ^2} 
+ \int_{-B}^{B}d\Lambda^{\prime}\,G (\Lambda,\Lambda^{\prime})\,2\pi\sigma (\Lambda^{\prime}) \, ,
\label{sigma2}
\end{equation}
where the kernel reads,
\begin{equation}
G (\Lambda,\Lambda^{\prime}) = - \frac{1}{\pi\,{\cal{C}}}\left({1\over 1 +  \left({\Lambda - \Lambda^{\prime}\over {\cal{C}}}\right)^2} - 
\frac{4}{\pi\,{\cal{C}}}\int_{-Q}^{Q}dk\,
{1\over \left(1 +  \left({2\Lambda-2k\over {\cal{C}}}\right)^2\right)\left(1 +  \left({2\Lambda^{\prime}-2k\over {\cal{C}}}\right)^2\right)}\right) \, .
\label{G}
\end{equation}
The $k$ integral in this expression can be solved analytically, with the result,
\begin{eqnarray}
G (\Lambda,\Lambda^{\prime}) & = & {2\over {\cal{C}}} A  \left({2\Lambda\over {\cal{C}}},{2\Lambda^{\prime}\over {\cal{C}}}\right)
\hspace{0.20cm}{\rm where}
\nonumber \\
A (r,r^{\prime}) & = & - \frac{1}{2\pi}\left({1\over 1 + \left({r -r^{\prime}\over 2}\right)^2}\right)
\nonumber \\
& \times & \left(1 - {1\over 2\pi}\sum_{\iota =\pm 1} (\iota)\left\{\arctan F_{\iota} (r)
+ \arctan F_{\iota} (r^{\prime}) + {\ln (1 + F_{\iota}^2 (r)) - \ln (1 + F_{\iota}^2 (r^{\prime}))
\over r-r^{\prime}}\right\}\right)\hspace{0.20cm}{\rm and}
\nonumber \\
F_{\iota} (r) & = &  r + \iota\,{2Q\over {\cal{C}}} \, .
\label{Gexp}
\end{eqnarray}

As given in Eq. (\ref{defrhosima}),
the charge distribution $2\pi\rho (k)$ and the spin distribution $2\pi\sigma (\Lambda)$ are such that
$2\pi\rho (k)  = {\partial q (k)\over \partial k}$ and 
$2\pi\sigma (\Lambda) = {\partial q (\Lambda)\over \partial\Lambda}$, respectively. Hence they are 
defined by the equations,
\begin{equation}
q = \int_0^{k_0 (q)}dk \,2\pi\rho (k)
\hspace{0.20cm}{\rm and}\hspace{0.20cm}
q' = \int_0^{\Lambda_0 (q')}d\Lambda \,2\pi\sigma (\Lambda) \, ,
\nonumber
\end{equation}
respectively. The parameters $Q$ and $B$ are then self-consistently defined by the equations,
\begin{equation}
2k_F = \int_0^{Q}dk \,2\pi\rho (k) 
\hspace{0.20cm}{\rm and}\hspace{0.20cm}
k_{F\downarrow} = \int_0^{B}d\Lambda \,2\pi\sigma (\Lambda) \, ,
\nonumber
\end{equation}
respectively.

It follows that,
\begin{equation}
2k_F = Q + \frac{1}{\pi} \int_{-B}^{B}d\Lambda\,2\pi\sigma (\Lambda)\, \arctan \left({2Q - 2\Lambda\over {\cal{C}}}\right) \, ,
\nonumber
\end{equation}
and
\begin{equation}
k_{F\downarrow} =  {1\over\pi}\int_{-Q}^{Q}dk\,2\pi\rho (k)\, \arctan \left({2B-2k\over {\cal{C}}}\right) 
-  \frac{1}{\pi} \int_{-B}^{B}d\Lambda\,2\pi\sigma (\Lambda)\, \arctan \left({B -\Lambda\over {\cal{C}}}\right) \, .
\nonumber
\end{equation}

Note that Eqs. (\ref{eq-cont1}) and (\ref{eq-cont}) are equivalent to the following equations, 
\begin{equation}
q = k_0 (q) + \frac{1}{\pi} \int_{-k_{F\downarrow}}^{k_{F\downarrow}}dq^{\prime}\arctan \left({2k_0 (q) - 2\Lambda_0 (q^{\prime})\over {\cal{C}}}\right) 
\hspace{0.20cm}{\rm for}\hspace{0.20cm}q \in ]-\infty,\infty[ \, ,
\label{eq-cont1-q}
\end{equation}
and
\begin{equation}
q' = {1\over\pi}\int_{-2k_F}^{2k_F}dq''\arctan \left({2\Lambda_0 (q') - 2k_0 (q'')\over {\cal{C}}}\right) 
- \frac{1}{\pi} \int_{-k_{F\downarrow}}^{k_{F\downarrow}}dq''\arctan \left({\Lambda_0 (q') - \Lambda_0 (q'')\over {\cal{C}}}\right) 
\hspace{0.20cm}{\rm for}\hspace{0.20cm} q' \in [-k_{F\uparrow},k_{F\uparrow}] \, ,
\label{eq-cont-q}
\end{equation}
respectively. Here the integrals run over the ground-state momentum occupancies 
$q'' \in [-k_{F\downarrow},k_{F\downarrow}]$ of the $s$ band and $q'' \in [-2k_F,2k_F]$ of the
$c$ band whereas Eqs. (\ref{eq-cont1-q}) and (\ref{eq-cont-q})
define the functions $\Lambda_0 (q')$ and $k_0 (q)$ for their whole momentum ranges
$q' \in [-k_{F\uparrow},k_{F\uparrow}]$ and $q \in ]-\infty,\infty[$, respectively. 

Each occupied $c$ band momentum $q$ is associated with one $c$ pseudoparticle and 
occupied $s$ band momentum $q'$ with one $s$ pseudoparticle. The range of the corresponding 
$c$ band holes and $s$ band holes is thus $\vert q\vert \in [2k_F,\infty]$ and 
$\vert q'\vert \in [k_{F\downarrow},k_{F\uparrow}]$, respectively. For the alternative rapidity variables the ground-state
momentum occupancy ranges refer to $\Lambda \in [-B,B]$ for the $s$ band and 
$k \in [-Q,Q]$ for the $c$ band whereas the corresponding full ranges are
$\Lambda \in [-\infty,\infty]$ and $k \in [-\infty,\infty]$, respectively. 

The distributions $\eta_c (\Lambda)$ and $\eta_{s} (\Lambda)$ in Eq. (\ref{varepsilon-c-s}) are solutions of the integral equations,
\begin{equation}
\eta_c (k)  = 2k + \frac{2}{\pi\,{\cal{C}}} \int_{-B}^{B}d\Lambda\,{\eta_{s} (\Lambda)\over 1 +  \left({2k - 2\Lambda\over {\cal{C}}}\right)^2} \, ,
\label{ceta}
\end{equation}
and
\begin{equation}
\eta_{s} (\Lambda) = {2\over\pi\,{\cal{C}}}\int_{-Q}^{Q}dk\,{\eta_c (k)\over 1 +  \left({2\Lambda-2k\over {\cal{C}}}\right) ^2} 
- \frac{1}{\pi\,{\cal{C}}} \int_{-B}^{B}d\Lambda^{\prime}\,{\eta_{s} (\Lambda^{\prime})\over 1 +  \left({\Lambda - \Lambda^{\prime}\over {\cal{C}}})\right)^2} \, .
\label{Jeta}
\end{equation}

From the use of Eq. (\ref{ceta}) in Eq. (\ref{Jeta}) one obtains the following single integral equation for  
the distribution $\eta_{s} (\Lambda)$,
\begin{equation}
\eta_{s} (\Lambda) = {2\over\pi\,{\cal{C}}}\int_{-Q}^{Q}dk\,{2k\over 1 +  \left({2\Lambda-2k\over {\cal{C}}}\right) ^2} 
+ \int_{-B}^{B}d\Lambda^{\prime}\,G (\Lambda,\Lambda^{\prime})\,\eta_{s} (\Lambda^{\prime}) \, ,
\label{Jeta2}
\end{equation}
where the kernel is that already given in Eq. (\ref{G}).

On the one hand, the dispersion $\varepsilon_c^0 (q)$ in Eq. (\ref{varepsilon}) reads $\varepsilon_c^0 (q) = {\bar{\varepsilon}}_c^{\,0} (k_0 (q))$
where the related dispersion ${\bar{\varepsilon}}_c^{\,0} (k)$ is given by,
\begin{equation}
{\bar{\varepsilon}}_c^{\,0} (k) = k^2 + \frac{1}{\pi} \int_{-B}^{B}d\Lambda\,\eta_{s} (\Lambda)
\arctan\left({2k - 2\Lambda\over {\cal{C}}}\right) \, .
\nonumber
\end{equation}
On the other hand, $\varepsilon_{s}^0 (q') = {\bar{\varepsilon}}_{s}^{\,0} (\Lambda_0 (q'))$ where
${\bar{\varepsilon}}_{s}^{\,0} (\Lambda) = \int_{\infty}^{\Lambda}d\Lambda'\,2t\,\eta (\Lambda')$, so that,
\begin{equation}
{\bar{\varepsilon}}_{s}^{\,0} (\Lambda) = {1\over\pi}\int_{-Q}^{Q}dk\,\eta_c (k)
\arctan\left({2\Lambda-2k\over {\cal{C}}}\right)
- \frac{1}{\pi} \int_{-B}^{B}d\Lambda^{\prime}\,\eta_{s} (\Lambda^{\prime})
\arctan\left({\Lambda - \Lambda^{\prime}\over {\cal{C}}}\right) \, . 
\nonumber
\end{equation}

The distributions given in Eqs. (\ref{ceta}) and (\ref{Jeta}) are such that,
\begin{equation}
\eta_c (k) = {\partial {\bar{\varepsilon}}_c (k) \over \partial k} = {\partial {\bar{\varepsilon}}_c^{\,0} (k) \over \partial k} 
\hspace{0.20cm}{\rm and}\hspace{0.20cm}
\eta_{s} (\Lambda) = {\partial {\bar{\varepsilon}}_{s} (\Lambda) \over \partial\Lambda} 
= {\partial {\bar{\varepsilon}}_{s}^{\,0} (\Lambda) \over \partial\Lambda} \, ,
\label{eta-epsolin}
\end{equation}
respectively.

The $c$ and $s$ group velocities, Eq. (\ref{vcvs}), can be written as,
\begin{equation}
v_c (q) = {\partial {\bar{\varepsilon}}_c (k) \over \partial k} {\partial k_0 (q)\over\partial q} 
\hspace{0.20cm}{\rm and}\hspace{0.20cm}
v_{s} (q') = {\partial {\bar{\varepsilon}}_{s} (\Lambda) \over \partial\Lambda} {\partial \Lambda_0 (q')\over\partial q'} \, .
\nonumber
\end{equation}
One then finds from the use of Eqs. (\ref{defrhosima}) and (\ref{eta-epsolin}) that,
\begin{equation}
v_c (q) = {\eta_c (k) \over 2\pi\rho (k)}\vert_{k= k_0 (q)} 
\hspace{0.20cm}{\rm and}\hspace{0.20cm}
v_{s} (q') = {\eta_{s} (\Lambda)\over 2\pi\sigma (\Lambda)}\vert_{\Lambda= \Lambda_0 (q')} \, ,
\label{vcsBAR}
\end{equation}
where the distributions $2\pi\rho (k)$ and $2\pi\sigma (\Lambda)$ are the solutions of Eqs. (\ref{rho})-(\ref{sigma2})
and the distributions $\eta_c (k)$ and $\eta_{s} (\Lambda)$ are the solutions of Eqs. (\ref{ceta})-(\ref{Jeta2}).

The $c$ and $s$ Fermi velocities then read,
\begin{equation}
v_c (\pm 2k_F) = \pm {\eta_c (Q) \over 2\pi\rho (Q)} 
\hspace{0.20cm}{\rm and}\hspace{0.20cm}
v_{s} (\pm k_{F\downarrow}) = \pm {\eta_{s} (B)\over 2\pi\sigma (B)} \, .
\label{vcsBARF}
\end{equation}

The expressions of the energy scales $2\mu$ and $2\mu_B\,h$ given in Eq. (\ref{mu-muBH0})
can be expressed as,
\begin{eqnarray}
2\mu = 2\varepsilon_c^0 (2k_F) - \varepsilon_{s}^0 (k_{F\downarrow}) & = &
2Q^2 + \frac{2}{\infty} \int_{-B}^{B}d\Lambda\,\eta_{s} (\Lambda)
\arctan\left({2Q - 2\Lambda\over {\cal{C}}}\right)
\nonumber \\
& + & {1\over\pi}\int_{-Q}^{Q}dk\,\eta_c (k)
\arctan\left({2k-2B\over {\cal{C}}}\right)
- \frac{1}{\pi} \int_{-B}^{B}d\Lambda^{\prime}\,\eta_s (\Lambda^{\prime})
\arctan\left({\Lambda^{\prime}-B\over {\cal{C}}}\right)
\nonumber \\
2\mu_B\,h = - \varepsilon_{s}^0 (k_{F\downarrow}) & = & 
{1\over\pi}\int_{-Q}^{Q}dk\,\eta_c (k)
\arctan\left({2k-2B\over {\cal{C}}}\right)
- \frac{1}{\pi} \int_{-B}^{B}d\Lambda^{\prime}\,\eta_s (\Lambda^{\prime})
\arctan\left({\Lambda^{\prime}-B\over {\cal{C}}}\right) \, .
\label{mu-muBH}
\end{eqnarray}

The particle $c$ energy dispersion bandwidth of the occupied $c$ Fermi sea reads,
\begin{eqnarray}
W_c^p & = & - \varepsilon_c (0) = - {\bar{\varepsilon}}_c (0) 
\nonumber \\
& = & Q^2 + \frac{1}{\pi} \int_{-B}^{B}d\Lambda\,\eta_{s} (\Lambda)
\left(\arctan\left({2\Lambda\over {\cal{C}}}\right) +
\arctan\left({2Q - 2\Lambda\over {\cal{C}}}\right)\right) \, .
\label{Wc-expr}
\end{eqnarray}

The energy bandwidth of the $s$ bands $\varepsilon_{s}^0 (q')$ and $\varepsilon_{s} (q')$
is given by,
\begin{eqnarray}
W_{s} & = & \varepsilon_{s}^0 (k_{F\uparrow}) - \varepsilon_{s}^0 (0) = \varepsilon_{s} (k_{F\uparrow}) - \varepsilon_{s} (0)
\nonumber \\
& = & {\bar{\varepsilon}}_{s}^{\,0} (\infty) - {\bar{\varepsilon}}_{s}^{\,0} (0)  = {\bar{\varepsilon}}_{s} (\infty) - {\bar{\varepsilon}}_{s} (0) \, .
\label{Ws}
\end{eqnarray}
Here $\varepsilon_{s}^0 (k_{F\uparrow})={\bar{\varepsilon}}_{s}^{\,0} (\infty) =0$ so that,
\begin{equation}
W_{s} = - {\bar{\varepsilon}}_{s}^{\,0} (0) = {1\over\pi}\int_{-Q}^{Q}dk\,\eta_c (k)
\arctan\left({2k\over {\cal{C}}}\right)
- \frac{1}{\pi} \int_{-B}^{B}d\Lambda^{\prime}\,\eta_s (\Lambda^{\prime})
\arctan\left({\Lambda^{\prime}\over {\cal{C}}}\right) \, . 
\label{Ws-expr}
\end{equation}
The $s$ energy bandwidth $W_{s}$ can be written as,
\begin{equation}
W_{s} = W_{s}^p + W_{s}^h 
\hspace{0.2cm}{\rm where}\hspace{0.2cm}
W_{s}^p = - \varepsilon_{s} (0) = - {\bar{\varepsilon}}_{s} (0) 
\hspace{0.2cm}{\rm and}\hspace{0.2cm}
W_{s}^h = \varepsilon_{s} (k_{F\uparrow}) = 2\mu_B\, h  \, .
\label{Wph-lim}
\end{equation}
$W_{s}^p$ and $W_{s}^h$ are here the energy bandwidths of the occupied and unoccupied $s$ Fermi sea, respectively. Hence,
\begin{equation}
W_{s}^p = {1\over\pi}\int_{-Q}^{Q}dk\,\eta_c (k)
\arctan\left({2k\over {\cal{C}}}\right)
- \frac{1}{\pi} \int_{-B}^{B}d\Lambda\,\eta_s (\Lambda)
\arctan\left({\Lambda\over {\cal{C}}}\right) - 2\mu_B\, h \, . 
\label{Ws-exprp}
\end{equation}
For all ${\cal{C}}>0$ values the energy scale $W_{s}^h$ is an increasing function of the magnetic field $h$ with the
following limiting behaviors,
\begin{equation}
\lim_{h\rightarrow 0}W_{s}^h = 0 
\hspace{0.20cm}{\rm and}\hspace{0.20cm}
\lim_{h\rightarrow h_c}W_{s}^h = 2\mu_B h_c \, ,
\label{Ws-ph}
\end{equation}
where the $2\mu_B h_c $ expression is given in Eq. (\ref{hc}).

The phase shifts contributing to functionals in Eqs. (\ref{OESFfunctional-m0}) and (\ref{OESFfunctional}) 
can be expressed in terms of the corresponding phase-shift functions of rapidity variables as,
\begin{eqnarray}
\Phi_{s,s}\left(\pm k_{F\downarrow},q'\right) & = & \bar{\Phi }_{s,s} \left(\pm  {2B\over {\cal{C}}},{2\Lambda_0 (q')\over {\cal{C}}}\right) \, ;
\hspace{0.40cm}
\Phi_{s,c}\left(\pm k_{F\downarrow},q\right) = \bar{\Phi }_{s,c} \left(\pm {2B\over {\cal{C}}},{2k_0 (q)\over {\cal{C}}}\right) 
\nonumber \\
\Phi_{c,c}\left(\pm 2k_F,q\right) & = & \bar{\Phi }_{c,c} \left(\pm {2Q\over {\cal{C}}},{2k_0 (q)\over {\cal{C}}}\right) \, ;
\hspace{0.40cm}
\Phi_{c,s}\left(\pm 2k_F,q'\right) = \bar{\Phi }_{c,s} \left(\pm {2Q\over {\cal{C}}},{2\Lambda_0 (q')\over {\cal{C}}}\right) \, ,
\label{PhiqFq}
\end{eqnarray}
The latter phase-shift functions are uniquely defined by solution of the coupled integral equations given below.
At zero spin density the $s$ particle phase shifts in Eq. (\ref{PhiqFq}) have the following simple expressions
due to the spin SU(2) symmetry,
\begin{eqnarray}
\Phi_{s,c}(\iota k_F,q) & = & - {\iota\over 2\sqrt{2}} 
\nonumber \\
\Phi_{s,s}(\iota k_F,q') & = & {\iota\over 2\sqrt{2}} \hspace{0.20cm}{\rm for} \hspace{0.20cm}q' \neq \iota k_F
\nonumber \\
& = & {\iota\over 2\sqrt{2}}(3-2\sqrt{2}) \hspace{0.20cm}{\rm for} \hspace{0.20cm}q' = \iota k_F
\hspace{0.20cm}{\rm at}\hspace{0.20cm}m=0
\hspace{0.20cm}{\rm and}\hspace{0.20cm}\iota = \pm 1 \, . 
\label{PhiscssFq}
\end{eqnarray}

The phase shifts on the right-hand side of the equations, Eq. (\ref{Phis-all-qq}), are functions of the rapidity-related variables
$r=2k/{\cal{C}}$ for the $c$ branch and $r=2\Lambda/{\cal{C}}$ for the $s$ branch. They are defined by the following integral equations,
\begin{eqnarray}
\bar{\Phi }_{s,s}\left(r,r'\right) & = & {1\over \pi}\arctan\left({r-r'\over 2}\right) 
- {1\over \pi^2}\int_{-2Q/{\cal{C}}}^{2Q/{\cal{C}}} d r'' {\arctan \left(r''-r'\right)\over{1+\left(r-r''\right)^2}} 
\nonumber \\
& + & \int_{-2B/{\cal{C}}}^{2B/{\cal{C}}} dr''\,A (r,r'')\,{\bar{\Phi}}_{s,s} (r'',r') \, ,
\label{Phissn-m}
\end{eqnarray}
\begin{equation}
\bar{\Phi }_{s,c}\left(r,r'\right) = -{1\over \pi}\arctan(r-r') + 
\int_{-2B/{\cal{C}}}^{2B/{\cal{C}}} dr''\,A(r,r'')\,{\bar{\Phi }}_{s,c} (r'',r') \, ,
\label{Phisc-m}
\end{equation}
\begin{equation}
\bar{\Phi }_{c,c}\left(r,r'\right) = {1\over\pi}\int_{-2B/{\cal{C}}}^{2B/{\cal{C}}} dr''{\bar{\Phi}_{s,c} (r'',r') \over {1+(r-r'')^2}} \, ,
\label{Phicc-m}
\end{equation}
and
\begin{equation}
\bar{\Phi }_{c,s}\left(r,r'\right) = -{1\over \pi}\arctan (r-r') + {1\over\pi}\int_{-2B/{\cal{C}}}^{2B/{\cal{C}}} dr''{\bar{\Phi}_{s,s} (r'',r') \over {1+(r-r'')^2}} \, .
\label{Phicsn-m}
\end{equation}
The kernel $A (r,r')$ appearing in Eqs. (\ref{Phissn-m}) and (\ref{Phisc-m}) is that given in Eq. (\ref{Gexp}). 

Alternative equations for the phase shifts $\bar{\Phi }_{s,s}\left(r,r'\right)$ and $\bar{\Phi }_{s,c}\left(r,r'\right)$ are,
\begin{eqnarray}
\bar{\Phi }_{s,s}\left(r,r'\right) & = & {1\over \pi}\arctan\left({r-r'\over 2}\right) 
+ {1\over\pi}\int_{-{2Q/{\cal{C}}}}^{{2Q/{\cal{C}}}} dr''{\bar{\Phi }_{c,s}\left(r'',r'\right)\over{1+(r-r'')^2}} 
- {1\over{2\pi}}\int_{-2B/{\cal{C}}}^{2B/{\cal{C}}} dr''{{\bar{\Phi}}_{s,s} (r'',r')\over{1+\left({r''-r\over 2}\right)^2}} \, ,
\label{Phissn-m2}
\end{eqnarray}
\begin{eqnarray}
\bar{\Phi }_{s,c}\left(r,r'\right) & = & -{1\over \pi}\arctan(r-r') 
+ {1\over\pi}\int_{-{2Q/{\cal{C}}}}^{{2Q/{\cal{C}}}} dr''{\bar{\Phi }_{c,c}\left(r'',r'\right)\over{1+(r-r'')^2}} 
- {1\over{2\pi}}\int_{-2B/{\cal{C}}}^{2B/{\cal{C}}} dr''{{\bar{\Phi}}_{s,c} (r'',r')\over{1+\left({r''-r\over 2}\right)^2}} \, .
\label{Phisc-m2}
\end{eqnarray}

The parameters in Eq. (\ref{x-aa}) are the entries of two $2\times 2$ matrices, one for each $j = 0, 1$ value,  
\begin{equation}
Z^1 = \left[\begin{array}{cc}
\xi^{1}_{c\,c} & \xi^{1}_{c\,s}  \\
\xi^{1}_{s\,c}  & \xi^{1}_{s\,s}  
\end{array}\right]
\hspace{0.20cm}{\rm and}\hspace{0.20cm}
Z^0 = ((Z^1)^{-1})^T = \left[\begin{array}{cc}
\xi^{0}_{c\,c} & \xi^{0}_{c\,s}  \\
\xi^{0}_{s\,c}  & \xi^{0}_{s\,s} 
\end{array}\right] \, .
\label{ZZ-gen}
\end{equation}

From manipulations of the phase-shift integral equations, Eqs. (\ref{Phissn-m})-(\ref{Phicsn-m}),
one finds that the entries of the $2\times 2$ matrix $Z^1$ are given by,
\begin{equation}
\xi_{c\,c}^1 = \xi_{c\,c}^1 \left({2Q\over {\cal{C}}}\right) \, ; \hspace{0.20cm}
\xi_{c\,s}^1 = \xi_{c\,s}^1 \left({2Q\over {\cal{C}}}\right) \, ; \hspace{0.20cm}
\xi_{s\,s}^1 = \xi_{s\,s}^1 \left({2B\over {\cal{C}}}\right) \, ; \hspace{0.20cm}
\xi_{s\,c}^1 = \xi_{s\,c}^1 \left({2B\over {\cal{C}}}\right) \, ,
\label{xi1all}
\end{equation}
where the functions on the right-hand side of these equations are the solutions
of the integral equations,
\begin{eqnarray}
\xi_{c\,c}^1 (r) & = & 1 + {1\over\pi}\int_{-2B/{\cal{C}}}^{2B/{\cal{C}}} dr'{\xi_{s\,c}^1 (r') \over {1+(r-r')^2}} 
\nonumber \\
\xi_{c\,s}^1 (r) & = & {1\over\pi}\int_{-2B/{\cal{C}}}^{2B/{\cal{C}}} dr'{\xi_{s\,s}^1 (r') \over {1+(r-r')^2}} 
\nonumber \\
\xi_{s\,s}^1 (r) & = & 1 + \int_{-2B/{\cal{C}}}^{2B/{\cal{C}}} dr'\,A(r,r')\,\xi_{s\,s}^1 (r') 
\nonumber \\
\xi_{s\,c}^1 (r) & = & {1\over\pi}\sum_{\iota =\pm 1} (\iota) \arctan \left(r + \iota{2Q\over {\cal{C}}}\right) + 
\int_{-2B/{\cal{C}}}^{2B/{\cal{C}}} dr'\,A(r,r')\,\xi_{s\,c}^1 (r') \, .
\label{xi-ss-qq}
\end{eqnarray}
Here $A (r,r')$ stands for the kernel whose expression is provided in Eq. (\ref{Gexp}).

For $m\rightarrow 0$ the matrices in Eq. (\ref{ZZ-gen}) are given by,
\begin{equation}
\lim_{m\rightarrow 0}\,Z^1 = \left[\begin{array}{cc}
\xi_{0} & \xi_{0}/2 \\
0 & 1/\sqrt{2} 
\end{array}\right]
\hspace{0.20cm}{\rm and}\hspace{0.20cm}
\lim_{m\rightarrow 0}\,Z^0 = \left[\begin{array}{cc}
1/\xi_{0} & 0 \\
-1/\sqrt{2} & \sqrt{2} 
\end{array}\right] \, .
\nonumber
\end{equation}
The dependence of the $m\rightarrow 0$ parameter $\xi_0$ on the particle density $n\in ]0,\infty[$ and ${\cal{C}}$
is given by $\xi_{0} = \xi_{0} \left(2Q/{\cal{C}}\right)$. The function $\xi_{0} (r)$ obeys  
the following integral equation,
\begin{equation}
\xi_{0} (r) = 1 + \int_{-2Q/{\cal{C}}}^{2Q/{\cal{C}}} d r' D (r-r')\,\xi_{0} (r')  
\hspace{0.20cm}{\rm where}\hspace{0.20cm}
D (r) = {1\over \pi}\int_0^{\infty} d\omega {\cos (\omega\,r)\over 1+ e^{2\omega}} \, .
\label{xi-0}
\end{equation}
The parameter $\xi_0$ has limiting values $\xi_0=\sqrt{2}$ for ${\cal{C}}\rightarrow 0$ and
$\xi_0=1$ for ${\cal{C}}\rightarrow\infty$.

On the one hand, for $n\in ]0,\infty[$ and $m>0$ such matrices read in the ${\cal{C}}\rightarrow 0$ limit, 
\begin{equation}
\lim_{{\cal{C}}\rightarrow 0}\,Z^1 = \left[\begin{array}{cc}
1 & 0 \\
1 & 1 
\end{array}\right]
\hspace{0.20cm}{\rm and}\hspace{0.20cm}
\lim_{m\rightarrow n}\,Z^0 = \left[\begin{array}{cc}
1 & -1 \\
0 & 1
\end{array}\right] \, ,  
\label{ZZ-gen-u0}
\end{equation}
On the other hand, if one takes the $m\rightarrow 0$ limit before ${\cal{C}}\rightarrow 0$ they rather read,
\begin{equation}
\lim_{{\cal{C}}\rightarrow 0}\lim_{m\rightarrow 0}\,Z^1 = \left[\begin{array}{cc}
\sqrt{2} & 1/\sqrt{2} \\
0 & 1/\sqrt{2} 
\end{array}\right]
\hspace{0.20cm}{\rm and}\hspace{0.20cm}
\lim_{{\cal{C}}\rightarrow 0}\lim_{m\rightarrow 0}\,Z^0 = \left[\begin{array}{cc}
1/\sqrt{2} & 0 \\
-1/\sqrt{2} & \sqrt{2} 
\end{array}\right] \, . 
\label{ZZ-gen-m0u0}
\end{equation}
The singular behavior of the phase-shift parameters, Eq. (\ref{x-aa}), reported in
Eqs. (\ref{ZZ-gen-u0}) and (\ref{ZZ-gen-m0u0}) means that at $m=0$ and for $m\rightarrow 0$
they have different values in the ${\cal{C}}\rightarrow 0$ limit. This does not affect though the
values of the physical quantities that depend on them.

\section{Simplified expressions for $m\rightarrow n$}
\label{EDPS}

In the $m\rightarrow n$ limit for which $B=0$ the quantities defined in Appendix \ref{UBAQ} in terms of equations
have in general simple analytical expressions. Specifically,
\begin{equation}
2\pi\rho (k) = 1\hspace{0.20cm}{\rm and}\hspace{0.20cm}
2\pi\sigma (\Lambda) = - {1\over\pi}\sum_{\iota =\pm1}(\iota)\arctan\left({2\Lambda -\iota 2\pi n\over {\cal{C}}}\right) \, ,
\nonumber
\end{equation}
\begin{equation}
\eta_c (k) = 2k
\hspace{0.20cm}{\rm and}\hspace{0.20cm}
\eta_s (\Lambda) = - {2\Lambda\over\pi}\sum_{\iota =\pm1}(\iota)\arctan\left({2\Lambda -\iota 2\pi n\over {\cal{C}}}\right) 
+ {{\cal{C}}\over 2\pi}\sum_{\iota =\pm1}(\iota)\ln \left(1 + \left({2\Lambda -\iota 2\pi n\over {\cal{C}}}\right)^2\right) \, ,
\nonumber
\end{equation}
\begin{eqnarray}
k_0 (q) & = & q \hspace{0.30cm}{\rm and}\hspace{0.30cm} q (k) = k
\nonumber \\
\Lambda_0 (q') & & {\rm inverse}\hspace{0.30cm}{\rm of}\hspace{0.30cm}
q (\Lambda) = {1\over 2}\eta_s  (\Lambda) + n\sum_{\iota =\pm1}\arctan\left({2\Lambda -\iota 2\pi n\over {\cal{C}}}\right) \, ,
\nonumber
\end{eqnarray}
\begin{eqnarray}
{\bar{\varepsilon}}_c^{\,0} (k) & = & k^2 \hspace{0.30cm}{\rm and}\hspace{0.30cm}
{\bar{\varepsilon}}_c (k) = k^2 - (\pi\,n)^2
\nonumber \\
{\bar{\varepsilon}}_{s}^{\,0} (\Lambda) & = &  
{1\over 4\pi}\sum_{\iota =\pm1}(\iota)\left({\cal{C}}^2 + (2\pi n)^2 - (2\Lambda)^2\right)
\arctan\left({2\Lambda -\iota 2\pi n\over {\cal{C}}}\right) 
\nonumber \\
& + & {\Lambda\, {\cal{C}}\over 2\pi}\sum_{\iota =\pm1}(\iota)\ln \left(1 + \left({2\Lambda -\iota 2\pi n\over {\cal{C}}}\right)^2\right)
+ n\, {\cal{C}}\hspace{0.30cm}{\rm and}\hspace{0.30cm}
{\bar{\varepsilon}}_{s} (\Lambda) = {\bar{\varepsilon}}_{s}^{\,0} (\Lambda) - {\bar{\varepsilon}}_{s}^{\,0} (0) \, ,
\label{varepsilonsB0}
\end{eqnarray}
\begin{eqnarray}
\varepsilon_c^{\,0} (q) & = & q^2 \hspace{0.30cm}{\rm and}\hspace{0.30cm}
\varepsilon_c (q)  = q^2 - (\pi\,n)^2
\nonumber \\
\varepsilon_{s}^{\,0} (q') & = &  
{1\over 4\pi}\sum_{\iota =\pm1}(\iota)\left({\cal{C}}^2 + (2\pi n)^2 - (2\Lambda_0 (q'))^2\right)
\arctan\left({2\Lambda_0 (q') -\iota 2\pi n\over {\cal{C}}}\right) 
\nonumber \\
& + & {\Lambda_0 (q')\, {\cal{C}}\over 2\pi}\sum_{\iota =\pm1}(\iota)\ln \left(1 + \left({2\Lambda_0 (q') -\iota 2\pi n\over {\cal{C}}}\right)^2\right)
+ n\, {\cal{C}}\hspace{0.30cm}{\rm and}\hspace{0.30cm}
\varepsilon_{s} (q') = \varepsilon_{s}^{\,0} (q') - \varepsilon_{s}^{\,0} (0) \, .
\label{varepsilonsqB0}
\end{eqnarray}

The related limiting behaviors of the energy scale $2\mu_B\,h_c$, Eq. (\ref{hc}), are,
\begin{eqnarray}
2\mu_B h_c & = & 2\pi n^2 \hspace{0.30cm}{\rm for}\hspace{0.30cm}n\ll 1
\nonumber \\
& = & \pi^2 n^2 \hspace{0.30cm}{\rm for}\hspace{0.30cm}n\gg 1 \, ,
\label{hclimitsn}
\end{eqnarray}
for finite interaction ${\cal{C}}$ and,
\begin{eqnarray}
2\mu_B h_c & = & 2\pi n^2 \hspace{0.30cm}{\rm for}\hspace{0.30cm}{\cal{C}}\gg 1
\nonumber \\
& = & \pi^2 n^2 \hspace{0.30cm}{\rm for}\hspace{0.30cm}{\cal{C}}\ll 1 \, ,
\label{hclimitsC}
\end{eqnarray}
or finite density $n$.

The phase shifts expressed in terms of rapidities in Eqs. (\ref{Phissn-m})-(\ref{Phicsn-m}) of Appendix \ref{UBAQ} simplify to,
\begin{eqnarray}
\bar{\Phi }_{s,s}\left(r,r'\right) & = & {1\over \pi}\arctan\left({r-r'\over 2}\right) 
- {1\over \pi^2}\int_{-{2\pi n\over {\cal{C}}}}^{2\pi n\over {\cal{C}}} d r'' {\arctan \left(r''-r'\right)\over{1+\left(r-r''\right)^2}} \, ,
\nonumber
\end{eqnarray}
\begin{equation}
\bar{\Phi }_{s,c}\left(r,r'\right) = -{1\over \pi}\arctan(r-r') \, ,
\nonumber
\end{equation}
\begin{equation}
\bar{\Phi }_{c,c}\left(r,r'\right) = 0 \, ,
\nonumber
\end{equation}
\begin{equation}
\bar{\Phi }_{c,s}\left(r,r'\right) = -{1\over \pi}\arctan (r-r') \, .
\nonumber
\end{equation}

Moreover, in the $m\rightarrow n$ limit the matrices in Eq. (\ref{ZZ-gen}) of Appendix \ref{UBAQ} read,
\begin{equation}
\lim_{m\rightarrow n}\,Z^1 = \left[\begin{array}{cc}
1 & 0 \\
\eta_0 & 1 
\end{array}\right]
\hspace{0.20cm}{\rm and}\hspace{0.20cm}
\lim_{m\rightarrow n}\,Z^0 = \left[\begin{array}{cc}
1 & -\eta_0 \\
0 & 1
\end{array}\right] \, ,  
\nonumber
\end{equation}
where,
\begin{equation}
\eta_0 = {2\over\pi}\arctan\left({2\pi n\over {\cal{C}}}\right) \, .
\nonumber
\end{equation}

\section{Limiting behaviors of the one-fermion branch lines spectra}
\label{AIBRL}

In the ${\cal{C}}\rightarrow 0$ limit the spectrum, Eqs. (\ref{OkudLAcc-m0}) and (\ref{kudLAcc-m0}), is given by,
\begin{eqnarray}
\omega_{c^+} (k) & = & - {1\over 2}(k-k_F)^2 + 2(k_F)^2 
\hspace{0.2cm}{\rm for}\hspace{0.2cm}k \in \left[-k_F,3k_F\right]\hspace{0.2cm}{\rm and}\hspace{0.2cm}\gamma = -1
\nonumber \\
& = & (k - 2k_F)^2 - k_F^2\hspace{0.2cm}{\rm for}\hspace{0.2cm}k \in \left[-\infty,-k_F\right]\hspace{0.2cm}{\rm and}\hspace{0.2cm}\gamma = + 1
\nonumber \\
& = & k^2 - k_F^2\hspace{0.2cm}{\rm for}\hspace{0.2cm}k \in \left[3k_F,\infty\right]\hspace{0.2cm}{\rm and}\hspace{0.2cm}\gamma = + 1 \, .
\label{c+BLDCzero-m0}
\end{eqnarray}
For ${\cal{C}}\rightarrow\infty$ it reads,
\begin{eqnarray}
\omega_{c^+} (k) & = & \gamma\left((k-k_F)^2 - (2k_F)^2\right) 
\hspace{0.2cm}{\rm for}\hspace{0.2cm} k \in \left[-k_F,3k_F\right]\hspace{0.2cm}{\rm and}\hspace{0.2cm}\gamma = -1 
\nonumber \\
& & \hspace{0.60cm}{\rm and}\hspace{0.2cm}{\rm for}\hspace{0.2cm}k \in \left[-\infty,-k_F\right]\,;\hspace{0.2cm}
k \in \left[3k_F,\infty\right]\hspace{0.2cm}{\rm and}\hspace{0.2cm}\gamma = + 1 \, .
\label{c+BLDCinfinite-m0}
\end{eqnarray}
To derive the expressions of the spectrum in Eqs.  (\ref{c+BLDCzero-m0}) and (\ref{c+BLDCinfinite-m0})
those of the $c$ energy dispersion for ${\cal{C}}\rightarrow 0$ and ${\cal{C}}\rightarrow\infty$
in Eqs. (\ref{varepsilon-c-s-C0}) and (\ref{varepsilon-c-s-inf}) for $m=0$, respectively, have been used.

In the ${\cal{C}}\rightarrow 0$ limit the up-spin and down-spin one-fermion $c^+$ branch-line spectra, Eqs. (\ref{OkudRLAcc}) 
and (\ref{OkudLAcc}), respectively, read,
\begin{eqnarray}
\omega_{c^+}^{\sigma} (k) & = & - \left(k - {(1-\gamma_{\sigma})\over 2}(k_{F\uparrow}-k_{F\downarrow})\right)^2
+ k_{F\uparrow}^2 \hspace{0.2cm}{\rm for}\hspace{0.2cm}
k \in \left[-k_{F\sigma},-k_{F\downarrow}+ {(1-\gamma_{\sigma})\over 2}(k_{F\uparrow}-k_{F\downarrow})\right]
\hspace{0.2cm}{\rm and}\hspace{0.2cm}\gamma = -1
\nonumber \\
& = & - {1\over 2}(k-k_{F{\bar{\sigma}}})^2 + k_{F\uparrow}^2 + k_{F\downarrow}^2
 \hspace{0.2cm}{\rm for}\hspace{0.2cm}
k \in \left[-k_{F\downarrow}+ {(1-\gamma_{\sigma})\over 2}(k_{F\uparrow}-k_{F\downarrow}),k_{F{\bar{\sigma}}}+2k_{F\downarrow}\right]
\hspace{0.2cm}{\rm and}\hspace{0.2cm}\gamma = -1
\nonumber \\
& = & - (k-k_{F{\bar{\sigma}}}-k_{F\downarrow})^2 + k_{F\uparrow}^2 
\hspace{0.2cm}{\rm for}\hspace{0.2cm}
k \in \left[k_{F{\bar{\sigma}}}+2k_{F\downarrow},k_{F{\bar{\sigma}}}+2k_F\right]
\hspace{0.2cm}{\rm and}\hspace{0.2cm}\gamma = -1
\nonumber \\
& = & (k-k_{F{\bar{\sigma}}}+k_{F\downarrow})^2 - k_{F\uparrow}^2 
\hspace{0.2cm}{\rm for}\hspace{0.2cm}
k \in [-\infty,-k_{F\sigma}] 
\hspace{0.2cm}{\rm and}\hspace{0.2cm}\gamma = +1
\nonumber \\
& = & (k-k_{F{\bar{\sigma}}}-k_{F\downarrow})^2 - k_{F\uparrow}^2 
\hspace{0.2cm}{\rm for}\hspace{0.20cm}
k \in [(2k_F+k_{F\bar{\sigma}}),\infty]
\hspace{0.2cm}{\rm and}\hspace{0.2cm}\gamma = +1 \, .
\label{c+BLDCzero}
\end{eqnarray}
For ${\cal{C}}\rightarrow\infty$ they are given by,
\begin{eqnarray}
\omega_{c^+}^{\sigma} (k) & = & \gamma \left((k-k_{F{\bar{\sigma}}})^2 - (2k_F)^2\right)
\hspace{0.2cm}{\rm for}\hspace{0.2cm} k \in \left[-k_{F\sigma},k_{F{\bar{\sigma}}}+2k_F\right] 
\hspace{0.2cm}{\rm and}\hspace{0.2cm}\gamma = - 1
\nonumber \\
& & \hspace{0.60cm}{\rm and}\hspace{0.2cm}{\rm for}\hspace{0.2cm}k \in  [-\infty,-k_{F\sigma}]\,;\hspace{0.2cm}
k \in [(2k_F+k_{F\bar{\sigma}}),\infty]\hspace{0.2cm}{\rm and}\hspace{0.2cm}\gamma = + 1 \, .
\label{c+BLDCinfinite}
\end{eqnarray}
To derive the expressions of the spectra in Eqs.  (\ref{c+BLDCzero}) and (\ref{c+BLDCinfinite})
those of the $c$ and $s$ energy dispersions for ${\cal{C}}\rightarrow 0$ and ${\cal{C}}\rightarrow\infty$
in Eqs. (\ref{varepsilon-c-s-C0}) and (\ref{varepsilon-c-s-inf}), respectively, have been used.

In the ${\cal{C}}\rightarrow 0$ limit the spectrum, Eqs. (\ref{OkudRs-m0}) and (\ref{kqsup-m0}), is given by,
\begin{eqnarray}
\omega_s (k) & = & - k^2 + k_F^2 
\hspace{0.2cm}{\rm for}\hspace{0.2cm}k \in [0,k_F] \hspace{0.2cm}{\rm and}\hspace{0.2cm}\gamma = -1
\nonumber \\
& = & - (k - 2k_F)^2 + k_F^2\hspace{0.2cm}{\rm for}\hspace{0.2cm}k \in [k_F,3k_F]\hspace{0.2cm}{\rm and}\hspace{0.2cm}\gamma = + 1 \, .
\label{sBLDCzero-m0}
\end{eqnarray}
To derive the expressions of the spectrum in Eq. (\ref{sBLDCzero-m0}) and $\omega_s (k)=0$
for ${\cal{C}}\rightarrow\infty$ those of the $s$ energy dispersion for ${\cal{C}}\rightarrow 0$ and ${\cal{C}}\rightarrow\infty$
in Eqs. (\ref{varepsilon-c-s-C0}) and (\ref{varepsilon-c-s-inf}) for $m=0$, respectively, have been used.

In the ${\cal{C}}\rightarrow 0$ limit the spectrum, Eqs. (\ref{OkudRs}) and (\ref{kqsup}), is given by,
\begin{eqnarray}
\omega_s^{\uparrow} (k) & = & - \gamma\left((k-2k_F)^2 - k_{F\downarrow}^2\right)
\hspace{0.2cm}{\rm for}\hspace{0.2cm}k \in [k_{F\downarrow},(2k_F+k_{F\downarrow})] \hspace{0.2cm}{\rm and}\hspace{0.2cm}\gamma = \pm 1
\nonumber \\
& = & \gamma\left(k^2 - k_{F\downarrow}^2\right)\hspace{0.2cm}{\rm for}\hspace{0.2cm}k \in [0,k_{F\uparrow}] 
\hspace{0.2cm}{\rm and}\hspace{0.2cm}\gamma = \pm 1 \, ,
\label{sBLDCzero}
\end{eqnarray}
where the subdomains of these $k$ intervals that correspond to one-fermion addition $(\gamma = +1)$ and
removal $(\gamma = -1)$ are given in Eqs. (\ref{kupRLAs}) and (\ref{OkdownRLAsoth}).
To derive the expressions of the spectrum in Eq. (\ref{sBLDCzero}) and $\omega_s (k)=0$
for ${\cal{C}}\rightarrow\infty$ those of the $s$ energy dispersion for ${\cal{C}}\rightarrow 0$ and ${\cal{C}}\rightarrow\infty$
in Eqs. (\ref{varepsilon-c-s-C0}) and (\ref{varepsilon-c-s-inf}), respectively, have again been used.

\section{Additional information on the spectra of the one-fermion removal boundary lines}
\label{AIRBL}

The expansions of the boundary line spectra, Eqs. (\ref{BLSPD-m0}) and (\ref{BLSPD}), given
in this Appendix involve the derivatives of the $\beta = c,s$ band group velocities in Eq. (\ref{vcvs}), 
\begin{equation}
a_{\beta} (q) = {\partial v_{\beta} (q)\over\partial q} \hspace{0.2cm}{\rm where}\hspace{0.2cm}
\beta =c,s \, ,
\label{abetaq}
\end{equation}
Such quantities obey the following inequalities,
\begin{eqnarray}
a_{c} (q) & > & 0 \hspace{0.2cm}{\rm for}\hspace{0.2cm}q \in [-\infty,\infty]
\nonumber \\
a_{s} (q') & > & 0 \hspace{0.2cm}{\rm for}\hspace{0.2cm}q' \in ]-q_s^0,q_s^0[
\, ; \hspace{0.2cm} a_{s} (\pm q_s^0) = 0 \, ; \hspace{0.2cm}
a_{s} (q') <0 \hspace{0.2cm}{\rm for}\hspace{0.2cm}\vert q'\vert  \in ]q_s^0,k_{F\uparrow}]
\hspace{0.2cm}{\rm when}\hspace{0.2cm}m<0\hspace{0.2cm}{\rm and}\hspace{0.2cm}{\cal{C}}> 0
\nonumber \\
a_{s} (q') & > & 0 \hspace{0.2cm}{\rm for}\hspace{0.2cm}
q' \in[-k_{F\uparrow},k_{F\uparrow}] \hspace{0.2cm}{\rm when}\hspace{0.20cm}(i)\hspace{0.20cm}m=0
\hspace{0.2cm}{\rm for}\hspace{0.2cm}{\cal{C}}\in [0,\infty]
\hspace{0.2cm}{\rm and}\hspace{0.20cm}(ii)\hspace{0.20cm}{\cal{C}}\rightarrow 0  
\hspace{0.2cm}{\rm for}\hspace{0.2cm}m\in [0,n]  \, .
\label{Ineacas}
\end{eqnarray}
Here $q_s^0$ is defined by the equation $a_{s} (q_s^0) =0$.

For excitation momentum $k$ in the vicinity and just above $k_{min}$ the small-$(k -k_{min})$
expansion of the boundary line spectrum, Eq. (\ref{BLSPD-m0}), reads,
\begin{eqnarray}
\omega_{BL} (k) & = & - \varepsilon_c (q_c^0) + v_s (k_{F})\,(k -k_{min})
-  {a_s (k_{F})\,a_c (q_c^0)\over 2(a_s (k_{F}) + a_c (q_c^0))}(k -k_{min})^2
\nonumber \\
& = &  - \varepsilon_c (q_c^0) + v_s (k_{F})\,(k + q_c^0 - k_{F} )
- {a_s (k_{F})\,a_c (q_c^0)\over 2(a_s (k_{F}) + a_c (q_c^0))}(k + q_c^0 - k_{F} )^2 \, .
\nonumber
\end{eqnarray}
Moreover, for $k$ near and both just below and above $2k_F$ it has the following behavior,
\begin{equation}
\omega_{BL} (k) = - \varepsilon_c (0)  - \varepsilon_s (0) 
-  {a_s (0)\,a_c (0)\over 2(a_s (0) + a_c (0))}(k -2k_F)^2 \, ,
\nonumber
\end{equation}
whereas for $k$ near and just below $k_{max}$ it is given by,
\begin{eqnarray}
\omega_{BL} (k) & = & - \varepsilon_c (q_c^0) - v_s (k_{F})\,(k -k_{max})
-  {a_s (k_{F})\,a_c (q_c^0)\over 2(a_s (k_{F}) + a_c (q_c^0))}(k -k_{max})^2
\nonumber \\
& = & - \varepsilon_c (q_c^0) - v_s (k_{F})\,(k -2k_F - q_c^0 - k_{F})
-  {a_s (k_{F})\,a_c (q_c^0)\over 2(a_s (k_{F}) + a_c (q_c^0))}(k -2k_F - q_c^0 - k_{F})^2 \, .
\nonumber
\end{eqnarray}
Here $a_{c} (q)$ and $a_{s} (q')$ are the derivatives of the $\beta = c,s$ band group velocities,
Eq. (\ref{abetaq}), and $-\varepsilon_c (0)>0$ and $-\varepsilon_s (0)>0$ are the energy bandwidths of 
the $c$ and $s$ band occupied Fermi seas defined in Eq. (\ref{Wc-expr})
and Eqs. (\ref{Wph-lim}) and (\ref{Ws-exprp}), respectively, 
of Appendix \ref{UBAQ} for $m=0$.

The reference $c$ band momentum $q_c^0$ and related momentum values $k_{min}$ and $k_{max}$
defined in Eq. (\ref{kminkmax-m0}) appearing in the above expressions have at zero spin density the following limiting values,
\begin{eqnarray}
\lim_{{\cal{C}}\rightarrow 0}q_c^0 & = & 2k_F\hspace{0.2cm}{\rm and}\hspace{0.2cm}
\lim_{{\cal{C}}\rightarrow\infty}q_c^0  = 0
\nonumber \\ 
\lim_{{\cal{C}}\rightarrow 0}k_{min} & = & - k_F
\hspace{0.2cm} {\rm and}\hspace{0.2cm}
\lim_{{\cal{C}}\rightarrow\infty}k_{min} = k_F
\nonumber \\ 
\lim_{{\cal{C}}\rightarrow 0}k_{max} & = & 5k_F
\hspace{0.2cm} {\rm and}\hspace{0.2cm}
\lim_{{\cal{C}}\rightarrow\infty}k_{max} = 3k_F \, .
\label{qkklimits-m0}
\end{eqnarray}

In the ${\cal{C}}\rightarrow 0$ limit the boundary line spectrum, Eq. (\ref{BLSPD-m0}), is given by,
\begin{eqnarray}
\omega_{BL} (k) & = & - {1\over 3}\,(k-2k_F)^2 + 3\,(k_F)^2
\hspace{0.2cm}{\rm for}\hspace{0.2cm}
k \in \left[-k_F,5k_F\right] \, ,
\nonumber
\end{eqnarray}
where the expression of the $c$ and $s$ energy dispersions for ${\cal{C}}\rightarrow 0$, 
Eq. (\ref{varepsilon-c-s-C0}) for $m=0$, have been used.

For ${\cal{C}}\rightarrow\infty$ such a boundary line spectrum reads,
\begin{equation}
\omega_{BL} (k) = (2k_F)^2 
\hspace{0.2cm}{\rm for}\hspace{0.2cm} k \in \left[k_F,3k_F\right] \, ,
\nonumber
\end{equation}
where the expression of the $c$ and $s$ energy dispersions for ${\cal{C}}\rightarrow\infty$, 
Eq. (\ref{varepsilon-c-s-inf}) for $m=0$, have been used.

The velocity relative to the physical excitation momentum $k$ 
of the one-fermion removal boundary line spectrum, Eq. (\ref{BLSPD-m0}), is given by,
\begin{equation}
{\partial\omega_{BL} (k)\over\partial k} = v_c (q_c^{BL}) = v_s (q_s^{BL}) \, .
\nonumber
\end{equation}
Particular reference values of that velocity read,
\begin{equation}
{\partial\omega_{BL} (k)\over\partial k}\vert_{k=k_{min}} = v_s (k_F) \, ; \hspace{0.2cm}
{\partial\omega_{BL} (k)\over\partial k}\vert_{k=2k_F} = 0 \, ; \hspace{0.2cm}
{\partial\omega_{BL} (k)\over\partial k}\vert_{k=k_{max}} = - v_s (k_F) \, .
\nonumber
\end{equation}

The following expansions of the up-spin and down-spin boundary line spectra, Eq. (\ref{BLSPD}), involve
both the $\beta = c,s$ group velocities and their derivatives $a_{\beta} (q)$, Eq. (\ref{abetaq}).
For excitation momentum $k$ near and just above $k_{min}^{\uparrow}$ the corresponding
small-$(k -k_{min}^{\uparrow})$ expansion of the up-spin boundary 
line spectrum, Eq. (\ref{BLSPD}) for $\sigma =\uparrow$, reads,
\begin{eqnarray}
\omega_{BL}^{\uparrow} (k) & = & - \varepsilon_c (q_c^0) + v_s (k_{F\downarrow})\,(k -k_{min}^{\uparrow})
-  {a_s (k_{F\downarrow})\,a_c (q_c^0)\over 2(a_s (k_{F\downarrow}) - a_c (q_c^0))}(k -k_{min}^{\uparrow})^2
\nonumber \\
& = &  - \varepsilon_c (q_c^0) + v_s (k_{F\downarrow})\,(k + q_c^0 - k_{F\downarrow})
- {a_s (k_{F\downarrow})\,a_c (q_c^0)\over 2(a_s (k_{F\downarrow}) - a_c (q_c^0))}(k + q_c^0 - k_{F\downarrow})^2 \, ,
\nonumber
\end{eqnarray}
whereas for $k$ near and just below $k_{max}^{\uparrow}$ it is given by,
\begin{equation}
\omega_{BL}^{\uparrow} (k) = - \varepsilon_c (0) + \varepsilon_{s} (k_{F\uparrow})
- {a_s (k_{F\uparrow})\,a_c (0)\over 2(a_s (k_{F\uparrow}) - a_c (0))}(k -k_{F\uparrow})^2 \, .
\nonumber
\end{equation}

For excitation momentum $k$ in the vicinity and just above $k_{min}^{\downarrow}$ the 
expansion of the down-spin boundary line spectrum, Eq. (\ref{BLSPD}) for $\sigma =\downarrow$, reads,
\begin{eqnarray}
\omega_{BL}^{\downarrow} (k) & = & - \varepsilon_c (q_c^0) + v_s (k_{F\downarrow})\,(k -k_{min}^{\downarrow})
-  {a_s (k_{F\downarrow})\,a_c (q_c^0)\over 2(a_s (k_{F\downarrow}) + a_c (q_c^0))}(k -k_{min}^{\downarrow})^2
\nonumber \\
& = &  - \varepsilon_c (q_c^0) + v_s (k_{F\downarrow})\,(k + q_c^0 - k_{F\uparrow} )
- {a_s (k_{F\downarrow})\,a_c (q_c^0)\over 2(a_s (k_{F\downarrow}) + a_c (q_c^0))}(k + q_c^0 - k_{F\uparrow} )^2 \, .
\nonumber
\end{eqnarray}
Moreover, for $k$ near and both just below and above $2k_F$ it has the following behavior,
\begin{equation}
\omega_{BL}^{\downarrow} (k) = - \varepsilon_c (0)  - \varepsilon_s (0) 
-  {a_s (0)\,a_c (0)\over 2(a_s (0) + a_c (0))}(k -2k_F)^2 \, ,
\nonumber
\end{equation}
whereas for $k$ near and just below $k_{max}^{\downarrow}$ the down-spin boundary line spectrum 
expansion is given by,
\begin{eqnarray}
\omega_{BL}^{\downarrow} (k) & = & - \varepsilon_c (q_c^0) - v_s (k_{F\downarrow})\,(k -k_{max}^{\downarrow})
-  {a_s (k_{F\downarrow})\,a_c (q_c^0)\over 2(a_s (k_{F\downarrow}) + a_c (q_c^0))}(k -k_{max}^{\downarrow})^2
\nonumber \\
& = & - \varepsilon_c (q_c^0) - v_s (k_{F\downarrow})\,(k -2k_F - q_c^0 - k_{F\downarrow})
-  {a_s (k_{F\downarrow})\,a_c (q_c^0)\over 2(a_s (k_{F\downarrow}) + a_c (q_c^0))}(k -2k_F - q_c^0 - k_{F\downarrow})^2 \, .
\nonumber
\end{eqnarray}

The limiting values of the $\sigma=\uparrow,\downarrow$ intervals $k\in [k_{min}^{\sigma},k_{max}^{\sigma}]$ read,
\begin{eqnarray}
k_{min}^{\uparrow} & = & - q_c^0 + k_{F\downarrow} \in [-k_{F\downarrow},k_{F\downarrow}]
\nonumber \\
k_{max}^{\uparrow} & = & k_{F\uparrow}\hspace{0.2cm}{\rm when}\hspace{0.2cm}m<0\hspace{0.2cm}{\rm and}\hspace{0.2cm}{\cal{C}}> 0
\nonumber \\
k_{max}^{\uparrow} & = & - q_c^{max} + k_{F\uparrow} \in [-k_{F\downarrow},k_{F\uparrow}] \hspace{0.2cm}{\rm when}\hspace{0.20cm}(i)\hspace{0.20cm}m=0\hspace{0.2cm}{\rm for}\hspace{0.2cm}{\cal{C}}\in [0,\infty]
\hspace{0.2cm}{\rm and}\hspace{0.20cm}(ii)\hspace{0.20cm}{\cal{C}}\rightarrow 0
\hspace{0.2cm}{\rm for}\hspace{0.2cm}m\in [0,n] \, .
\nonumber \\
k_{min}^{\downarrow} & = & 2k_F + k_{F\downarrow} + q_c^0 \in [2k_F + k_{F\downarrow},2k_F + 3k_{F\downarrow}]
\nonumber \\
k_{max}^{\downarrow} & = & 4k_F-k_{min}^{\downarrow} = k_{F\uparrow} - q_c^0 \in [k_{F\uparrow}-2k_{F\downarrow},k_{F\uparrow}] \, .
\label{kminkmax}
\end{eqnarray}

At finite spin density that $c$ band momentum $q_c^0$ and the related momentum values $k_{min}^{\sigma}$ and $k_{max}^{\sigma}$ 
given in Eq. (\ref{kminkmax}) appearing in the above expressions
have the following limiting values where $\sigma$ is the projection that the interval 
$k\in [k_{min}^{\sigma},k_{max}^{\sigma}]$ when specified and otherwise the limiting values refer to both 
$\sigma=\uparrow,\downarrow$,
\begin{eqnarray}
\lim_{{\cal{C}}\rightarrow 0}q_c^0 & = & 2k_{F\downarrow}\hspace{0.2cm} {\rm and}\hspace{0.2cm}
\lim_{{\cal{C}}\rightarrow\infty}q_c^0 = 0
\nonumber \\
\lim_{{\cal{C}}\rightarrow 0}q_c^{max} & = & 2k_F\hspace{0.2cm} {\rm and}\hspace{0.2cm}
\lim_{{\cal{C}}\rightarrow\infty}q_c^{max} = 0\hspace{0.2cm} {\rm for}\hspace{0.2cm}\sigma = \uparrow \, ,
\nonumber \\
\lim_{{\cal{C}}\rightarrow 0}k_{min}^{\uparrow} & = & - k_{F\downarrow}
\hspace{0.2cm} {\rm and}\hspace{0.2cm}
\lim_{{\cal{C}}\rightarrow\infty}k_{min}^{\uparrow} = k_{F\downarrow} 
\nonumber \\ 
\lim_{{\cal{C}}\rightarrow 0}k_{max}^{\uparrow} & = & k_{F\uparrow} \hspace{0.2cm}{\rm when}
\hspace{0.2cm}m<0\hspace{0.2cm}{\rm and}\hspace{0.2cm}{\cal{C}}> 0
\nonumber \\ 
\lim_{{\cal{C}}\rightarrow 0}k_{max}^{\uparrow} & = & - k_{F\downarrow}\hspace{0.2cm}{\rm when}
\hspace{0.20cm}(i)\hspace{0.20cm}m=0\hspace{0.2cm}{\rm for}\hspace{0.2cm}{\cal{C}}\in [0,\infty]
\hspace{0.2cm}{\rm and}\hspace{0.20cm}(ii)\hspace{0.20cm}{\cal{C}}\rightarrow 0
\hspace{0.2cm}{\rm for}\hspace{0.2cm}m\in [0,n]
\nonumber \\ 
\lim_{{\cal{C}}\rightarrow\infty}k_{max}^{\uparrow} & = & k_{F\uparrow}
\nonumber \\ 
\lim_{{\cal{C}}\rightarrow 0}k_{min}^{\downarrow} & = & k_{F\uparrow} - 2k_{F\downarrow}
\hspace{0.2cm} {\rm and}\hspace{0.2cm}
\lim_{{\cal{C}}\rightarrow\infty}k_{min}^{\downarrow} = k_{F\uparrow}
\nonumber \\ 
\lim_{{\cal{C}}\rightarrow 0}k_{max}^{\downarrow} & = & 2k_F + 3k_{F\downarrow}
\hspace{0.2cm} {\rm and}\hspace{0.2cm}
\lim_{{\cal{C}}\rightarrow\infty}k_{max}^{\downarrow} = 2k_F + k_{F\downarrow} \, .
\label{qkklimits}
\end{eqnarray}

The reference $c$ band momentum values $q_c^0$ and $q_c^{max}$ appearing in Eq. (\ref{kminkmax})
are related to reference $s$ band momentum values $q_s^0$ and $q_s^1$, the latter existing
only for $\sigma =\uparrow$. Those reference $c$ and $s$ band reference momentum values are
defined by the following velocities and $s$ band velocity derivative
$a_{s} (q') = \partial v_{\beta} (q')/\partial q'$ relations where
$\sigma$ is projection that the interval $k\in [k_{min}^{\sigma},k_{max}^{\sigma}]$ refers to,
\begin{eqnarray}
v_c (q_c^0) & = & v_s (k_{F\downarrow}) \hspace{0.2cm}{\rm for}\hspace{0.2cm}\sigma =\uparrow,\downarrow
\nonumber \\
v_c (q_c^0) & = & v_s (q_s^1) \hspace{0.2cm}{\rm and}\hspace{0.2cm}a_s (q_s^1) < 0
\hspace{0.2cm}{\rm for}\hspace{0.2cm}\sigma =\uparrow
\hspace{0.2cm}{\rm when}\hspace{0.2cm}m<0\hspace{0.2cm}{\rm and}\hspace{0.2cm}{\cal{C}}> 0
\nonumber \\
v_c (q_c^{max}) & = & v_s (q_s^0) \hspace{0.2cm}{\rm and}\hspace{0.2cm}a_s (q_s^0) = 0
\hspace{0.2cm}{\rm for}\hspace{0.2cm}\sigma =\uparrow
\hspace{0.2cm}{\rm when}\hspace{0.2cm}m<0\hspace{0.2cm}{\rm and}\hspace{0.2cm}{\cal{C}}> 0
\nonumber \\
v_c (q_c^{max}) & = & v_s (k_{F\uparrow}) 
\hspace{0.2cm}{\rm for}\hspace{0.2cm}\sigma =\uparrow
\hspace{0.2cm}{\rm when}\hspace{0.20cm}(i)\hspace{0.20cm}m=0
\hspace{0.2cm}{\rm for}\hspace{0.2cm}{\cal{C}}\in [0,\infty]
\hspace{0.2cm}{\rm and}\hspace{0.20cm}(ii)\hspace{0.20cm}{\cal{C}}\rightarrow 0
\hspace{0.2cm}{\rm for}\hspace{0.2cm}m\in [0,n]
\nonumber \\
v_c (q_c^{max}) & = & v_s (k_{F\uparrow}) 
\hspace{0.2cm}{\rm for}\hspace{0.2cm}\sigma =\downarrow \, ,
\nonumber
\end{eqnarray}
where $k_{F\downarrow}<q_s^0<q_s^1<k_{F\uparrow}$.

In the ${\cal{C}}\rightarrow 0$ limit the boundary line spectra, Eq. (\ref{BLSPD}), read,
\begin{eqnarray}
\omega_{BL}^{\uparrow} (k) & = & k_{F\uparrow}^2 - k_{F\downarrow}^2
\hspace{0.2cm}{\rm at}\hspace{0.2cm}k = - k_{F\downarrow}
\nonumber \\
\omega_{BL}^{\downarrow} (k) & = & - {1\over 3}\,(k-2k_F)^2 +
k_{F\uparrow}^2 + 2\,(k_{F\downarrow})^2
\hspace{0.2cm}{\rm for}\hspace{0.2cm}
k \in \left[2k_F - 3k_{F\downarrow},2k_F + 3k_{F\downarrow}\right] \, .
\label{BLspecCzero}
\end{eqnarray}
Note that the up-spin one-fermion removal boundary line collapses in the ${\cal{C}}\rightarrow 0$ limit 
to a single $(k,\omega)$-plane point.

For ${\cal{C}}\rightarrow\infty$ such boundary line spectra are given by,
\begin{eqnarray}
\omega_{BL}^{\uparrow} (k) & = & (2k_F)^2 
\hspace{0.2cm}{\rm for}\hspace{0.2cm} k \in \left[k_{F\downarrow},k_{F\uparrow}\right] 
\nonumber \\
\omega_{BL}^{\downarrow} (k) & = & (2k_F)^2 
\hspace{0.2cm}{\rm for}\hspace{0.2cm} k \in \left[k_{F\uparrow},2k_F + k_{F\downarrow}\right] \, .
\label{BLspecCinfinite}
\end{eqnarray}
To obtain the expressions of the spectra in Eqs.  (\ref{BLspecCzero}) and (\ref{BLspecCinfinite})
those of the $c$ and $s$ energy dispersions for ${\cal{C}}\rightarrow 0$ and ${\cal{C}}\rightarrow\infty$
in Eqs. (\ref{varepsilon-c-s-C0}) and (\ref{varepsilon-c-s-inf}), respectively, have been used.

The velocities relative to the physical excitation momentum $k$ 
of the boundary lines spectra, Eq. (\ref{BLSPD}), read,
\begin{equation}
{\partial\omega_{BL}^{\sigma} (k)\over\partial k} = v_c (q_c^{BL}) = v_s (q_s^{BL}) \, .
\nonumber
\end{equation}
Specific reference values of those velocities are given by,
\begin{eqnarray}
{\partial\omega_{BL}^{\sigma} (k)\over\partial k}\vert_{k=k_{min}^{\sigma}} & = & v_s (k_{F\downarrow}) \, ;
\hspace{0.2cm}
{\partial\omega_{BL}^{\uparrow} (k)\over\partial k}\vert_{k=k_{F\uparrow}} = 0
\hspace{0.2cm}{\rm when}\hspace{0.2cm}m<0\hspace{0.2cm}{\rm and}\hspace{0.2cm}{\cal{C}}> 0
\nonumber \\
{\partial\omega_{BL}^{\downarrow} (k)\over\partial k}\vert_{k=2k_F} & = & 0 \, ;
\hspace{0.2cm}
{\partial\omega_{BL}^{\downarrow} (k)\over\partial k}\vert_{k=k_{max}^{\downarrow}} = - v_s (k_{F\downarrow}) \, .
\nonumber
\end{eqnarray}

Related energy scales read,
\begin{eqnarray}
\omega_{BL}^{\uparrow} (k_{F\uparrow}) & = & - \varepsilon_c (0) + \varepsilon_s (k_{F\uparrow}) =
- \varepsilon_c (0) + 2\mu_B\,h
\hspace{0.2cm}{\rm when}\hspace{0.2cm}m<0\hspace{0.2cm}{\rm and}\hspace{0.2cm}{\cal{C}}> 0 
\nonumber \\
\omega_{BL}^{\downarrow} (2k_F) & = & - \varepsilon_c (0) - \varepsilon_s (0) \, ,
\nonumber
\end{eqnarray}
where $-\varepsilon_c (0)>0$ and $-\varepsilon_s (0)>0$ are the energy bandwidths of 
the $c$ and $s$ band occupied Fermi seas, respectively, and the magnetic energy $2\mu_B\,h = \varepsilon_{s} (k_{F\uparrow})$,
Eq. (\ref{mu-muBH0}), equals the energy bandwidth of the $s$ band unoccupied Fermi sea.
Such energy bandwidths are defined in Eqs. (\ref{Wc-expr})-(\ref{Ws-ph}) of Appendix \ref{UBAQ}.

\end{document}